\documentclass[prb,twocolumn,superscriptaddress,letterpaper]{revtex4}

\usepackage{amsmath,amsfonts}
\usepackage{graphicx}
\usepackage{color}
\usepackage{enumerate}

\newcommand{\e}{\mathrm{e}}
\newcommand{\be}{\mathbf{e}}
\newcommand{\bI}{\mathbf{I}}
\newcommand{\bS}{\mathbf{S}}
\newcommand{\bh}{\mathbf{h}}
\newcommand{\br}{\mathbf{r}}
\newcommand{\bB}{\mathbf{B}}
\newcommand{\mean}[1]{\langle #1 \rangle}
\newcommand{\bmean}[1]{\bigl\langle #1 \bigr\rangle}
\newcommand{\bracket}[2]{\langle #1 | #2 \rangle}

\newcommand{\up}{\uparrow}
\newcommand{\dw}{\downarrow}

\begin{document}


\title{
Nuclear magnetism and electron order in 
interacting one-dimensional conductors
}
\author{Bernd Braunecker}
\affiliation{Department of Physics, University of Basel, 
             Klingelbergstrasse 82, 4056 Basel, Switzerland}

\author{Pascal Simon}
\affiliation{Laboratoire de Physique des Solides, CNRS UMR-8502,
             Universit\'{e} de Paris Sud, 91405 Orsay Cedex, France}

\author{Daniel Loss}
\affiliation{Department of Physics, University of Basel, 
             Klingelbergstrasse 82, 4056 Basel, Switzerland}

\date{\today}



\begin{abstract}
The interaction between localized magnetic moments and the electrons of 
a one-dimensional conductor can lead to an ordered phase in which the magnetic moments
and the electrons are tightly bound to each other. 
We show here that this occurs when a lattice of nuclear spins is
embedded in a Luttinger liquid.
Experimentally available examples of such a system are single wall carbon nanotubes
grown entirely from $^{13}$C and GaAs-based quantum wires. In these systems the 
hyperfine interaction between the nuclear spin and the conduction electron spin is very weak, 
yet it triggers a strong feedback reaction that results in an ordered phase consisting
of a nuclear helimagnet that is inseparably bound to an electronic density wave combining
charge and spin degrees of freedom. This effect can be interpreted as a 
strong renormalization of the nuclear Overhauser field and is a unique signature of 
Luttinger liquid physics.
Through the feedback the order persists up into the millikelvin 
range. A particular signature is the reduction of the electric 
conductance by the universal factor 2.
\end{abstract}


\maketitle


\section{Introduction}

The interaction between localized magnetic moments and delocalized electrons
contains the essential physics of many modern condensed matter systems. 
It is on the basis of nuclear magnets,\cite{froehlich:1940} heavy fermion materials
of the Kondo-lattice type,\cite{tsunetsugu:1997} and
ferromagnetic semiconductors.\cite{ohno:1992,ohno:1998,dietl:1997,koenig:2000}
In this work we focus on the interplay between strong electron-electron interactions
and the magnetic properties of the localized moments. 
Low-dimensional electron conductors are ideal systems to examine this physics:
The nuclear spins of the ions of the crystal (or suitably substituted isotopes)
form a lattice of localized moments;
these spins couple to the conduction electron spin through the hyperfine interaction;
and the confinement of the electrons in a low-dimensional structure enhances the 
importance of the electron-electron interactions.

In previous work we have studied the magnetic properties of the nuclear spins 
embedded in a two-dimensional (2D) electron gas of a GaAs heterostructure,\cite{simon:2007,simon:2008}
and in $^{13}$C substituted single-wall carbon nanotubes\cite{braunecker:2009} (SWNTs)
as a specific example of a one-dimensional (1D) conductor.
In this work we focus on 1D more generally than in Ref. \onlinecite{braunecker:2009}:
We cover not only the case of SWNTs but also of GaAs-based quantum wires or 
different (yet not in detail discussed) 1D conductors, under the assumption that 
the electrons are in the Luttinger liquid (LL) state as a result of their interactions.
In these systems the coupling between the nuclear spins and the conduction electrons
has remarkable consequences.

\begin{figure*}
	\includegraphics[width=1.5\columnwidth]{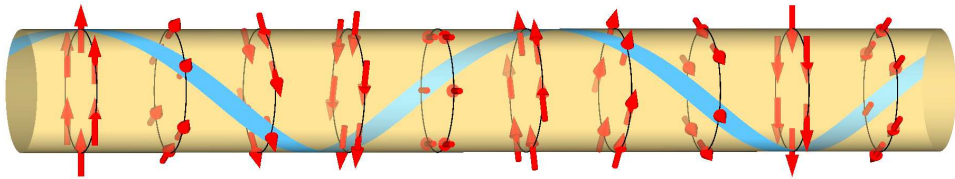}
	\caption{%
	    Illustration of the helical nuclear magnetism 
        of the single wall $^{13}$C nanotube (SWNT), which is triggered by the RKKY 
		interaction over the electron system (not shown). 
        The nuclear spins (red arrows) order ferromagnetically on a cross-section 
        of the SWNT and rotate along the SWNT axis with a period $\pi/k_F = \lambda_F/2$ in the
        spin $xy$ plane (chosen here arbitrarily orthogonal to the SWNT axis).
		The blue ribbon is a guide to the eye for the helix.
        The feedback of this nuclear magnetic field strongly renormalizes the electron
		system through the opening of a partial gap, due to a strongly renormalized Overhauser field, 
		and so modifies the RKKY interaction.
		Through this strong coupling of electron and nuclear systems the combined order
		persists up into the millikelvin range. As a particular consequence the electric conductance
        of the SWNT is reduced by a factor of precisely 2.
        \label{fig:nanotube}}
\end{figure*}

\begin{figure*}
	\includegraphics[width=1.9\columnwidth]{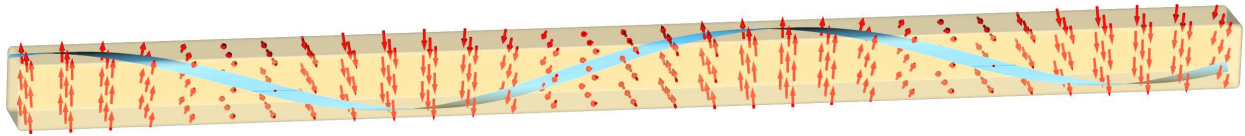}
	\caption{%
		Illustration of the helical nuclear magnetism for a GaAs-based quantum wire. 
		Compared with the SWNT (Fig. \ref{fig:nanotube}) the number of ferromagnetically locked
		spins on a cross-section through the wire is much larger, and the wavelength $\pi/k_F$ of the 
		helical rotation along the wire is longer.
		The feedback effect remains otherwise the same and the combined electron and 
		nuclear order persists into the millikelvin range as well.
        \label{fig:quantumwire}}
\end{figure*}

Indeed, below a cross-over temperature $T^*$ (in the millikelvin 
range for the considered systems) the nuclear spins form a spiral, 
a helimagnet (see Figs. \ref{fig:nanotube} and \ref{fig:quantumwire}), 
caused by the effective Ruderman-Kittel-Kasya-Yoshida (RKKY) interaction induced by the electron system. 
The ordered nuclear spins create an Overhauser field that acts back on the electron spins.
This feedback is essential: It enhances an instability of the electron conductor
toward a density wave order, and the electronic states are restructured.
A gap appears in one half of the low-energy modes and leads to
a partial electron spin polarization that follows the nuclear spin helix.
The gap can be interpreted as a strong renormalization
of the Overhauser field, and so as a strong renormalization of the hyperfine coupling 
constant for the gapped collective electron modes.
The remaining gapless electron modes in turn further strengthen the RKKY coupling between
the nuclear spins. The transition temperature $T^*$ of the nuclear spins can therefore 
lie much above the temperature that would be found without the feedback (called $T_0^*$ below).
In SWNTs, for instance, the feedback leads to an enhancement of $T^*$ by about four orders of magnitude.

This means that below $T^*$ there is a temperature range where the nuclear order and 
the electron order exist only through their mutual stabilization.
The nuclear spins and the electrons form a combined ordered phase, 
even though the energy and time scales in both systems differ by orders of magnitude.
We remark that the order is unstable in the thermodynamic limit due to 
long-wavelength fluctuations. For any realistic sample length $L$, however, 
those fluctuations are cut off, the order extends over the whole system, and 
$T^*$ is in fact independent of $L$.

We discuss this physics here specifically for $^{13}$C SWNTs and GaAs quantum 
wires because such systems have become available for experiments recently:
SWNTs with a purity of $^{13}$C up to 99\% have been reported in Refs.
\onlinecite{simon_f:2004,ruemmeli:2007,churchill:2009a,churchill:2009b}.
The cleaved edge overgrowth method\cite{pfeiffer:1997,auslaender:2002} 
has made it furthermore possible to produce quantum wires on the edge of 
a GaAs heterostructure with LL parameters as low as\cite{steinberg:2008}
$K_c = 0.4 - 0.5$.
For both systems, we predict a feedback-generated cross-over temperature 
$T^*$ that lies in the range of 10 -- 100 mK.

This work is related to several studies found in the literature:
NMR experiments\cite{singer:2005} on $^{13}$C 
enriched SWNTs grown inside regular SWNTs\cite{simon_f:2004,ruemmeli:2007}
have revealed the existence of a large gap of about 30 K
in the spin response. 
While the microscopic origin of this gap seems still be unresolved, 
the NMR response could be well modeled with a partially gapped 
Tomonaga-Luttinger model.\cite{dora:2007,dora:2008}
Interestingly our microscopic theory predicts a spin excitation gap 
for a part of the transverse electronic susceptibility, although we 
consider isolated SWNTs and obtain a gap with a smaller 
magnitude below 1 K.

The coupling between a quantum spin chain and a LL was 
studied in Ref. \onlinecite{zachar:1996}, and it was shown that this system
can acquire gaps as well. Such a system is very different 
from our model in that it involves a single chain of small quantum spins with anisotropic 
coupling to the electrons. Such an anisotropy appears spontaneously in our case,
and built-in anisotropy has a very different effect as discussed in Sec. \ref{sec:anisotropic}.
A spin gap also appears in a LL in the presence of Rashba spin-orbit
interactions.\cite{gritsev:2005,gangadharaiah:2008}
LLs with a gap in the spin sector are known as Luther-Emery liquids,\cite{luther:1974} 
and the partially gapped LL in our model has indeed a strong resemblance to such a system.
Yet the gap opens not only in the spin sector but involves a combination of 
electronic spin and 
charge degrees of freedom, therefore, in addition breaks the usual spin-charge separation of a LL.
The RKKY interaction at zero temperature 
was calculated for LLs in Ref. \onlinecite{egger:1996}
and for the case including Rashba spin-orbit interactions in Ref. \onlinecite{schulz:2009}.
The use of the hyperfine interaction of $^{13}$C to couple spin and valley 
quantum numbers in carbon-based quantum dots was explored in Ref. \onlinecite{palyi:2009}.

Very recent spin blockade measurements\cite{churchill:2009a} on quantum dots formed by 
$^{13}$C SWNTs suggest that the hyperfine interaction constant $A_0$ is by $10^2$ -- $10^3$
larger than what is expected from $^{60}$C data\cite{pennington:1996} or
band structure theory.\cite{fischer:2009}
However, this interaction strength is inferred from the comparison with models that were
originally designed for GaAs quantum dots, and so the precise value of $A_0$
requires further investigation.\cite{trauzettel:2009}

The observation of LL physics has been reported for various 
1D conductors such as carbon nanotubes,\cite{bockrath:1999,yao:1999,bachtold:2001}
GaAs quantum wires,\cite{auslaender:2002,tserkovnyak:2002,tserkovnyak:2003,auslaender:2005,steinberg:2008,jompol:2009}
bundles of NbSe$_3$ nanowires,\cite{slot:2004}
polymer nanofibers,\cite{aleshin:2004}
atomic chains on insulating substrates,\cite{segovia:1999}
MoSe nanowires,\cite{venkataraman:2006}
fractional quantum Hall edge states,\cite{chang:1996}
and very recently in a bulk material, conjugated polymers at high 
carrier densities.\cite{yuen:2009}
If there is a coupling to localized magnetic moments, 
we expect that the effect described in this paper should be detectable 
in these systems as well,
with the exception of the chiral LLs of fractional quantum Hall edge states
because they lack the backscattering between left and right moving modes that is crucial for the
effect. To overcome this restriction, two edges with counterpropagating modes would have
to be brought close together by a constriction.

Recently much progress has been made in producing and tuning the properties of 
carbon nanotubes as quantum wires or quantum 
dots,\cite{tans:1997,bockrath:1997,bockrath:1999,kong:2000,minot:2004,jarillo:2004,mason:2004,simon_f:2004,%
biercuk:2005,cao:2005,sapmaz:2006,onac:2006,cottet:2006,graeber:2006a,graeber:2006b,jorgensen:2006,%
meyer:2007,ruemmeli:2007,deshpande:2008,kuemmeth:2008,churchill:2009a,churchill:2009b,steele:2009}
and ultraclean SWNTs are now available.\cite{kuemmeth:2008,steele:2009}

The outline of the paper is as follows: 
In the next section we state the conditions for the discussed physics and
present the main results. A detailed account of the theory is then given:
In Sec. \ref{sec:model} we derive the effective model.
The nuclear order and its stability without the feedback is discussed in 
Sec. \ref{sec:nucl_order}. The feedback and its consequences are examined in 
Sec. \ref{sec:feedback}. 
In Sec. \ref{sec:self_consistency} we discuss the self-consistency of the theory.
The effect of the renormalization above the cross-over temperature is 
outlined in Sec. \ref{sec:above_T^*}.
In Sec. \ref{sec:nanotubes} we show that the single-band description we 
have used in the preceding sections is appropriate for SWNTs, which 
normally require a two-band model. 
We shortly conclude in Sec. \ref{sec:conclusions}.
The Appendices contain the technical details.
The numerical parameters we use and derive for the SWNTs and GaAs quantum wires are listed
in Table \ref{tab:values}. 
For a brief overview, we refer the reader to Sec. \ref{sec:results}.


\section{Conditions and main results}
\label{sec:results}

We summarize in this section the conditions and the main results of our work.
This allows us to give an overview of the physics to the reader without going 
into the technical details and conceptual subtleties. These are then discussed in the 
subsequent sections.

Two conditions for the described physics are important: 
A 1D electric conductor in the LL state confined in a single transverse mode (in the 
directions perpendicular to the 1D conductor axis),
and a three-dimensional (3D) nuclear spin lattice embedded in this
1D conductor. Higher transverse modes are split off by a large energy gap $\Delta_t$.
The coupling of the nuclear spins to the electrons is
weighted by the transverse mode, which eventually leads to a ferromagnetic
locking of the $N_\perp \gg 1$ nuclear spins on a cross-section in the transverse direction.
Consequently these ferromagnetically locked nuclear spins behave as a single effective 
large spin, allowing us to use a semiclassical description 
(see Sec. \ref{sec:confinement_1D} and \ref{sec:interpretation_NI}).
This picture can be different for strongly anisotropic systems, where 
the coupling to the electron spin favors a different spin locking 
(see Sec. \ref{sec:anisotropic}). 
However, the physics described here remains valid as long as this locked
configuration has a nonzero average magnetization.

In addition, we treat only systems in the RKKY (Ref. \onlinecite{RKKY})
regime, which is indeed the natural limit for the electron--nuclear spin coupling. 
This regime is characterized by energetics such that the
characteristic time scales between the slow nuclear and the fast electron dynamics decouple.
This makes it possible to derive an effective instantaneous interaction between the nuclear 
spins, the RKKY interaction, which is transmitted by the electron gas.
If $A_0$ is the hyperfine coupling constant between a nuclear spin and an electron
state localized at the nuclear spin site, and if $E_F$ denotes the typical energy 
scale of the electron system, the RKKY regime is determined by the condition
$A_0 / E_F \ll 1$. This condition is naturally fulfilled in GaAs-based low-dimensional
conductors\cite{paget:1977} where $A_0 / E_F \sim 10^{-2}$ (for both Ga and As ions)
and in carbon nanotube systems
grown entirely from the $^{13}$C isotope (which has a nuclear spin $I=1/2$) 
where\cite{pennington:1996,fischer:2009}
$A_0 / E_F \sim 10^{-6}$.
Recent measurements on $^{13}$C nanotube 
quantum dots\cite{churchill:2009a} suggest a much higher value though,
but still such that $A_0 /E_F < 10^{-3}$.
An adjustment of this value, however, might be necessary
because it relies on models that were not specifically tailored 
for $^{13}$C nanotubes.\cite{trauzettel:2009} 
To clarify this discrepancy between the reported values of $A_0$
further experimental and theoretical work is required. 
We can speculate though that a similar renormalization of $A_0$ as presented 
in this paper can also occur for the quantum dot system, and hence mimic
a larger value of $A_0$.
Due to this, the band structure value $A_0/E_F \sim 10^{-6}$ 
for the bare, unrenormalized hyperfine interaction strength is used in this work.

The RKKY energy is minimized when the nuclear spins
align in the helimagnetic order. Through the separation of time scales and due to the 
large effective nuclear spins, this order can then be treated as a static nuclear
magnetic field acting on the electrons. Most remarkably, this interaction is
relevant in the renormalization group sense for the electron system and leads to 
the opening of a gap in one half of the electron excitation spectrum. 
This gap can be interpreted as a strong increase of the nuclear Overhauser field
in the direction defined by the nuclear helimagnet, while the hyperfine coupling 
in the orthogonal directions remains unrenormalized. The gap in the electron system 
is the result of the strong binding of collective electron spin modes
to the nuclear magnetization.
The resulting RKKY interaction is then mostly carried by the remaining gapless
electron modes and becomes much stronger. This leads to a further strong stabilization of the nuclear helimagnet. 
Through this feedback
the combined order remains stable up into experimentally accessible temperatures
(see below).

The strong renormalization is in fact possible due to an instability of the LL
toward a density wave order,\cite{giamarchi:2004} which is signaled by the divergence of the
electron susceptibilities at momentum $2k_F$. The same divergence is responsible
for ordering the nuclear spins, and so the back-action of the Overhauser field on the electrons
enhances the instability for a part of the electron degrees of freedom. 
This results in the partial order in the electron system.
Due to this, the effect of the feedback is strong even for very weak $A_0$.

We emphasize that this feedback is a pure LL effect and absent in Fermi liquids. 
It leads to a number of experimental signatures (described below) that may
be used to unambiguously identify a LL without the need of fitting power laws to 
measured response functions. Let us also mention that an alternative test of the LL theory has been proposed
for strongly interacting 1D current rectifiers.\cite{feldman:2005,braunecker:2005,braunecker:2007}
Here a pure LL signature is found in form of a specific asymmetric bump in the $I-V$ curve.

Table \ref{tab:values} lists the physical parameters we use for the numerical
estimates for the GaAs quantum wires and the $^{13}$C SWNTs.
\begin{table*}
	\begin{ruledtabular}
	\begin{tabular}{l|r|r}
		Physical quantity
		&GaAs Quantum Wire%
			\footnote{From Refs. \onlinecite{yacoby:1997,pfeiffer:1997,auslaender:2002,auslaender:2005,steinberg:2008}.}
		&$^{13}$C Single Wall Nanotube%
			\footnote{From Refs. \onlinecite{egger:1997,kane:1997b,egger:1998,pennington:1996,saito:1998,fischer:2009}.}\\
		\hline
		\hline
		Hyperfine (on-site) coupling constant $|A_0|$ & 1 K ; 90 $\mu$eV & 7 mK ; 0.6 $\mu$eV
		\\
		Nuclear spin $I$   & $3/2$ & $1/2$
		\\
		Electron spin $S=1/2$   & &
		\\
		Fermi vector $k_F$ & $1 \times 10^8$ m$^{-1}$ & $4 \times 10^8$ m$^{-1}$
		\\
		Fermi wavelength $\lambda_F = 2\pi/k_F$ & 63 nm & 17 nm
		\\
		Electron density $n_{el}=2k_F/\pi$ & $0.6 \times 10^8$ m$^{-1}$ & $2.4 \times 10^8$ m$^{-1}$
		\\
		Fermi velocity $v_F$ & $2 \times 10^5$ m s$^{-1}$ & $8 \times 10^5$ m s$^{-1}$ 
		\\
		Fermi (kinetic) energy $E_F = \hbar v_F k_F/2$ & $7$ meV & $0.1$ eV
		\\
		Lattice spacing $a$   & $5.65$ \AA & $2.46$ \AA
		\\
		Nuclear spin density (1D) $n_I = 1/a$   & $1.8 \times 10^9$ m$^{-1}$ & $4.1 \times 10^{9}$ m$^{-1}$
		\\
		Electron fraction per nuclear spin $n_{el}/n_I$	&0.04 $\approx 1/28$ &0.06 $\approx 1/17$
		\\
		System length $L$   & $2 - 40$ $\mu$m & $2$ $\mu$m
		\\
		Number of sites in transverse direction $N_\perp$ & $\sim$ $50 \times 50$ & $\sim$ 50
		\\
		Luttinger liquid parameter (charge) $K_c$ & $0.5$ & $0.2$%
			\footnote{\label{fn:SWNT}%
			See Sec. \ref{sec:nanotubes} for the use of $K_c, K_s$ within the 2-band
			description of SWNTs.}
		\\
		Luttinger liquid parameter (spin)   $K_s$ & 1 & 1\textsuperscript{\ref{fn:SWNT}}
		\\
		\hline
		\hline
		Approximate bandwidth
		$\Delta_a = \hbar v_F /a$ & 0.23 eV & 2.1 eV 
		\\
		Longitudinal level spacing
		$\Delta_L$ & 3--70 $\mu$eV & 260 $\mu$eV
		\\
		Transverse level spacing (subband splitting)
		$\Delta_t$ & $>20$ meV     & 0.65 eV
		\\
		\hline
		\hline
		Exponent $g  = g_{x,y,z}$ (single-band expression) & 0.75 & 0.6\textsuperscript{\ref{fn:SWNT}}
		\\
		Exponent $g  = g_{x,y,z}$ (SWNT 2-band expression for $T_0^*$)&& 0.8\textsuperscript{\ref{fn:SWNT}} 
		\\
		Exponent $g' = g'_{x,y}$  & 0.67 & 0.33\textsuperscript{\ref{fn:SWNT}}
		\\
		Exponent $g'_{z}$         & 0.33 & 0.17\textsuperscript{\ref{fn:SWNT}}
		\\
		\hline
		\hline
		Cross-over temperature (without feedback)
		$T^*_0$   & 53 mK ; 5 $\mu$eV     & 2 $\mu$K ; 0.2 neV 
		\\
		Cross-over temperature (with feedback)
		$T^*$     & 75 mK ; 7 $\mu$eV     & 11 mK ; 1 $\mu$eV
		\\
		Renormalized hyperfine coupling constant (in the direction&&
		\\
		of the nuclear spin polarization)
		$A^* = A_0 (\xi/a)^{1-g}$     & $\sim$ 4.6 K ; $\sim$ 400 $\mu$eV          & $\sim$ 0.25 K ; $\sim$ 22 $\mu$eV
		\\
		Upper bound for cross-over temperature $B^* = S A^* n_{el}/n_I$
		& $\sim 80$ mK ; $\sim$ 7 $\mu$eV & $\sim 10$ mK ; $\sim 1$ $\mu$eV
		\\
		Electron spin polarization (fraction) $\mean{\bS}/S$
		&0.05 & $5 \times 10^{-5}$
		\\
		Electron spin polarization (on-site) $\mean{\bS_i}/S = (n_{el}/n_I) \mean{\bS}/S$
		&0.002 & $3 \times 10^{-6}$
		\\
		Correlation length for gapped electrons $\xi$ &  $\xi = \xi_\infty = 0.2$ $\mu$m & $\xi = L = 2$ $\mu$m
		\\
	\end{tabular}
	\end{ruledtabular}
	\caption{Physical parameters for GaAs Quantum Wires and 
	$^{13}$C single wall nanotubes used in this paper, 
	and the derived quantities discussed in the text.
	Note that the values of the derived quantities have an $O(1)$ uncertainty,
	which is unavoidable in any Luttinger liquid theory due to the required ultraviolet cutoff
	[see also the discussion before Eq. \eqref{eq:Delta}]. 
	Within this uncertainty we have $k_B T^* \sim B^*$.
	Since $\xi = L$ for SWNTs, increasing $L$ (within
	limits) also increases $A^*$ and $B^*$.
	Anisotropic hyperfine interactions can slightly reduce $A^*$ and $B^*$
	(see Sec. \ref{sec:anisotropic}).
	Energy is converted into temperature by the relation 
	1 eV $\widehat{=}$ 11604.5 K.
	\label{tab:values}
	}
\end{table*}
With these values we find that the feedback effect is most remarkable for the $^{13}$C SWNTs where the 
cross-over temperature for the nuclear helimagnet without the feedback, $T^*_0$
[Eq. \eqref{eq:T^*_0}], 
would be close to a microkelvin. Through the feedback, however, $T^*_0$ is replaced by 
the correct $T^*$ [Eq. \eqref{eq:T^*}], which reaches into the experimentally 
accessible \emph{millikelvin} temperatures.
\footnote{
Note that we have chosen $E_F > 0$ in Table \ref{tab:values}, corresponding to
electron carriers in the SWNT, while experimentally often hole carriers (with $E_F<0$) 
are used. SWNTs with a linear dispersion relation (armchair type) are particle-hole 
symmetric and so the results exposed here remain unchanged (see also Ref. \onlinecite{jarillo:2004}). 
Spin-orbit interactions would break this symmetry, but recent experiments have shown\cite{kuemmeth:2008} 
that they are weak compared with $E_F$ in SWNTs. 
Moreover, it has been shown that under normal conditions 
the LL state is stable under the inclusion of, for instance, 
Rashba spin-orbit interactions.\cite{gritsev:2005,schulz:2009}
The effect from the coupling to the nuclear spins is much more dramatic and overcomes 
the spin-orbit interaction, which we therefore neglect in this work.
}

\begin{figure}
	\includegraphics[width=\columnwidth]{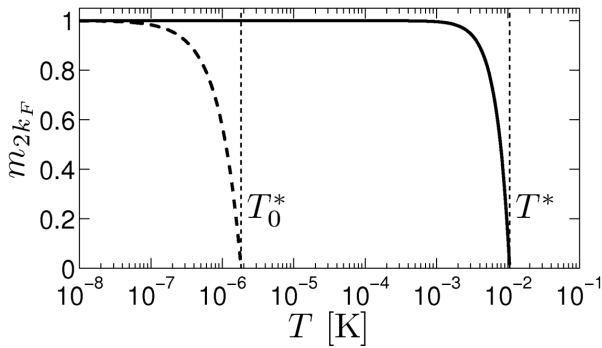}
	\caption{
	Magnetization $m_{2k_F}$ as a function of $T$ for $^{13}$C single wall nanotubes.
	Dashed line: without the feedback [Eq. \eqref{eq:m_T*_0}];
	solid line: including the feedback [Eq. \eqref{eq:m_T*}].
	The vertical lines mark the cross-over temperatures written next to them
	[Eqs. \eqref{eq:T^*_0} and \eqref{eq:T^*}].
	$T^*_0$ is evaluated with the 2-band model of Sec. \ref{sec:nanotubes}, while
	we use the quantitatively correct 1-band description for $T^*$. 
	Note that compared with Ref. \onlinecite{braunecker:2009} we use a larger
	$A_0$ (according to the results of Ref. \onlinecite{fischer:2009}) and so 
	obtain a slightly larger $T^*$. On the other hand $T^*_0$ is smaller because 
	the exponent $g$ is larger the 2-band model.
	\label{fig:m_T_carbon}}
\end{figure}

\begin{figure}
	\includegraphics[width=\columnwidth]{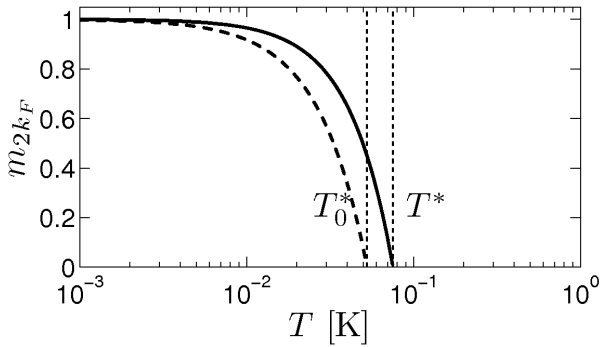}
	\caption{
	Magnetization $m_{2k_F}$ as a function of $T$ for GaAs quantum wires.
	Dashed line: without the feedback [Eq. \eqref{eq:m_T*_0}];
	solid line: including the feedback [Eq. \eqref{eq:m_T*}].
	The vertical lines mark the cross-over temperatures written next to them
	[Eqs. \eqref{eq:T^*_0} and \eqref{eq:T^*}].
	Since the exponent $g$ changes only little through the renormalization
	($g \to g'$), the values of $T^*_0$ and $T^*$ are of the same order of magnitude.
	\label{fig:m_T_GaAs}}
\end{figure}

\begin{figure}
	\includegraphics[width=\columnwidth]{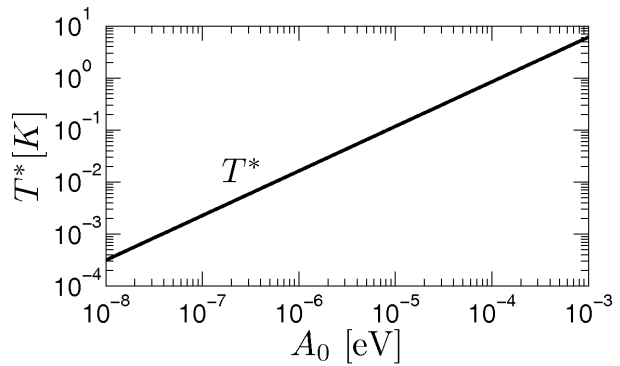}
	\caption{
	Cross-over temperature $T^*$ [Eq. \eqref{eq:T^*}] as a function of the hyperfine
	coupling constant $A_0$ for $^{13}$C single wall nanotubes. 
	$T^*$ follows a power-law $T^* \propto A_0^{2/(3-g'_{x,y})} = A_0^{0.86}$,
	and is plotted up to the self-consistency limit $T^* \approx v_F / L k_B = 3$ K.
	The values about $A_0 \sim 10^{-4}$ eV correspond to those deduced
	in Ref. \onlinecite{churchill:2009a}.
	Note that through the whole range we have $k_B T^* \sim B^*$ [Eq. \eqref{eq:B^*}].
	\label{fig:T_A_carbon}}
\end{figure}

\begin{figure}
	\includegraphics[width=\columnwidth]{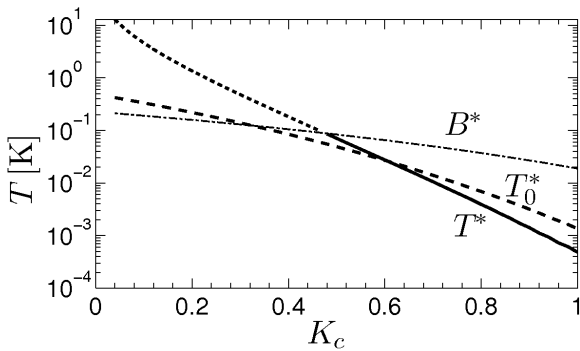}
	\caption{
	Cross-over temperatures $T^*_0$ [dashed line, Eq. \eqref{eq:T^*_0}], 
	$T^*$ [full line, Eq. \eqref{eq:T^*}], 
	and the bound $B^*$ [dash-dotted line, Eq. \eqref{eq:B^*}]
	for GaAs quantum wires
	as functions of the interaction strength, expressed by the Luttinger liquid parameter $K_c$
	(keeping $K_s = 1$).
	The noninteracting limit is $K_c = 1$, smaller $K_c>0$ correspond to increasingly 
	stronger repulsive interactions. 
	The dotted line is the continuation of $T^*$ beyond the energy scale set by $B^*$,
	for which the validity of the theory becomes uncertain.
	The curves of $T^*$ and $T^*_0$ cross at $K_c \approx 0.6$ because the RKKY interaction $J'_{2k_F}$ 
	defining $T^*$ has a prefactor that is by $1/2$ smaller than in the case without the 
	feedback.
	Since nuclear spin order unavoidably leads to the feedback, $T^*$ defines the 
	cross-over temperature for the order even when $T^*_0 > T^*$.
	Note that close to $K_c = 1$, Eqs. \eqref{eq:T^*_0} and \eqref{eq:T^*} diverge because 
	the cutoff $\delta$ in Eq. \eqref{eq:chi^>_T>0} was neglected in the further evaluation 
	of the RKKY interaction. This is valid for $K_c$ smaller than $\sim 0.8$. 
	Reintroducing the cutoff self-consistently close to $K_c=1$ regularizes
	the divergence and leads to the displayed curves (see Appendix \ref{sec:susc}).
	\label{fig:T_Kc_GaAs}}
\end{figure}

For GaAs quantum wires the effect on the cross-over temperature is much less pronounced
due to a less dramatic modification of the LL parameters. Yet through the
larger ratio $A_0/E_F$ we already have $T^*_0 \sim 50$ mK, which increases
through the feedback to $T^* \sim 80$ mK. 
Figures \ref{fig:m_T_carbon} through \ref{fig:T_Kc_GaAs} 
show the dependences of these temperatures
and of the nuclear magnetization on the variation of different system parameters.
Note that the large values of $T^*_0$ and $T^*$ for the GaAs quantum wires are 
due to the small value of $K_c = 0.5$ we use from Ref. \onlinecite{steinberg:2008} for 
high quality quantum wires.
The more common quantum wires with weaker electron-electron interactions and $K_c \approx 0.8$ lead
to $T^* \sim 1$ mK, as shown in Fig. \ref{fig:T_Kc_GaAs}.
In this figure we also show the energy scale $B^*$ [Eq. \eqref{eq:B^*}], which we use as
an upper bound for $k_B T^*$, below which our approach is controlled.
Further self-consistency conditions are discussed in Sec. \ref{sec:self_consistency}.

As a general rule, stronger electron-electron interactions (i.e. smaller LL parameters
$K_c$) lead to larger $T^*$ in a much more pronounced way than a larger value 
of the hyperfine constant $A_0$.
The explicit dependence can be read off from Eq. \eqref{eq:T^*}, and is given by
\begin{equation} \label{eq:T^*_intro}
	k_B T^*
	= I |A_0| D' \left(\frac{\Delta_a}{I |A_0|}\right)^{\frac{1-2g'}{3-2g'}},
\end{equation}
where $k_B$ is the Boltzmann constant, 
$I$ the nuclear spin, 
$\Delta_a = \hbar v_F/a$ is on the order of the bandwidth (with $v_F$ the Fermi velocity and
$a$ the lattice constant),
$D'$ [Eq. \eqref{eq:D'}] a nonuniversal dimensionless
constant of about $D' \sim 0.2 - 0.3$ for the values in Table \ref{tab:values}, and
$g' = 2K_c/[K_s(K_c+K_s^{-1})]$ [Eq. \eqref{eq:g'_xy}],
where $K_c, K_s$ are the LL parameters associated with charge and spin fluctuations,
respectively.
Note that SWNTs require a 2-band description and so four different LL
parameters\cite{egger:1997,kane:1997b}
($K_c = K_{cS} = 0.2$, $K_{cA} = K_{sS} = K_{sA} = 1$; see Sec. \ref{sec:nanotubes}).
While we take this into account when we neglect the feedback,
we show in Sec. \ref{sec:nanotubes} that the single-band description with $K_c = 0.2, K_s = 1$
is quantitatively valid when the feedback is taken into account, and therefore can be used
for the determination of  $T^*$, and the renormalized hyperfine constant $A^*$ below.

We stress that $T^*$ is independent of the system length $L$ for any realistic sample
(provided that we have $L \gg k_F^{-1}$ such that the LL theory is applicable).
For very large $L$ there would be a cross-over where $T^*$ is replaced by a $L$-dependent
quantity such that $T^* \to 0$ as $L \to \infty$. This cross-over, however, occurs only at
$L$ that lie orders of magnitude above realistic sample lengths
(see Sec. \ref{sec:infinite_systems} and Appendix \ref{sec:small_level_spacing}).

The order parameter for the nuclear helimagnet is the $2k_F$ Fourier component of the 
magnetization, which has close to $T^*$ the behavior of a generalized Bloch law [Eq. \eqref{eq:m_T*}]
\begin{equation} \label{eq:m_T_intro}
	m_{2k_F} = 1 - \left(\frac{T}{T^*}\right)^{3-2g'}.
\end{equation}
This magnetization may be detectable by magnetic sensors with a spatial resolution 
smaller than the period of the helix $\lambda_F/2 \sim$ 10 -- 30 nm such as, for instance,
magnetic resonance force microscopy.\cite{mamin:2007,degen:2009}

The nuclear spin ordering acts back on the electron system and leads to a 
strong renormalization of the hyperfine interaction between the nuclear spins and 
a part of the electron modes. We can capture this renormalization by the replacement
$A_0 \to A^*$ of the hyperfine constant. We emphasize though that this replacement 
also requires a reinterpretation of the role of $A^*$: It no longer describes the local
coupling between a nuclear spin and an electron at a lattice site, but the coupling of
a nuclear spin to a fraction of the collective electron modes in the LL. 
The modified $A^*$ has thus a similar interpretation as the dressing of 
impurity scattering\cite{kane:1992,furusaki:1993} in a LL that no longer corresponds 
to the backscattering of a single particle but to the
generation of collective density waves near the impurity site.
The renormalization is expressed by
[Eq. \eqref{eq:A^*}]
\begin{equation}
	A^* = A_0 \left(\xi/a\right)^{1-g},
\end{equation}
where $g = (K_c + K_s^{-1})/2$
and $\xi = \min(L,\lambda_T,\xi_\infty)$ is a correlation length.
Here $\lambda_T = \hbar v_F / k_B T$ is the 
thermal length, and $\xi_\infty = a (\Delta_a/ I A_0 m_{2k_F})^{1/(2-g)}$ the correlation
length for an infinite system. 
We stress that that $\xi$ cannot exceed $L$ or $\lambda_T$. 
An uncritical use of $\xi = \xi_\infty$ exceeding $L$
or $\lambda_T$ can lead to self-consistency violations of the theory as explained in 
Sec. \ref{sec:feedback_el}.
Note that for noninteracting electrons (including Fermi liquids) $g=1$ and so $A_0$
remains unrenormalized. The increase $A_0 \to A^*$ is a direct consequence of
LL physics.

For the systems under consideration we have $I |A^*| \gg k_B T^*$ and hence the 
electron spin modes following the nuclear helimagnet (ferromagnetically or antiferromagnetically 
according to the sign of $A_0$) are pinned into a spatially rotating spin density wave. 
This affects, however, only one half of the low-energy electron degrees of freedom. The remaining 
electron spins remain in their conducting LL state. 
They do no longer couple to the ordered nuclear spins, yet couple to 
fluctuations out of the ordered nuclear phase with the \emph{unrenormalized} hyperfine
coupling constant $A_0$. Those conducting electrons carry then the dominant RKKY interaction,
which has a modified form leading to the stabilization of the combined order up to the 
renormalized temperature $T^*$.

On the other hand, the gapped electrons have an excitation gap that is directly given 
by the renormalized Overhauser field, given by $B_{xy}^* = I m_{2k_F} A^* / 2$.
Therefore, the nuclear magnetization $m_{2k_F}$ can be directly determined by measuring
the electronic excitation gap through, for instance, tunneling into the 
system.\cite{auslaender:2002,auslaender:2005,steinberg:2008,jompol:2009}
Since $A^*$ itself can depend on $m_{2k_F}$ through 
the correlation length $\xi_\infty$, we have [Eq. \eqref{eq:B_T}]
\begin{equation}
	B_{xy}^* \propto
	\begin{cases}
	m_{2k_F}  =
	\left[ 1 - \left(\frac{T}{T^*}\right)^{3-2g'} \right] 
	& \text{for $m_{2k_F} < m_\infty$,}
\\
	m_{2k_F}^{\frac{3-g}{2-g}} 
	=
	\left[ 1 - \left(\frac{T}{T^*}\right)^{3-2g'} \right]^{\frac{3-g}{2-g}}
	&\text{otherwise},
	\end{cases}
\end{equation}
where $m_\infty$ is such that $\xi = \xi_\infty$ for all $m_{2k_F} > m_\infty$.
Notice that for SWNTs, we have $m_\infty >1$ and so we always have 
$B_{xy}^* \propto m_{2k_F}$.
If, however, $m_\infty < 1$, the cross-over between the two different scaling
behaviors of $B_{xy}^*$ can be tuned by varying either $L$ or $\lambda_T$, depending
on which one is smaller.

\begin{figure}
	\includegraphics[width=1\columnwidth]{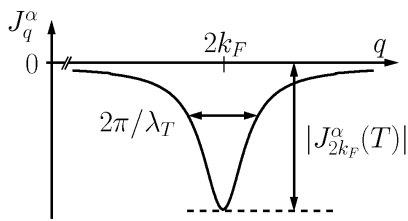}
	\caption{
	Sketch of the RKKY interaction $J_q^\alpha$ [Eq. \eqref{eq:J_q}] 
	or equivalently the spin susceptibility $\chi_\alpha(q)$ [Eq. \eqref{eq:susc_app}].
	\label{fig:J}
	}
\end{figure}

Experiments detecting this phase can rely on two more effects.
First, the freezing out of one half of the conducting channels of the electron system 
leads to a drop of the electrical conductance by precisely the factor 2 (see 
Refs. \onlinecite{braunecker:2009} and \onlinecite{braunecker:2009EPAPS}).
Second, the breaking of the electron spin SU(2) symmetry through the spontaneous appearance
of the nuclear magnetic field leads to the emergence of anisotropy in the electron 
spin susceptibility
(see Ref. \onlinecite{braunecker:2009EPAPS} and Appendix \ref{sec:susceptibility}).
The susceptibilities are defined by Eq. \eqref{eq:susc_def} and 
evaluated in Appendix \ref{sec:susceptibility}. From Eqs. \eqref{eq:chi_x_feedback}
and \eqref{eq:chi_z_feedback} we find that for momenta $q$ close to $2k_F$
and at $T=0$
\begin{align}
	\chi_{x,y}(q) &\propto (q-2k_F)^{-2(1-g')},
\\
	\chi_z(q)     &\propto (q-2k_F)^{-2(1-g'_{z})},
\end{align}
with $g'_z = (K_c K_s^{-1} + K_c K_s)/2(K_c+K_s^{-1})$ [Eq. \eqref{eq:g'_z}].
For $K_s = 1$ we have $g'_z = g'/2$.
At temperatures $T>0$, these power-law singularities are broadened 
at $q < \pi/\lambda_T$ [Eq. \eqref{eq:susc_app}]. The qualitative shape of these 
susceptibilities is shown in Fig. \ref{fig:J}.


\section{Model and effective model}
\label{sec:model}

\subsection{Model}

We consider a system of conduction electrons and nuclear spins
expressed by the Kondo-lattice type Hamiltonian
\begin{equation} \label{eq:H}
	H = 
	H_{el}^{1D} 
	+ \sum_{i} A_0 \bS_i \cdot \bI_i 
	+ \sum_{ij,\alpha,\beta} v_{ij}^{\alpha\beta} I_i^\alpha I_j^\beta.
\end{equation}
We have chosen here a tight-binding description, where the indices $i,j$ 
run over the three-dimensional (3D) lattice sites $\br_i, \br_j$ of
the nuclear spins, with lattice constant $a$.
The hyperfine coupling between the nuclear and electron spin on site $i$
is expressed by the constant $A_0$, the electron spin operator 
$\bS_i = (S_i^x,S_i^y,S_i^z)$,
and the nuclear spin operator $\bI_i = (I_i^x,I_i^y,I_i^z)$.
For GaAs we have $I= 3/2$ and for $^{13}$C this spin is $I = 1/2$.
We shall generally set $\hbar = 1$ in this paper and reintroduce it only for important 
results. 
We assume here an isotropic hyperfine interaction. The case of anisotropy is 
discussed in Sec. \ref{sec:anisotropic}.

The Hamiltonian $H_{el}^{1D}$ describes the 1D electrons (confined in a single
transverse mode) and is given in detail below. In addition to the transverse 
confinement, we assume that the 1D system has a length $L$ on the order of 
micrometers that may be the natural system length or be imposed by gates
(see Table \ref{tab:values}).
In contrast to the usual Kondo-lattice
model, $H_{el}^{1D}$ contains the here crucial electron-electron interactions.

The last term in Eq. \eqref{eq:H} denotes the direct dipolar interaction between 
the nuclear spins, or for the terms with $i=j$ the quadrupolar splitting 
of the nuclear spins ($\alpha,\beta = x,y,z$).
Keeping this term would make the analysis of this model cumbersome as we 
would have to solve a full 3D interacting problem. Yet 
those interactions are associated with the smallest energy scales
in the system. The dipolar interaction has been estimated to be on the order of\cite{paget:1977}
$10^{-11}$ eV $\sim 100$ nK. 
For all ions considered here, the quadrupolar splitting is 
absent in $^{13}$C and is otherwise the largest for As with 
a magnitude \cite{abragam:1961,salis:2001,bowers:2006} $\sim 10^{-10}$ eV $\sim 1$ $\mu$K.
These scales are overruled by the much stronger effective
RKKY interaction derived below and, in particular, are much smaller than
the temperatures we consider and that are experimentally accessible. 
This allows us to neglect the dipolar and quadrupolar terms henceforth.

The model \eqref{eq:H} does not yet contain the confinement of the 
electrons into a 1D conductor. Since we neglect the dipolar interaction
we can focus only on those nuclear spins that lie within the 
support of the transverse confining mode. This leads to a first 
simplification that is considered right below.


\subsection{Confinement into 1D}
\label{sec:confinement_1D}

In this work we consider only conductors, in which the electrons are 
confined in a single transverse mode $\phi_\perp$. 
Higher transverse harmonics are split off by an energy $\Delta_t$.
For SWNT, this transverse level spacing is determined by\cite{kane:1997a}
$\Delta_t = 2\pi v_F \hbar/|\mathbf{C}|$, where $\mathbf{C}$ is the chiral 
vector describing the wrapping of the nanotube. For a
$\mathbf{C} = (n,n)$ armchair nanotube we have $|\mathbf{C}| = 3 n a_{CC}$
(with $a_{CC} = a/\sqrt{3}$ the distance between carbon ions),
and so $\Delta_t = (2\pi/\sqrt{3}) \Delta_a/n $. Due to the 
armchair structure, there are 4 nuclear spins per $3 a_{CC}$ on the cross section
so that the number of sites on the cross-section is
$N_\perp = 4 n$. 
For our choice $N_\perp \approx 50$ we then find $n \approx 12$, 
leading to $\Delta_t \approx 0.65$ eV. 
For the cleaved edge overgrown GaAs quantum wires, the transverse subband 
splitting has been reported\cite{pfeiffer:1997} to exceed 20 meV.
These large values allow us to focus on the lowest transverse mode (subband) only.

It is then advisable to switch from the 3D tight binding basis
into a description that reflects the confinement into the lowest transverse mode: 
Let $\phi_{t_0},\phi_{t_1},\dots$
label a full set of 2D orthonormal transverse single-electron wavefunctions
such that $\phi_{t_0} = \phi_\perp$, and let $\phi_i$ be the 3D tight binding
Wannier functions centered at lattice sites $\br_i$. We decompose the position
vector $\br$ into a longitudinal part $r_{||}$ and the 2D vector along the
transverse direction $\br_{\perp}$. Similarly we decompose the 3D lattice index
$i$ into the parts $i_{||}$ and $i_{\perp}$. We then write 
$\phi_i(\br) = \phi_{i_{||}}(r_{||}) \phi_{i_\perp}(\br_\perp)$,
and perform the basis change 
of the electron operators $c_{i,\sigma}^\dagger$ as
\begin{equation}
	c_{i,\sigma}^\dagger = 
	\bracket{t_0=\perp}{i_\perp} c_{i_{||},\perp,\sigma}^\dagger 
	+ \bracket{t_1}{i_\perp} c_{i_{||},t_1,\sigma}^\dagger
	+ \dots
\end{equation}
with $\sigma = \up, \dw$,
$c_{i_{||},t_n}$ the electron operators corresponding to longitudinal coordinate $i_{||}$ and
transverse mode $t_n$, and
\begin{equation}
	\bracket{t_n}{i_\perp} 
	= 
	\int d\br_{\perp} \phi_{t_n}^*(\br_{\perp}) \phi_{i_\perp}(\br_{\perp}),
\end{equation}
for normalized wave functions $\phi_{t_n}$ and $\phi_{i_\perp}$.
With the condition that the electrons are confined in the $\phi_{t_0} = \phi_\perp$ 
mode, averages over the operators $c_{i_{||},t_n,\sigma}^\dagger$ vanish for $n\ge 1$.
This allows us to drop those operators from the beginning, and to use the projected electron
operators $c_{i,\sigma}^\dagger = \bracket{\perp}{i_\perp} c_{i_{||},\perp,\sigma}^\dagger$.
The 2D Wannier wavefunctions $\phi_{i_\perp}$ have their support over a surface 
$a^2$ centered at a lattice site, while $\phi_{\perp}$ extends over $N_\perp$ sites in the 
transverse direction. The normalization imposes that $|\phi_{i_\perp}(\br)|^2 \propto 1/a^2$
and $|\phi_\perp(\br)|^2 \propto 1/N_\perp a^2$ for $\br$ within the support of these wavefunctions.
Consequently $\bracket{\perp}{i_\perp} = C_{i_\perp}/\sqrt{N_\perp}$ with 
$C_{i_\perp}$ a dimensionless constant that is close to 1 in the support of 
$\phi_\perp$ and vanishes for those $i_\perp$ where $\phi_\perp = 0$
(possible phases of the $C_{i_\perp}$ can be absorbed in the electron operators
$c_{i_{||},\perp,\sigma}^\dagger$). 

The electron spin operator is quadratic in the electron creation and annihilation
operators and we obtain $\bS_i = \bS_{i_{||},\perp} C_{i_\perp}^2/N_\perp$.
The $i_\perp$ dependence in the Hamiltonian \eqref{eq:H} can then be summed out
by defining the new composite nuclear spins
\begin{equation} \label{eq:Itilde}
	\tilde{\bI}_{i_{||}} = \sum_{i_\perp} C_{i_\perp}^2 \bI_{i_{||},i_\perp},
\end{equation}
so that the Hamiltonian becomes
\begin{equation} \label{eq:H_1D} 
	H = H^{1D} = H_{el}^{1D} 
	+ \sum_{i_{||}} \frac{A_0}{N_\perp} \bS_{i_{||},\perp} \cdot \tilde{\bI}_{i_{||}}.
\end{equation}
This result is remarkable in that the complicated 3D Hamiltonian \eqref{eq:H}
is equivalent to the purely 1D system \eqref{eq:H_1D}, describing the coupling of
the 1D electron modes to a chain of effectively \emph{large} spins 
$\tilde{\bI}_{i_{||}}$. Indeed, since $\sum_{i_\perp} C_{i_\perp}^2 = N_\perp$
is imposed by the normalization, the composite spin has a length
$0 \le \tilde{I} \le I N_\perp$. As shown below, the maximal alignment 
$\tilde{I} = I N_\perp$ is energetically most favorable for the RKKY interaction,
such that in the ordered phase the composite spin can be treated as an effective 
spin of length $I N_\perp$.
The prefactor $1/N_\perp$ to the hyperfine
constant $A_0$ expresses the reduction of the on-site hyperfine interaction
by spreading out the single-electron modes over the $N_\perp$ sites.


\subsection{Interpretation of the composite nuclear spins $\tilde{\bI}_{i_{||}}$}
\label{sec:interpretation_NI}

It is important to stress that $N_\perp$ is large:
For SWNT, $N_\perp$ denotes the number of lattice sites
around a circular cross-section and is on the order of $N_\perp \sim 50$.
GaAs quantum wires have a confinement of about $50$ lattice sites in both 
transverse directions and so $N_\perp \sim 2500$.
In both cases $N_\perp$ is large enough such that the physics of pure
(small) quantum spins does not appear. Accordingly, we shall treat the 
nuclear spin fluctuations below within a semiclassical spin-wave approach
corresponding to an expansion in $1/I N_\perp$.

\begin{figure}
	\includegraphics[width=0.7\columnwidth]{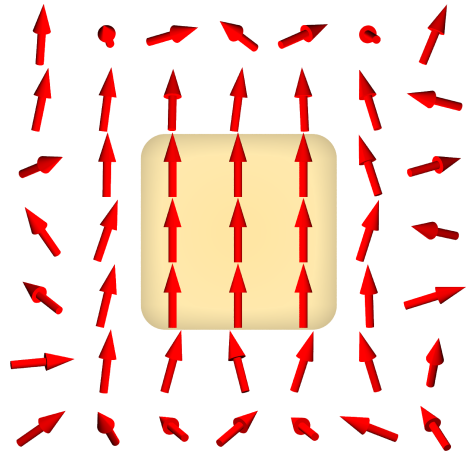}
	\caption{%
		Illustration of the cross-section through the 1D conductor.
		The $N_\perp \gg 1$ nuclear spins within the support of the 
		transverse confinement of the electron wave function (central colored 
		region) are ferromagnetically locked and behave 
		like a single large spin $\tilde{I} = N_\perp I$.
		The nuclear spins outside this support do not interact with the 
		electron system and are generally disordered because their
		direct dipolar interaction is very weak.
		\label{fig:cross_section}
	}
\end{figure}

We note moreover that the interaction with the electron spin acts only on
the $\tilde{\bI}_{i_{||}}$ mode. 
Since $C_{i_\perp} \approx 1$ over most of the support of $\phi_\perp$,
we have $\tilde{\bI}_{i_{||}} \approx \sum_{i_\perp} \bI_{i_{||},i_\perp}$.
Hence, all individual nuclear spins on a cross-section couple identically
to the electrons. The RKKY interaction acts only on $\tilde{\bI}_{i_{||}}$
and so any nuclear order minimizing the RKKY interaction energy 
is imposed simultaneously and identically on all nuclear spins on a cross-section. 
This leads to the ferromagnetic locking of these nuclear spin shown in 
Fig. \ref{fig:cross_section}. 
Otherwise said, since the electrons couple only to the transverse Fourier
mode $\tilde{\bI}_{i_{||}}$ describing the ferromagnetic alignment,
any order due to the interaction with the electrons can only maximize this
Fourier component and so lead to the ferromagnetic locking.

For SWNT, where $C_{i_\perp} \equiv C$ through rotational invariance of 
$\phi_\perp$ on the circular cross-section, the electrons couple only to this
ferromagnetic transverse component. 
For GaAs quantum wires deviations from the ferromagnetic
component are concentrated at the boundary of the confinement described by 
$\phi_\perp$. The nuclear spins in this boundary layer are more weakly coupled 
to the electron spin, with an amplitude reduced by the factor
$C_{i_\perp}^2 \sim |\phi_{\perp}(\br_{i_\perp})|^2 < 1$, and hence are 
more sensitive to fluctuations. 

In the following we shall interpret $\tilde{\bI}_{i_{||}}$ mainly as this
ferromagnetic component. For GaAs quantum wires
this either means that we consider only very well confined wires where $C_{i_\perp} = 1$ 
over essentially the whole transverse cross-section, or that we restrict our attention
to only those nuclear spins where $C_{i_\perp} = 1$, and so effectively slightly reduce
what we call the cross-section of the wire.

The case of an anisotropic hyperfine interaction is discussed in Sec. \ref{sec:anisotropic}.
In this case spin configurations different from the ferromagnetic alignment are possible.
The described physics, however, remains valid as long as these configurations produce a 
finite nuclear magnetic field. 

%


\subsection{Effective Hamiltonians}

From the above considerations we have seen that the original 3D Hamiltonian
is through the confinement of the conduction electrons equivalent to the
1D Hamiltonian \eqref{eq:H_1D}, which we rewrite here as
\begin{equation} \label{eq:H_1Da} 
	H = H^{1D} = H_{el}^{1D} 
	+ \sum_{i} \frac{A_0}{N_\perp} \bS_{i} \cdot \tilde{\bI}_{i},
\end{equation}
where $i=i_{||}$ now runs over the 1D sites of a 1D lattice of length $L$
with lattice constant $a$.
$\bS_i \equiv \bS_{i_{||},\perp}$ is now a 1D electron spin operator, and 
$\tilde{\bI}_i = \sum_{i_\perp} \bI_{i,i_\perp}$ is the ferromagnetic
component of the $N_\perp$ nuclear spins on the cross-section, as discussed before.
Note that even the reduced $A_0/N_\perp$ remains much larger than the 
neglected dipolar interaction.

The fact that $A_0 \ll E_F$ and, in particular, $A_0/N_\perp \ll E_F$,
shows that the time and energy scales between the electron and nuclear 
spin systems decouple.
[A more thorough investigation of this condition can be found in Sec. \ref{sec:validity_RKKY}.]
This characterizes the RKKY regime, in which 
the nuclear spins are coupled through an effective interaction carried
over the electron system. Indeed, a change of $\tilde{\bI}_i$ induces a 
local electron spin excitation. The response of the electron gas to this
local perturbation is the electron spin susceptibility $\chi_{ij}$, describing
the propagation of the effect of the local perturbation from site $i$ to 
site $j$. At site $j$ the electrons can couple then to the nuclear
spin $\tilde{\bI}_j$, hence inducing the effective interaction. 
The strict separation of time scales implies that this interaction can 
be considered as instantaneous, and so only the static electron susceptibility
$\chi_{ij}(\omega \to 0)$ is involved in the interaction.
This interaction can be derived in detail, for instance, through a Schrieffer-Wolff
transformation followed by an integration over the electron degrees of freedom
as in Refs. \onlinecite{simon:2007,simon:2008}. The result
is the effective Hamiltonian for the nuclear spins
\begin{equation} \label{eq:H_eff_n}
	H^{\text{eff}}_{n}
	= \sum_{ij,\alpha \beta} \frac{J_{ij}^{\alpha\beta}}{N_\perp^2} \tilde{I}_i^\alpha \tilde{I}_j^\beta,
\end{equation}
with $\alpha,\beta = x,y,z$ and the RKKY interaction
\begin{equation} \label{eq:J_def}
	J_{ij}^{\alpha\beta} = \frac{A_0^2}{2} a^2 \chi_{ij}^{\alpha\beta},
\end{equation}
where $\chi_{ij}^{\alpha\beta}$ is the static electron spin susceptibility
\begin{equation} \label{eq:susc_def}
	\chi_{ij}^{\alpha\beta} = 
	-\frac{i}{a^2} \int_0^\infty dt \, \e^{-\eta t} \mean{[S^\alpha_i(t) , S^\beta_j(0)]},
\end{equation}
for an infinitesimal $\eta > 0$ and the average determined by $H_{el}$.
Note the $1/a^2$ in this definition, which allows us to pass to the continuum
limit  $\chi_{ij}^{\alpha\beta} \to \chi^{\alpha\beta}(r)$ without further complication
(see Appendix \ref{sec:susc}).
We assume that the total spin is conserved in the system, and so 
$\chi^{\alpha\beta}(r) = \delta_{\alpha\beta}\chi_\alpha(r)$, 
$J^{\alpha\beta}(r) = \delta_{\alpha\beta}J^\alpha(r)$.
In momentum space we obtain
\begin{equation} \label{eq:H_eff_n_q}
	H^{\text{eff}}_{n}
	= \frac{1}{N} \sum_{q, \alpha} \frac{J_{q}^{\alpha}}{N_\perp^2} \tilde{I}_{-q}^\alpha \tilde{I}_q^\alpha,
\end{equation}
with $N = L/a$, the Fourier transform $\tilde{I}^\alpha_q = \sum_i \e^{- i q r_i} \tilde{I}^\alpha_i$,
and
\begin{equation} \label{eq:J_q_def}
	J_q^{\alpha}
	= \frac{A_0^2}{2} a \chi_\alpha(q)
	= \frac{A_0^2}{2} a \int dr \, \e^{-i q r} \chi_\alpha(r).
\end{equation}
In this derivation of the RKKY interaction we have tacitly assumed that the electrons
are unpolarized. This will be no longer the case once the feedback coupling between
electrons and nuclear spins has been taken into account. The necessary modification
to Eq. \eqref{eq:H_eff_n_q} is discussed in Sec. \ref{sec:feedback_nucl} and leads
to the Hamiltonian \eqref{eq:H_n^eff_pol}. 

This feedback is driven by the Overhauser
field generated by the nuclear spins acting back on the electrons. To model 
this we rely again on the separation of time scales, which allows us to 
treat the Overhauser field as a static external field for the electrons.
Hence, a mean field
description of the nuclear Overhauser field is very accurate. This
leads to the effective Hamiltonian for the electron system
\begin{equation} \label{eq:H_eff_el}
	H^{\text{eff}}_{el} = H^{1D}_{el} + \sum_{i} \bh_i \cdot \bS_i,
\end{equation}
with $\bh_i = \frac{A_0}{N_\perp} \mean{\tilde{\bI}_i}$, and where
the expectation value is taken with respect to $H^{\text{eff}}_{n}$.
The Hamiltonians $H^{\text{eff}}_{el}$ and $H^{\text{eff}}_{n}$, and so
the properties of the electron and nuclear subsystems self-consistently
depend on each other.


\subsection{Electron Hamiltonian}

The confinement of the conduction electrons in the single transverse mode $\phi_\perp$
makes the electron Hamiltonian strictly 1D. Through this dimensional reduction
electron-electron interactions have a much stronger effect than in higher
dimensions. In particular, they lead to a departure from the Fermi liquid 
paradigm, and often the Luttinger liquid (LL) concept, based on the 
Tomonaga-Luttinger model, is the valid starting point to characterize 
the system properties. 

We therefore consider a 1D system of length $L$ with electron-electron interactions
that are effectively short-ranged due to screening by gates. Such a system 
allows a description by the Tomonaga-Luttinger model, given by the Hamiltonian\cite{giamarchi:2004}
\begin{align}
	&H_{el}^{1D} 
\nonumber
\\
	&= 
	\sum_{\sigma=\up,\dw} \int dr 
	v_F 
	\left[ 
		\psi^\dagger_{L\sigma}(r) i\nabla\psi_{L \sigma}(r)
		-
		\psi^\dagger_{R\sigma}(r) i\nabla\psi_{R \sigma}(r)
	\right]
\nonumber
\\
	&+
	\sum_{\sigma\sigma'} \int dr dr' U(r-r') 
	\psi^\dagger_\sigma(r) \psi^\dagger_{\sigma'}(r') \psi_{\sigma'}(r') \psi_\sigma(r),
\label{eq:H_TL}
\end{align}
where $\psi_{\ell\sigma}(r)$ are the operators for 
left moving ($\ell = L = -$, with momenta close to $-k_F$) 
and right moving ($\ell = R = +$, with momenta close to $+k_F$)
electrons with spin $\sigma= \up = +$
and $\sigma = \dw = -$. 
The positions $r,r'$ run over a system of length $L$ such that $L \gg \pi /k_F$, 
where $2k_F/\pi = n_{el}$ is the electron density of the 1D conductor.
The operator
$\psi_\sigma(r) = \psi_{L\sigma}(r) + \psi_{R\sigma}(r)$ is the conventional
electron operator.
The potential $U(r-r')$ describes the screened electron-electron interaction.

For simplicity, we consider here a single-band description of the 1D conductor.
This is not correct for carbon nanotubes, which require a 2-band model. 
In Sec. \ref{sec:nanotubes}, however, we show that the main conclusions
and results are quantitatively determined by the single-band model, 
so that we can 
avoid the more complicated 2-band description henceforth.

In the Hamiltonian \eqref{eq:H_TL} we have assumed a linear fermionic dispersion relation
for the left and right moving electrons, $\epsilon_q = \pm v_F (q \mp k_F)$
(setting the chemical potential $\mu = 0$). 
If this assumption is valid, 
we have with the bosonization technique\cite{giamarchi:2004,gogolin:1998} a powerful
tool to evaluate the properties of the electron system to, in principle,
arbitrary strength of the electron-electron interactions, leading to the 
LL description.

We assume here that this theory holds. Possible deviations are discussed in Sec. \ref{sec:validity_LL}.
The derivation of the bosonic theory can then be done in the 
standard way\cite{giamarchi:2004} by 
expressing the fermion operators $\psi_{\ell\sigma}$ 
in terms of boson fields $\phi_{\ell\sigma}$ as
\begin{equation} \label{eq:psi_phi}
	\psi_{\ell\sigma}(r) 
	= \frac{\eta_{\ell\sigma}}{\sqrt{2\pi a}} \e^{i \ell k_F r} \e^{i \ell \phi_{\ell\sigma}(r)},
\end{equation}
with $\eta_{\ell\sigma}$ the Klein factor removing a $(\ell,\sigma)$ 
particle from the system and 
\begin{equation} \label{eq:phi_phi_theta}
	\phi_{\ell\sigma} =
	\frac{1}{\sqrt{2}}[\phi_c - \ell\theta_c + \sigma (\phi_s - \ell\theta_s)].
\end{equation}
Here $\phi_{c,s}$ are boson fields such that $-\nabla\phi_{c,s}\sqrt{2}/\pi$ measure
the charge and spin fluctuations in the system, respectively.
The boson fields $\theta_{c,s}$
are such that $\nabla \theta_{c,s}/\pi$ are canonically conjugate to $\phi_{c,s}$.
The Hamiltonian \eqref{eq:H_TL} can then be rewritten in these
boson fields as
\begin{equation} \label{eq:H_LL}
	H_{el}^{1D}
	= \sum_{\nu = c,s}
	\int \frac{dr}{2\pi} \left[
		\frac{v_\nu}{K_\nu} (\nabla\phi_\nu(r))^2 + v_\nu K_\nu (\nabla\theta_\nu(r))^2
	\right],
\end{equation}
where $K_{c,s}$ are the LL parameters for the charge and spin density fluctuations,
and $v_{c,s}$ are charge and spin density wave velocities.
The electron-electron interactions are included in this Hamiltonian 
through a renormalization of $K_{c,s}$ and $v_{c,s}$. The noninteracting case
is described by $K_c = K_s = 1$ and $v_c = v_s = v_F$. Repulsive electron-electron interactions
lead to $0 < K_c < 1$. 
If the spin SU(2) symmetry is preserved $K_s = 1$ otherwise
$K_s > 1$. The case $K_s < 1$ would open a gap in the spin sector\cite{giamarchi:2004,gogolin:1998} 
and is not considered here.
For ideal LLs one has $v_{c,s} = v_F / K_{c,s}$.
With Eq. \eqref{eq:H_LL} we have furthermore assumed that $k_F$ is not
commensurate with the lattice spacing.
Altogether, this allowed us to drop irrelevant backscattering and Umklapp scattering 
terms in Eq. \eqref{eq:H_LL}.

\subsection{RKKY interaction}

The calculation of the RKKY interaction, i.e. the calculation of the 
electron spin susceptibility, is standard in the LL theory.
We have outlined its derivation in Appendices \ref{sec:susceptibility} and \ref{sec:RKKY_real_space} 
(the real space form 
of the RKKY interaction at $T=0$ has been derived before in Ref. \onlinecite{egger:1996}), and from
Eq. \eqref{eq:susc_app} together with Eq. \eqref{eq:J_q_def} we obtain,
for $q > 0$,
\begin{align} 
	J_q^{\alpha}(g_\alpha,v_F) 
	\approx 
	&-
	\frac{A_0^2}{\Delta_a} C(g_\alpha)
	\left(\frac{\Delta_a}{k_B T}\right)^{2-2g_\alpha}
\nonumber\\
	&\times	
	\left|\frac{\Gamma(\frac{g_\alpha}{2}- i \frac{\lambda_T}{4\pi}(q-2k_F))}
	           {\Gamma(\frac{2-g_\alpha}{2}- i \frac{\lambda_T}{4\pi}(q-2k_F))}\right|^2.
\label{eq:J_q}
\end{align}
In this expression we have neglected an additional small term depending 
on $q+2k_F$ and the small forward-scattering contribution. 
We note, however, that $J_{-q}^\alpha = J_{q}^\alpha$.
Much of $J_q^\alpha$ depends on the quantities
\begin{equation} \label{eq:g}
	g = g_{x,y} = (K_c + 1/K_s)/2,
	\quad
	g_z = (K_c + K_s)/2.
\end{equation}
For SWNTs these definitions must be modified due to the existence of two bands, which we 
use for the case when the feedback to the electron system is neglected.
From the discussion in Sec. \ref{sec:nanotubes} we have
\begin{equation} \label{eq:g_SWNT}
	g = g_{x,y,z} = (K_c + 3)/4.
\end{equation}
We stress, however, that the single-band values \eqref{eq:g} are quantitatively correct
also for SWNTs when we take the feedback into account (see below).

For $K_s = 1$ it follows from $K_c \le 1$ that $g_\alpha \le 1$.
The prefactor in Eq. \eqref{eq:J_q} is given by
\begin{equation} \label{eq:C}
	C(g)
	= 
	\frac{\sin(\pi g)}{2} \Gamma^2(1-g) \left(2\pi\right)^{2g-4},
\end{equation}
we have introduced the thermal length
\begin{equation} \label{eq:lambda_T}
	\lambda_T = \hbar v_F / k_B T,
\end{equation}
and the energy scale
\begin{equation} \label{eq:Delta_a}
	\Delta_a = \hbar v_F / a,
\end{equation}
which is on the order of the bandwidth.

A sketch of $J^\alpha_q$ is shown in Fig. \ref{fig:J}. This interaction 
has a pronounced minimum at $q =\pm 2k_F$ with a width $\sim \pi/\lambda_T$, and 
a depth of
\begin{equation} \label{eq:J_2kF}
	J^\alpha_{2k_F} = 
	- \frac{A_0^2}{\Delta_a} C(g_\alpha) 
	\left(\frac{\Delta_a}{k_B T}\right)^{2-2g_\alpha}
	\left|\frac{\Gamma(g_\alpha/2)}{\Gamma(1-g_\alpha/2)}\right|^2.
\end{equation}
For $|q-2k_F|  > \pi / \lambda_T$ the zero temperature form of this 
curve becomes quickly accurate: For $q>0$ and combining Eqs. \eqref{eq:susc_app_T0}
and \eqref{eq:J_q_def} we find
\begin{equation}  \label{eq:J_q_T=0}
	J^\alpha_q 
	\approx - \frac{A_0^2}{\Delta_a} \frac{\sin(\pi g_\alpha)}{8 \pi^2}
	\left|\frac{2}{a(q - 2k_F)}\right|^{2-2g_\alpha},
\end{equation}
which can also be verified by letting $T \to 0$ in Eq. \eqref{eq:J_q}.

The singular behavior at $\pm 2k_F$ defines the real space form $J^\alpha(r)$.
The latter can be obtained by Fourier transforming Eq. \eqref{eq:J_q} or 
by time integrating Eq. \eqref{eq:chi(rt)_T=0}.
The latter is done explicitly in Appendix \ref{sec:RKKY_real_space}. From Eq. \eqref{eq:chi(r)_app}
we then find
\begin{align}
	J^\alpha(r) 
	&= 
	- \frac{A_0^2 \lambda_T\sin(\pi g_\alpha)}{8 \pi^2 a \hbar v_F} 
	 \cos(2k_F x)
	\left(\frac{\pi a/\lambda_T}{\sinh(\pi |r|/\lambda_T)}\right)^{2g_\alpha}
\nonumber \\
	&\times 
	F\left(1/2, g_\alpha; 1; -\sinh^{-2}(\pi |r|/\lambda_T)\right),
\label{eq:J(r)_full}
\end{align}
where $F$ is the Gaussian hypergeometric function, defined in 
Eq. \eqref{eq:hypergeom}.
For $|r| \ll \lambda_T$, which is the case most of the time for the systems considered here,
and for $g>1/2$ we obtain the asymptotic behavior 
[see Eq. \eqref{eq:chi(r)_asymp_app} and Ref. \onlinecite{egger:1996}]
\begin{equation} \label{eq:J(r)}
	J^\alpha(r) 
	\sim \cos(2k_F r) (a/|r|)^{2g_\alpha-1}.
\end{equation}
Stronger electron-electron interactions lead to smaller $g_\alpha$ and so to
an RKKY interaction that extends to longer distances. 
Since the order discussed below is due to the long-range part of the RKKY interaction,
this leads to a better stabilization of the order and consequently to a higher
cross-over temperature.

Let us note that Eq. \eqref{eq:J(r)} cannot be extended to $g_\alpha \le 1/2$,
where it would describe an unphysical growth of $J(r)$ with distance. 
This regime is actually not reached for conventional LLs with $K_c > 0$ and $K_s =1$,
yet with the feedback below we obtain renormalized $g'_\alpha < 1/2$.
Eq. \eqref{eq:J(r)} then is regularized by further temperature-dependent
corrections coming 
from the expansion of the hypergeometric function (see Appendix \ref{sec:RKKY_real_space})
or at low temperatures by cutoffs such as the system length.
Since the relevant temperatures
determined below are such that $L < \lambda_T$ or at most $L \sim \lambda_T$,
the values $g'_\alpha < 1/2$ then lead to a RKKY interaction
that decays only little over the whole system range.


\section{Nuclear order}
\label{sec:nucl_order}

\subsection{Helical magnetization}

We have seen above that the ferromagnetic locking of the $N_\perp$ nuclear
spins in the direction across the 1D conductor leads to a 1D nuclear spin
chain of composite nuclear spins $\tilde{\bI}_i$ with maximal size $I N_\perp \gg 1$. 
This allows us to treat the nuclear subsystem semiclassically:\cite{simon:2007,simon:2008}
Pure quantum 
effects such as, for instance for antiferromagnetic chains, the Haldane gap for integer quantum 
spin chains\cite{haldane:1983,affleck:1989} or Kondo lattice physics\cite{tsunetsugu:1997} are absent, 
since their effect vanishes exponentially with increasing spin length.

In the present case the starting point is the classical ground state of 
the nuclear spins described by the Hamiltonian \eqref{eq:H_eff_n}.
The RKKY interaction $J_q$ reaches its minimum at $q=\pm 2k_F$, and the 
ground state energy is minimized by fully polarized $\tilde{I}_i = I N_\perp$
describing a helix with periodicity wave vector $2k_F$.
The corresponding ground states fall into two classes of different helicity,
\begin{equation} \label{eq:I_gs}
	\tilde{\bI}_i' = I N_\perp [\hat{\be}_x \cos(2k_F r_i) \pm \hat{\be}_y \sin(2k_F r_i) ],
\end{equation}
where $\hat{\be}_{x,y}$ are orthogonal unit vectors defining the spin $x,y$
directions. Through the spontaneous selection of the directions $\hat{\be}_{x,y}$
any rotational symmetry of the Hamiltonian \eqref{eq:H_eff_n} is 
broken.

The simultaneous existence of both helicities cannot occur for these
classical ground states because their superposition would result 
in a single helix, but with reduced amplitude. 
A coherent quantum superposition of such states, on the other hand,
can be excluded because each state involves the ordering of the 
large effective spins $\tilde{I} = I N_\perp$ over the whole system length $L$,
and hence $10^{5}$--$10^{8}$ individual spins.
This symmetry breaking and so the selection of a single 
helicity is in fact crucial for the 
feedback effect described in Sec. \ref{sec:feedback}. 
A single helix leads to a partially gapped electron system, while 
a superposition of the two helicities would result in an entirely 
gapped electron system. The physics of the latter case is interesting on its
own, but does not occur in the present case.

Transitions between both helical classes would involve a reorientation
of the entire nuclear system (and through the feedback 
of the electron system as well).
Such a transition, as well as the spontaneous emergence of domain walls, can be 
excluded because its energy cost scales with the system size due to the
long-range RKKY interactions.
The low-energy fluctuations about the ground state 
are either rotations of the entire nuclear spin system as a whole 
or magnons. 

Rotations of the whole system do not reduce the local magnetization,
yet they may lead to a zero time average. Due to the the aforementioned 
separation of energy scales between the nuclear and electron system, however,
the momentary nuclear spin configurations acts like a static, nonzero field
on the electrons. Our analysis, therefore, is not influenced by these 
modes. Moreover a pinning of those modes, since they involve the rigid 
rotation of the entire system, is very likely. 

More important are magnons, which describe the low energy fluctuations
to order $1/IN_\perp$. The magnon spectrum for the nuclear helimagnet 
is derived in Appendix \ref{sec:magnon_spectrum} (see also Ref. \onlinecite{simon:2008}). 
For the isotropic or anisotropic (when the feedback on the electrons is considered) RKKY 
interaction there exists a gapless magnon band with dispersion
given by Eq. \eqref{eq:omega_q_1},
\begin{equation} \label{eq:omega_q}
	\omega_q = 2 (IN_\perp) ( J^x_{2k_F-q} - J^x_{2k_F} ) / N_\perp^2.
\end{equation}
Let $m_i = \mean{\tilde{\bI}_i} \cdot \tilde{\bI}_i' / (IN_\perp)^2$ measure the component
of the average magnetization along $\tilde{\bI}_i'$, normalized to $-1 \le m_i \le 1$.
The Fourier component $m_{2k_F}$ then acts as an order parameter for the
nuclear helical order.
We can choose $0 \le m_{2k_F} \le 1$ by rotating the axes
$\hat{\be}_{x,y}$ if necessary. 
Magnons reduce this magnetization as follows\cite{simon:2007,simon:2008}
\begin{equation} \label{eq:m}
	m_{2k_F} = 1 - \frac{1}{IN_\perp} \frac{1}{N} \sum_{q \neq 0}
	\frac{1}{\e^{\omega_q / k_B T} - 1},
\end{equation}
where the sum represents the average magnon occupation number, and the
momenta $q$ run over the first Brillouin zone
$q \in [-\pi/a,\pi/a]$.


\subsection{Absence of order in infinite-size systems}
\label{sec:infinite_systems}

In the thermodynamic limit $L \to \infty$ the sum in 
Eq. \eqref{eq:m} turns into an integral that diverges as $L/a = N$
due to the $q \to 0$ magnon occupation numbers, showing that long-wavelength modes destabilize the long-range
order. It is noteworthy that this is not a consequence of the Mermin-Wagner theorem.\cite{mermin:1966}
The Mermin-Wagner theorem forbids long-range order in isotropic 
Heisenberg systems 
in low dimensions with sufficiently short-ranged interactions. 
An extension to the long-ranged RKKY interactions has been recently conjectured\cite{bruno:2001}
for the case of a free electron gas. 
The theorem thus cannot be applied for systems where the long-range RKKY interaction 
is modified by electron-electron interactions. Indeed, we have shown in previous
work\cite{simon:2007,simon:2008} that in this case long-range order of nuclear spins
embedded in 2D conductors becomes possible. In the present 1D case, however, the divergence
of the magnon occupation number at $q\to 0$ provides a direct example where long-range 
nuclear magnetic order is impossible in the $L \to \infty$ limit.

Realistically we always deal with samples of a finite length $L$ though.
The singularity at $q \to 0$ is cut off at momentum $\pi/L$, and the 
$q=0$ is absent in samples that are not rings. This means that the sum in 
Eq. \eqref{eq:m} is finite, and a finite magnetization is possible at 
low enough temperatures. We shall actually see below that even though 
the cutoff at $\pi/L$ plays a significant role for the 
stability of the order in realistic systems, the magnetization 
$m_{2k_F}$ is fully determined by $L$-independent quantities. 
This is much in contrast to what we would anticipate from the 
$L \to \infty$ limit.


\subsection{Order in finite-size systems}
\label{sec:finite_systems}

The energy representation of the momentum cutoff at $\pi/L$ is 
the level spacing
\begin{equation} \label{eq:DeltaL}
	\Delta_L = \hbar v_F / L.
\end{equation}
This level spacing must be carefully compared with any other temperature
scale characterizing the magnetization $m_{2k_F}$ of Eq. \eqref{eq:m}.

In particular, if $k_B T < \Delta_L$, the momentum quantization $\pi /L$ is
larger than the inverse thermal length, $\pi/ \lambda_T$. Since the 
width of the minimum of $J^x_{2k_k+q}$ is on the order of $\pi/\lambda_T$
(see Fig. \ref{fig:J}),
the first possible magnon energy $\omega_{\pi/L}$ is already
very large and close to the maximal value $2 I |J^x_{2k_F}|/N_\perp$.
If we define a temperature $T_{M0}$ at which 
$\omega_{\pi/L} / k_BT \approx 2I |J^x_{2k_F}|/N_\perp k_B T = 1$, 
we have for $T > T_{M0}$
\begin{equation}
	m_{2k_F}(T) 
	= 1 - \frac{1/IN_\perp}{\e^{(\frac{T_{M0}}{T})^{3-2g}}-1}
	= 1 - \frac{1}{IN_\perp} \left(\frac{T}{T_{M0}}\right)^{3-2g},
\end{equation}
with $g= g_{x,y} = (K_c + K_s^{-1})/2$ from Eq. \eqref{eq:g} for GaAs quantum wires
and $g = (K_c + 3)/4$ for SWNTs (see Sec. \ref{sec:nanotubes}).
We eventually write the magnetization as
\begin{equation} \label{eq:m_T*_0}
	m_{2k_F} = 1 - \left(\frac{T}{T^*_0}\right)^{3-2g},
\end{equation}
where we have defined the temperature $T^*_0$ by 
\begin{align}
	k_B T^*_0 
	&= (I N_\perp)^{\frac{1}{3-2g}} k_B T_{M0}
	= 2I^2 |J^x_{2k_F}|
\nonumber\\
	&= \left[ 
		2 I^2 C(g)
		\frac{A_0^2}{\Delta_a} 
		\left(\frac{\Delta_a}{k_B T^*_0}\right)^{2-2g} 
		\frac{\Gamma^2(g/2)}{\Gamma^2(1-g/2)} 
	\right]^{\frac{1}{3-2g}} 
\nonumber\\
	&= I |A_0| D \left(\frac{\Delta_a}{I |A_0|}\right)^{\frac{1-2g}{3-2g}},
\label{eq:T^*_0}
\end{align}
where $C(g)$ is defined in Eq. \eqref{eq:C}, $\Delta_a = \hbar v_F / a$
[Eq. \eqref{eq:Delta_a}], 
and where $D$ is the dimensionless constant
\begin{equation} \label{eq:D}
	D = \left[\sin(\pi g) \Gamma^2(1-g) (2\pi)^{2g-4} \frac{\Gamma^2(g/2)}{\Gamma^2(1-g/2)}\right]^{\frac{1}{3-2g}}.
\end{equation}
Eq. \eqref{eq:m_T*_0} can be considered as a generalized Bloch law for the
nuclear magnetization with an exponent $3-2g$ that depends on the electron-electron
interactions.

The arguments above are based on the assumption $k_B T < \Delta_L$ so that
Eq. \eqref{eq:m_T*_0} is in principle only valid if $k_B T^*_0 \ll \Delta_L$.
In Appendix \ref{sec:small_level_spacing} we show, however, that Eq. \eqref{eq:m_T*_0}
remains valid far into the range $k_B T > \Delta_L$. In particular, it remains
valid for GaAs quantum wires, where we find that $k_B T_0^*, k_B T^* \sim \Delta_L$,
where $T^*$ is the resulting cross-over temperature when taking into account the 
feedback onto the electron system [Eq. \eqref{eq:T^*}].

This means that the system length $L$ has no influence on the 
nuclear magnetization for any realistic SWNT and
GaAs quantum wire system.


\subsection{$T^*_0$ sets the only available energy scale}

\begin{figure}
	\includegraphics[width=0.9\columnwidth]{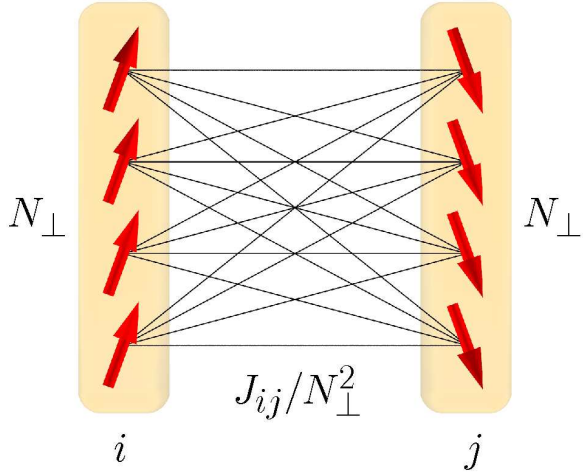}
	\caption{%
	Illustration of the RKKY coupling between two large spins composed
	of $N_\perp$ individual spins at sites $i$ and $j$. 
	The hyperfine interaction is reduced by distributing
	a single electron over $N_\perp$ nuclear spins, $A = A_0 / N_\perp$, resulting in 
	an RKKY interaction $J_{ij} / N_\perp^2$. 
	This reduction by $1/N_\perp^2$ is compensated because $N_\perp^2$ nuclear spins
	are mutually coupled through the same RKKY interaction $J_{ij}$.
	\label{fig:connections}
	}
\end{figure}

It may seem surprising that $T^*_0$ does no longer depend on $N_\perp$.
Yet we need to recall that the reduction of the RKKY interaction $J$ by $1/N_\perp^2$
is compensated through the coupling of two composite spins containing each
$N_\perp$ spins $I$. The coupling energy therefore depends on 
$(J/N_\perp^2) \times (I N_\perp) \times (I N_\perp) = I^2 J$, which no longer
contains $N_\perp$ (see Fig. \ref{fig:connections}). The cross-over temperature 
(since the $L$ dependence has 
been ruled out) can then only depend on the energy scales that characterize
$I^2 J$. Since $J$ is described by the width and the depth of its
minimum, which depend both on $T$, there is only one characteristic
temperature that can be self-consistently identified, by setting
$I^2 |J_{2k_F}(T)| = k_B T$. The result is $T^*_0$ 
[up to the factor 2 in Eq. \eqref{eq:T^*_0}],
which consequently must set the scale for the cross-over temperature,
independently of the chosen approach, be it magnons (as here), mean field
or more refined theories.

In the next section we see, however, that $T^*_0$ is strongly 
renormalized by a feedback coupling between the electron and nuclear spin 
systems, which modifies the shape of $J_q$ itself. The feedback in addition introduces a second
scale through a partial electron spin polarization that acts like a 
spatially inhomogeneous Zeeman interaction on the nuclear spins.


\section{Feedback effects}
\label{sec:feedback}

\subsection{Feedback on electrons}
\label{sec:feedback_el}

We have seen that the electrons enforce a helical ordering of the nuclear spins, and
we have assumed that this helical ordering defines the spin $(x,y)$ plane
[see Eq. \eqref{eq:I_gs}]. In the following, we analyze the feedback of the nuclear spin ordering
on the electrons using the Renormalization Group (RG) approach. 
Since the dynamics of the nuclear spins is much slower than that of the electrons, 
we can safely assume that the main effect of the nuclear spins is well captured
by a spatially rotating static magnetic field of the form:
$\bB(r)=B_x\cos(2k_F r)\hat{\be}_x + B_y\sin(2k_F r)\hat{\be}_y$,
with $B_x=B_y=  I A_0 m_{2k_F}$. Note that we explicitly (and arbitrarily) 
choose the counterclockwise helicity for the helical ordering.
The effective Hamiltonian for the electron system 
then reads $H_{el}^{\text{eff}}=H_{el}^{1D}+H_{Ov}$ with
$H_{el}^{1D}$ given by Eq. \eqref{eq:H_LL} and
\begin{equation}
	H_{Ov}=\int dr \, \bB(r) \cdot \bS(r),
\end{equation}
is the coupling to the nuclear Overhauser field.

Using the standard bosonization formulas,\cite{giamarchi:2004} $H_{Ov}$ is expressed as
\begin{eqnarray}\label{eq:hboso}
	H_{Ov}
	&=&
	\int \frac{dr}{2\pi a} \, B_{xy}
	\left[
		\cos\bigl(\sqrt{2}(\phi_c+\theta_s)\bigr)
	\right.
\nonumber\\ 
	&+& 
	\left.
		\cos\bigl(\sqrt{2}(\phi_c-\theta_s)-4k_F r\bigr)
	\right].
\end{eqnarray}
where 
\begin{equation}
	B_{xy}= I A_0 m_{2k_F}/2,
\end{equation}
and where we have not written the forward scattering part because it has no influence.
The last term is oscillating and is generally incommensurate except for $4k_Fa=2\pi n$, with $n$ integer.
This special case would correspond to a fine-tuning of $E_F$ to about $1.7$ eV for carbon nanotubes
or $0.2$ eV for GaAs quantum wires, which is quite
unrealistic for the systems we consider here. 
We therefore assume in the following that $4k_Fa\ne 2\pi n$, and hence drop the last incommensurate term
in Eq. \eqref{eq:hboso}. 
The remaining $\cos(\sqrt{2}(\phi_c+\theta_s))$ term has the scaling dimension
$(K_c + K_s^{-1})/2$. For the systems under consideration
this operator is always relevant, so that the interaction term is always driven to the strong coupling regime.
To see this more clearly we change the basis of boson fields by introducing the boson fields 
$\phi_+,\phi_-,\theta_+,\theta_-$, defined by%
\footnote{The signs in Eqs. \eqref{eq:phi_s} and \eqref{eq:theta_c} correct the signs in the published version
of this paper [Phys. Rev. B {\bf 80}, 165119 (2009)] and agree with the EPAPS document (Ref. \onlinecite{braunecker:2009EPAPS}) 
of Ref. \onlinecite{braunecker:2009}.}
\begin{eqnarray}
	\phi_c 
	&=& \frac{\sqrt{K_c}}{\sqrt{K}} 
	\left[ \sqrt{K_c} \phi_+ - \frac{1}{\sqrt{K_s}} \phi_- \right],
\label{eq:phi_c}\\
	\phi_s
	&=&  \frac{\sqrt{K_s}}{\sqrt{K}} 
	\left[  \frac{1}{\sqrt{K_s}} \theta_+ + \sqrt{K_c} \theta_- \right],
\label{eq:phi_s}
\\
	\theta_c
	&=&  \frac{1}{\sqrt{K_c K}} 
	\left[   \sqrt{K_c} \theta_+ - \frac{1}{\sqrt{K_s}} \theta_- \right],
\label{eq:theta_c}
\\
	\theta_s
	&=& \frac{1}{\sqrt{K_s K}} 
	\left[ \frac{1}{\sqrt{K_s}} \phi_+ + \sqrt{K_c}\phi_- \right],
\label{eq:theta_s}
\end{eqnarray}
where we have set 
\begin{equation}
	K = K_c + K_s^{-1}.
\end{equation}
The new fields obey the standard commutation relations 
$[\phi_\kappa(x),\theta_{\kappa'}(y)]=i\pi\delta_{\kappa,\kappa'}\mathrm{sign}(x-y)$ 
with $\kappa, \kappa'=+,-$. In this basis, the electron Hamiltonian reads 
\begin{eqnarray}
	&&H_{el} 
	=
	\int \frac{dr}{2\pi} 
	\left\{ \sum_{\kappa=\pm} 
		v_\kappa \Bigl[(\nabla \phi_\kappa)^2 + (\nabla \theta_\kappa)^2 \Bigr]
	\right.
\nonumber \\
		&&+   
		\frac{B_{xy}}{a}\cos (\sqrt{2K}\phi_+)
\label{eq:H_pm}
\\ 
		&&+\left.
		(v_c-v_s)\frac{1}{K}\sqrt{\frac{K_c}{K_s}}
		\Bigl[(\nabla \theta_+)(\nabla \theta_-)-(\nabla \phi_+)(\nabla \phi_-)\Bigr]
	\right\},\nonumber
\end{eqnarray}
with 
\begin{eqnarray}
	v_+ &&= (v_cK_c+v_sK_s^{-1})/K\\
	v_- &&= (v_cK_s^{-1}+v_sK_c)/K.
\label{eq:v-}
\end{eqnarray}
If $v_c=v_s$, the electron Hamiltonian separates into two independent parts 
$H_{el}=H_{el}^+ + H_{el}^-$ where $H_{el}^+$ is the standard sine-Gordon Hamiltonian, while 
$H_{el}^-$ is a free bosonic Hamiltonian. The cosine term is relevant and generates a gap in the `$+$' 
sector.
If $v_c\ne v_s$ the terms in $\nabla \theta_+\nabla \theta_-$ and $\nabla \phi_+\nabla \phi_-$ are marginal
and are much less important than the strongly relevant cosine term. 
In a first approximation, we neglect these terms. We will come back to this point 
in Sec. \ref{sec:rg_marginal} below.
The RG equation for $B_{xy}$ then reads\cite{giamarchi:2004}
\begin{equation} \label{eq:B_xy_RG}
	\frac{d B_{xy}}{dl}
	= \left(2-\frac{K_c}{2}-\frac{1}{2K_s}\right)B_{xy}
	= (2-g) B_{xy},
\end{equation}
where $l$ is the running infrared cut-off and
\begin{equation}
	g = K/2 = (K_c + K_s^{-1})/2.
\end{equation}
We use this $g$ for both GaAs quantum wires and SWNTs,
in contrast to the $g$ of Eq. \eqref{eq:g_SWNT} that must be used for 
SWNTs in the absence of the feedback.
As explained in Sec. \ref{sec:nanotubes}, this 
is due to the fact that the feedback acts on each of the two Dirac cones
of the the SWNT dispersion relation separately, and hence effectively splits the 
two bands of the SWNT into separate single-band models within each cone.
 
Under the RG flow, $B_{xy}(l)$ grows exponentially as does the associated 
correlation length $\xi = a(l) = a \e^{l}$. The flow stops when either
$\xi$ exceeds $L$ or $\lambda_T$, or when the dimensionless coupling constant\cite{giamarchi:2004}
$y(l) = B_{xy}(l)/\Delta_a(l)$, with
$\Delta_a(l) = \hbar v_F / a(l)$ 
becomes of order 1. From the latter condition we obtain a correlation length
\begin{equation}
	\xi = \xi_\infty = a [y(0)]^{-1/(2-g)} = a [B_{xy}/\Delta_a]^{-1/(2-g)}.
\end{equation}
We emphasize that with the cutoff criterion $y(l) \sim 1$ the magnitude of the 
resulting $B_{xy}$ has an $O(1)$ uncertainty. In fact, we use here a different cutoff 
criterion as in Ref. \onlinecite{braunecker:2009},
namely $\Delta_a$ instead of $E_F$. While the use of both cutoffs is generally justified
for the perturbative RG scheme used here (the cutoff must be on the order of the bandwidth),
we notice that when using $E_F$ we obtain for the GaAs quantum wires 
too large values for $B_{xy}$ that exceed $E_F$. 
This is unphysical as it would imply that more electrons are polarized
than are contained in the system, and so it just means that $B_{xy}$ becomes
comparable to $E_F$. The consistency of the RG scheme then requires that the RG flow 
must be stopped earlier, and the natural
scale in the kinetic part of the Hamiltonian \eqref{eq:H_pm} is set by $\Delta_a$.\cite{giamarchi:2004}
Due to the resulting smaller gap $B_{xy}$, however, for SWNTs the correlation length $\xi$ would exceed 
the system length $L$ at the new cutoff scale (while $\xi \approx L$ in Ref. \onlinecite{braunecker:2009}). 
Hence, for SWNTs the flow is cut off even earlier at $L$.

\subsection{Renormalized Overhauser field and gap for electron excitations}
\label{sec:rg_results} 

Independently of the precise form of the correlation length $\xi$
we can always write 
$y(l) = y(0) (\xi/a)^{2-g}$. Since furthermore $\Delta_a(0)/\Delta_a(l) = a/\xi$, 
we obtain the following result for the gap, i.e. the renormalized
Overhauser field $B_{xy}^* = B_{xy}(l)$,
\begin{equation} \label{eq:Delta}
	B_{xy}^* = B_{xy} \ (\xi/a)^{1-g},
\end{equation}
with 
\begin{equation}
	\xi = \min\left\{L,\lambda_T,\xi_\infty = a [B_{xy}/\Delta_a]^{-1/(2-g)}\right\}.
\end{equation}
Since $B_{xy} \propto A_0$ we can translate this directly into 
a renormalized hyperfine interaction constant
\begin{equation} \label{eq:A^*}
	A^* = A_0 \ (\xi/a)^{1-g}.
\end{equation}
It is important to notice that even though $A^*$ can be called a ``renormalized hyperfine interaction''
it no longer can be interpreted in the same way as $A_0$. 
It does not describe the on-site interaction between a nuclear spin and an electron spin,
but results from the reaction of the entire electron system to the ordered nuclear spin system. 
We can see this as analogous to the strong growth 
of an impurity backscattering potential in a LL,\cite{kane:1992,furusaki:1993} 
which then no longer corresponds to the 
coupling between the impurity and an electron but involves a collective screening response 
by the electron system.

The values of $A^*$ are listed in Table \ref{tab:values}. 
Since $I A^* \gg k_B T$, for all temperatures within the ordered phase, we find 
that precisely one half of the degrees of freedom, the $\phi_+$ fields, are gapped,
while the $\phi_-$ fields remain in the gapless LL state.
As we have shown in Ref. \onlinecite{braunecker:2009EPAPS}, this has the direct 
consequence that the electrical conductance through the 1D system drops by the 
factor of precisely 2. Since the gap is identical to the nuclear Overhauser field,
and so proportional to the nuclear magnetization $m_{2k_F}$, it in addition allows
to directly measure the nuclear magnetization through a purely electronic quantity,
the gap $B_{xy}^*$. Using Eq. \eqref{eq:m_T_intro} for $m_{2k_F}$, which shall be 
proved explicitly in Eq. \eqref{eq:m_T*} below, we can rewrite the gap as
\begin{equation}
	B_{xy}^* = \frac{I A^*}{2} \left[ 1 - \left(\frac{T}{T^*}\right)^{3-2g'} \right],
\end{equation}
with $T^*$ given by Eq. \eqref{eq:T^*_intro}, or Eq. \eqref{eq:T^*} below.
For SWNTs $A^*$ is independent of $m_{2k_F}$, and so the gap $B_{xy}^*$ is directly 
proportional to the magnetization $m_{2k_F}$. 
For the GaAs quantum wires, $A^*$ is independent of $m_{2k_F}$ for small 
magnetizations such that $\xi_{\infty} > L$, and the correlation length is set by $L$. 
For a magnetization $m_{2k_F} > m_{\infty} \equiv (\Delta_a/I A_0) (L/a)^{2-g}$ we have
$\xi_\infty < L$ and $A^*$ becomes a function of $m_{2k_F}$.
The gap therefore follows the curve
\begin{equation} \label{eq:B_T}
	B_{xy}^* \propto
	\begin{cases}
	m_{2k_F}  =
	\left[ 1 - \left(\frac{T}{T^*}\right)^{3-2g'} \right] 
	& \text{for $m_{2k_F} < m_\infty$,}
\\
	m_{2k_F}^{\frac{3-g}{2-g}} 
	=
	\left[ 1 - \left(\frac{T}{T^*}\right)^{3-2g'} \right]^{\frac{3-g}{2-g}}
	&\text{otherwise}.
	\end{cases}
\end{equation}
Notice that the value $m_\infty$ for the cross-over magnetization 
can be tuned by the 
system length $L$. If $\lambda_T < L$, the $L$ in the definition of $m_\infty$
is replaced by $\lambda_T$.

The physical meaning of the gapped field $\phi_+$ is best seen by 
rewriting it in terms of the original boson fields $\phi_{\ell\sigma}$
using Eq. \eqref{eq:phi_phi_theta},
\begin{equation}
	\phi_+=(\phi_c + \theta_s)/\sqrt{K}=(\phi_{R\dw} + \phi_{L\up})/\sqrt{2K}.
\end{equation}
A gap in the `+' sector means that a linear combination of the spin $\dw$ electron right movers and
spin $\up$ electron left movers is gapped. This combination is pinned by the nuclear helical state.
This can be seen as the analog of a spin/charge density wave order except that it involves a mixture
of charge and spin degrees of freedom.
As shown in Sec. \ref{sec:electron_polarization}, this pinned density wave 
corresponds to an electron spin polarization following the nuclear helical order.

\subsection{Corrections by the marginal terms}
\label{sec:rg_marginal} 

The results of the RG analysis above
remain almost unchanged if we take into account the terms 
$\nabla\theta_+ \nabla\theta_-$ and $\nabla \phi_+\nabla \phi_-$. To support this assertion, 
we have checked numerically that
that $y = B_{xy}/\Delta_a$ reaches its cutoff scale, while the other coupling
constants remain almost unchanged.
Following Ref. \onlinecite{giamarchi:2004}, we then expand the cosine term up to second order. This provides a mass 
term $\propto B_{xy}$ for the $\phi_+$ mode. Within this approximation, the $\phi_+,\theta_+$ bosonic fields can be
exactly integrated out in the quadratic action.
This leaves us with an effective Hamiltonian $H_{\text{eff}}^-$ for the fields $\phi_-,\theta_-$, which has 
precisely the same form as $H_{el}^-$ up to some irrelevant terms,
\begin{equation} \label{eq:H_eff_-}
	H_{\text{eff}}^- =  v_-^{\text{eff}}\int \frac{dr}{2\pi} 
	\left[ \frac{1}{K_-^{\text{eff}}}(\nabla \phi_-(r))^2 + K_-^{\text{eff}}(\nabla \theta_-(r))^2\right],
\end{equation}
but with the renormalized parameters
$K_-^{\text{eff}}=\sqrt{1-\Gamma}$ and $ v_-^{\text{eff}}=v_-\sqrt{1-\Gamma}$, with
\begin{equation}
	\Gamma = \frac{K_c}{4K_s}\frac{(v_c-v_s)^2}{(v_cK_c+v_sK_s^{-1})(v_cK_s^{-1}+v_sK_c)}.
\end{equation}
For the systems under consideration
the factor $\sqrt{1-\Gamma}$ is slightly less than 1 and this renormalization 
has indeed no quantitative consequences. 


\subsection{Electron spin polarization}
\label{sec:electron_polarization}

The pinning of the $\phi_+$ modes leads to a partial polarization of the 
electron spins. This polarization follows the helix of the nuclear Overhauser
field and is parallel (for a ferromagnetic $A_0$) or antiparallel (for an
antiferromagnetic $A_0$) to the nuclear spin polarization. 

Within the LL theory, the form of this polarization can be found very easily:
The forward scattering contribution of $S_z \propto \nabla\phi_s$ has a zero
average. There remain the backscattering parts given by the $O_{SDW}^\alpha$
operators in Eqs. \eqref{eq:O_SDW_x} -- \eqref{eq:O_SDW_z}.
Since averages over exponents consisting of single boson fields vanish
in the LL theory,
$\mean{\e^{i \phi_-}} = \mean{\e^{i \theta_-}} = 0$ (up to finite size corrections
on the order of $a/L$), 
only the contribution to the spin density average consisting uniquely  
of the $\e^{\pm i \sqrt{2K} \phi_+}$ operators are nonzero because $\phi_+$ is pinned
at the minimum of the cosine term in Eq. \eqref{eq:H_pm}.
For $A_0 > 0$ this minimum is at $\sqrt{2K} \phi_+ = -\pi$ and for 
$A_0<0$ at $\sqrt{2K} \phi_+ = 0$. 
The two backscattering parts depending only on $\phi_+$ are
\begin{equation}
	\mean{\psi_{L\up}^\dagger(r) \psi_{R\dw}(r)}
	= \frac{\e^{2i k_F r}}{2\pi a} \e^{i \sqrt{2K} \phi_+}
	= -\frac{\e^{2i k_F r}}{2\pi a} \mathrm{sign}(A_0),
\end{equation}
and the conjugate 
$\mean{\psi_{R\dw}^\dagger(r) \psi_{L\up}(r)}$.
This leads to the LL result for the electron spin polarization density 
(with $S=1/2$)
\begin{equation}
	\mean{\bS(r)}_{LL}
	= -S \frac{\mathrm{sign}(A_0)}{\pi a}
	\begin{pmatrix}
		\cos(2k_F r) \\ \sin(2k_F r) \\ 0 
	\end{pmatrix}.
\end{equation}
The correct prefactor, the polarization density, cannot be obtained
from the LL theory, which only provides the dimensionally correct prefactor $1/\pi a$. 
This unphysical result is a direct consequence
of neglecting bandwidth and band curvature effects in the LL theory, 
which can lead to a violation of basic conservation laws. Concretely we obtain
here an electron polarization $\propto 1/a$ which largely exceeds the electron 
density $\propto k_F$ in the system.
To cure this defect, we use the following heuristic argument in the spirit
of Fr\"{o}hlich and Nabarro:\cite{froehlich:1940} 
The process of opening the gap $B_{xy}^*$ in the $\phi_+$ field is carried 
mostly by the electrons within the interval $B_{xy}^*$ about the Fermi energy
(using a free-electron interpretation). 
Hence, the polarization is on the order of $B_{xy}^*/E_F$. 
Note that the factor $1/2$ in $B_{xy}^* = I |A^*| m_{2k_F}/2$ can now be interpreted
as showing that only one half of the electron modes is gapped.
When going to the tight-binding description this amplitude must in addition
be weighted by the ratio of electron to nuclear spin densities $n_{el}/n_I$, expressing
that the number of electrons is much smaller than the number of nuclear spins $\tilde{\bI}$
in SWNT's and GaAs quantum wires.
This leads to our estimate of the local electron polarization
\begin{equation} \label{eq:S_i_average}
	\mean{\bS_i}
	\approx -\frac{I S A^* m_{2k_F}}{2 E_F}  \frac{n_{el}}{n_I}
	\begin{pmatrix}
		\cos(2k_F r_i) \\ \sin(2k_F r_i) \\ 0 
	\end{pmatrix}.
\end{equation}
This argument gives us furthermore an upper bound to the effective
hyperfine coupling constant, $I A^* < E_F/2$, telling that at most 
half of the electrons can be polarized. Further bounds and self-consistency 
checks are discussed in Sec. \ref{sec:self_consistency}.

For the chosen systems, this electron polarization is small. Assuming $m_{2k_F}=1$,
we find for GaAs quantum wires $|\mean{\bS_i}|/S = 2 \times 10^{-3}$ and
for SWNTs $|\mean{\bS_i}|/S = 3 \times 10^{-6}$
(see Table \ref{tab:values}).


\subsection{Feedback on nuclear spins}
\label{sec:feedback_nucl}

The partial helical polarization of the electrons naturally modifies
the susceptibilities and so the RKKY interaction. In addition, 
according to the principle \emph{actio $=$ reactio}, the polarized electrons
together with the renormalized coupling constant $A^*$ create a
magnetic field that acts back on the nuclear spins, hence polarizing them.
Therefore the stabilization of the nuclear order has now two ingredients:
The minimum of the RKKY interaction as before (yet with a modified shape),
and the Zeeman-like (but helimagnetic) polarization by $A^* \mean{\bS_i}$.

We stress that these are two different energy scales and,  
in particular, the Zeeman-like energy does not provide an upper bound
to the RKKY scale. Indeed, both expressions are of order $A_0^2$
because $\mean{\bS} \sim A^*/E_F$ as shown just above, and so they 
do not follow from a first and a second order perturbative expansion.
In addition, we have seen in the previous section that the electron polarizations
are indeed very small. Therefore we shall
see below that the RKKY interaction is dominated by the gapless $\phi_-$ 
modes, and hence involves different electrons than $\mean{\bS}$.
The bounds for the validity of the perturbation theory are, in fact,
imposed differently. As this is an important criterion of controllability
of the theory, we analyze it in Sec. \ref{sec:validity_RKKY}.
We show there that the perturbative expansion is indeed justified for
SWNTs and GaAs quantum wires.

The modified Hamiltonian for the nuclear spins then becomes
\begin{equation} \label{eq:H_n^eff_pol}
	H_n^{\text{eff}}
	= \sum_i \frac{A^* \mean{\bS_i} }{N_\perp} \cdot \tilde{\bI}_i
	+ \sum_{ij} \frac{J_{ij}^{\prime \alpha}}{N_\perp^2} \tilde{\bI}_i \cdot \tilde{\bI}_j.
\end{equation}
In the derivation of the modified RKKY interaction $J'$ we suppress any 
occurrence of $\mean{\bS_i}$ because 
such terms are of order $O(A_0^3)$ and are neglected in the perturbative 
expansion. Fluctuations involving the gapped fields $\phi_+$ and $\theta_+$ 
have furthermore a much reduced amplitude due to the gap $IA^*$,
and can be neglected compared with the RKKY interaction carried by the 
gapless modes $\phi_-$ and $\theta_-$ only. Fluctuations of the gapped fields
become in fact important only at temperatures $k_B T > I A^*$, which is
much larger than the characteristic temperatures of the ordered phase (see Table \ref{tab:values}).
This allows us 
to neglect any occurrence of $\phi_+$ and $\theta_+$ in the RKKY interaction.
The details of the modification of the susceptibilities are worked out 
in Appendix \ref{sec:susc_fb}.

The result is a susceptibility, and so a $J_q$, of a gapless LL described by $\phi_-$ and $\theta_-$
of the same form as Eq. \eqref{eq:J_q} with 
modified exponents $g_\alpha \to g_\alpha'$ that are determined by the prefactors of the $\phi_-$ and 
$\theta_-$ fields in the transformations \eqref{eq:phi_c}--\eqref{eq:theta_s}, 
and a modified velocity $v_F \to v_-$.
Since the nuclear Overhauser field singles out the spin $(x,y)$ plane over
the $z$ direction, anisotropy appears between $\chi_x=\chi_y$
and $\chi_z$. 
This is mainly expressed in different exponents $g_{x,y}' \neq g_z'$, but also 
in that the amplitudes of $\chi_{x,y}$ are only $1/2$ of that of $\chi_z$ because
one half of the correlators determining $\chi_{x,y}$ depend only on $\phi_+$, 
while all correlators for $\chi_z$ depend on $\phi_-$ and $\theta_-$.

From the results of the detailed calculation in Appendix \ref{sec:susc_fb}
we then see that the new RKKY interaction $J^{\prime\alpha}_q$ has precisely the same form 
as Eq. \eqref{eq:J_q} with the replacements
\begin{equation} \label{eq:J'_q}
	J_q^{\prime x,y} = J_q^{x,y}(g'_{x,y},v_-)/2,
	\quad
	J_q^{\prime z} = J_q^z(g'_z,v_-),
\end{equation}
and the exponents
\begin{align}
	g' = g'_{x,y} 
	&= 2 K_c/ K_s (K_c+K_s^{-1}),
\label{eq:g'_xy}
\\
	g'_{z} 
	&= (K_c K_s^{-1}+ K_c K_s) / 2 (K_c+K_s^{-1}),
\label{eq:g'_z}
\end{align}
satisfying $g'_{x,y,z} < g$ and $g'_z < g'_{x,y}$ (for $K_s=1$ we have 
precisely $g'_z = g'_{x,y}/2$)
for the nanotube and quantum wire systems.
Note that this single-band result is also quantitatively valid for the
SWNTs, as explained in Sec. \ref{sec:nanotubes}.

For the exponents we have quite generally $g'_{x,y} > g'_{z}$. Together
with the difference in amplitudes we see that $|J^{\prime x,y}_q| < |J^{\prime z}_q|$.
Naively this would mean that the system could gain RKKY energy by aligning the 
nuclear spins along the $z$ axis. 
However, this would destroy the feedback 
effect, and so lead to a large overall cost in energy. 
The helical order in the $(x,y)$ plane is therefore protected against 
fluctuations in the $z$ direction.
Since this is another important self-consistency check, 
a detailed analysis can be found in Sec. \ref{sec:stability_planar_order}.


\subsection{Modification of the cross-over temperature}
\label{sec:modification_Tc}

The analysis above shows that the ground state magnetization of the nuclear spins remains
a nuclear spin helix confined in the spin $(x,y)$ plane even when the feedback is taken into account.
The order parameter for the helical order remains $m_{2k_F}$.
Thermally excited magnons reduce this order parameter in the same way as before. 
As long as the gap $B_{xy}^*$ remains much larger than $k_B T$ the gapped modes
$\phi_+$ and $\theta_+$ remain entirely frozen out, and the reaction of the 
electron system is described by only the ungapped modes $\phi_-$ and $\theta_-$,
leading to the RKKY interaction $J^{\prime \alpha}_q$ as derived just above in 
Sec. \ref{sec:feedback_nucl}.
The evaluation of the magnon occupation number leading to Eq. \eqref{eq:m_T*_0}
remains otherwise identical upon the replacement $J_{2k_F} \to J^{\prime \alpha}_{2k_F}$
with $\alpha = x,y$. 
Since $|J^{z}_q| > |J^{x,y}_q|$, there is now a second magnon band $\omega^{(2)}_q$
with negative energies at $q \approx 2k_F$ (see Appendix \ref{sec:magnon_spectrum}). 
This usually means that the assumed
ground state is unstable. But in the same way as stated in the previous paragraph, the feedback
protects the ordered state against such destabilizing fluctuations. The details 
are again worked out in Sec. \ref{sec:stability_planar_order}, and as a consequence
we can neglect this second magnon band $\omega^{(2)}_q$ entirely.

Let us now look at the influence of the electron polarization on the nuclear spins.
The effective magnetic field created by the electrons is $B_{el} = |A^* \mean{\bS_i}|$.
With the polarizations estimated in Eq. \eqref{eq:S_i_average} we obtain
for GaAs quantum wires $B_{el} \sim 0.3$ $\mu$eV $\sim 4$ mK, 
and for SWNTs $B_{el} \sim 36$ neV $\sim 0.4$ $\mu$K. 
Both scales are very small compared with the values of $T^*$ we shall obtain 
from the modified RKKY interaction right below.
Hence, we can entirely neglect these magnetic fields, and so the gap they generate
in the magnon spectrum.

The magnon band is thus of the same type as Eq. \eqref{eq:omega_q},
and repeating the analysis of Sec. \ref{sec:finite_systems}
we obtain a magnetization of the form
\begin{equation} \label{eq:m_T*}
	m_{2k_F} = 1 - \left(\frac{T}{T^*}\right)^{3-2g'},
\end{equation}
with the cross-over temperature [combining Eqs. \eqref{eq:T^*_0} and \eqref{eq:J'_q}]
\begin{equation}\label{eq:T^*}
	k_B T^*
	= I |A_0| D' \left(\frac{\Delta_a}{I |A_0|}\right)^{\frac{1-2g'}{3-2g'}},
\end{equation}
where $g'$ is given by Eq. \eqref{eq:g'_xy}, again $\Delta_a = \hbar v_F /a$ [Eq. \eqref{eq:Delta_a}], 
and
\begin{align}
	&D' = 
	\left(\frac{v_-}{v_F}\right)^{\frac{1-2g'}{3-2g'}}
\nonumber \\
	&\times
	\left[ 
	\frac{\sin(\pi g') \Gamma^2(1-g')(2\pi)^{2g'-4}}{2} 
	\frac{\Gamma^2(g'/2)}{\Gamma^2(1-g'/2)}
	\right]^{\frac{1}{3-2g'}},
\label{eq:D'}
\end{align}
with $v_-$ from Eq. \eqref{eq:v-}.
These expressions replace the magnetization $m_{2k_F}$ in Eq. \eqref{eq:m_T*_0} and 
the temperature $T^*_0$ in Eq. \eqref{eq:T^*_0}.
As noted in Sec. \ref{sec:rg_results} the $m_{2k_F}$ can be directly detected
by measuring the electron excitation gap $B_{xy}^*$.


\section{Self-consistency conditions and generalizations}
\label{sec:self_consistency}

The strong renormalization of the system properties through the feedback
between the electron and nuclear systems below $T^*$ requires a reexamination
of the underlying conditions. We start with discussing the validity of the 
LL theory, which forms the starting point of our analysis.
The validity of the renormalized RKKY treatment is examined next.
This is followed by the investigation of the stability of the nuclear helimagnet
to a macroscopic realignment of the nuclear spins in the $z$ direction
that seems to be favored by the anisotropy of the modified RKKY.
We show that the nuclear helimagnet is stabilized
through the feedback. Finally, we show that intrinsic anisotropy in
the hyperfine interaction does not change our conclusions as long as
it maintains a finite magnetization along a cross-section through the 1D conductor.
The validity of using a single-band model for SWNTs is discussed in 
Sec. \ref{sec:nanotubes}.

\subsection{Validity of Luttinger liquid theory}
\label{sec:validity_LL}

The LL theory defined by Eq. \eqref{eq:H_LL} is an exact theory for 1D 
electron conductors with a perfectly linear electron dispersion relation.
The eigenstates of such a system are bosonic density waves.
Including electron-electron interactions does not change the nature of 
these eigenstates, but leads mainly to a renormalization of the LL parameters 
$K_c$ and $K_s$. 
Realistically, however, the dispersion relation is
not perfectly linear and restricted to a finite bandwidth, and 
electron-electron interactions can have a more substantial influence.
In such a situation the LL theory remains valid as long as (for the considered 
physical quantities) the bosonic 
density waves remain close to the true eigenstates and decay only over
a length scale exceeding the system length.

Deviations from LL behavior induced by the electron band curvature close
to $\pm k_F$ was investigated in Refs. \onlinecite{khodas:2007,glazman:2009a,glazman:2009b}.
Let us encode this curvature in a mass $m^*$ such that the 
electron dispersion reads $\epsilon_q = v_F q + q^2 / 2 m^*$, where $q$
is measured from $\pm k_F$. Defining then the parameter
$\varepsilon = (\omega - \epsilon_q)/(q^2 / 2 m^*)$ it was 
shown\cite{glazman:2009a}
that deviations from LL behavior become important at $|\varepsilon| \lesssim 1$
provided that $|q| \ll k_F$. 
In our case, the electron correlation functions are evaluated in the static limit $\omega = 0$, 
and by setting $q_m = 2 v_F m^*$ this condition becomes $|q| \gtrsim q_m$.

For armchair SWNTs we estimate\cite{manuel}
$m^* \approx 0.2 m_0$ (with $m_0$ the bare electron mass) within a few 0.1 eV about the 
Dirac points. This mass is very large, reflecting the
almost perfect linear dispersion of the armchair SWNTs. Accordingly 
this leads to a $q_m \approx 3$ nm$^{-1} \gg k_F$, or to an energy scale of about 2 eV.
Hence, $q_m \gg |q|$ for any $|q| \ll k_F$. 
For GaAs quantum wires the effective mass at the $\Gamma$ point is\cite{adachi:1985}
$m^* = 0.067 m_0$, and so
$q_m = 2 \times m_0 v_F / \hbar = 2.3 \times 10^{8} $m$^{-1}$.
This is slightly larger than $k_F$ and again we find that
$q_m \gg |q|$ for any $|q| \ll k_F$.
Therefore, for both systems the curvature-induced deviations from the
LL theory are negligible.

A different curvature-induced deviation from standard LL theory occurs at very low electron densities,
leading to the so-called incoherent LL
(see e.g. Refs. \onlinecite{matveev:2004a,matveev:2004b,fiete:2007,deshpande:2008}
and references therein).
At these densities the Coulomb energy $E^{\text{pot}}$ largely overrules the kinetic energy 
$E^{\text{kin}}$ of the electrons (expressed by a ratio $R = E^{\text{pot}} / E^{\text{kin}} \gg 1$),
and the electrons order in a Wigner crystal, while the electron spins form a Heisenberg chain.
Such a system still allows a bosonized description,\cite{matveev:2004a,matveev:2004b} 
yet with a large splitting between spin and
charge excitation energies. Realistic temperatures lie above the 
spin excitation energies, but can lie below the charge excitation energies.
The charge fluctuations then remains in a LL state, while the spin fluctuations
have an incoherent behavior.
The ratio $R$ depends much on the band mass $m$ of the system. 
For a quadratic dispersion we have
$E^{\text{kin}} = k_F^2/2m \propto n_{el}^2/2m$. 
For a potential energy $E^{\text{pot}} = n_{el} e^2/\epsilon$ (with $e$ the electron charge
and $\epsilon$ the dielectric constant) we obtain
$R = E^{\text{pot}}/E^{\text{kin}} \propto m/n_{el}$.
The incoherent LL regime can therefore be reached in systems with a large mass and 
a low electron density.\cite{deshpande:2008} In the GaAs/AlGaAs heterostructure of the quantum 
wires we have\cite{adachi:1985} $\epsilon \approx 12$, and $E^{\text{kin}} = E_F$.
With the values from Table \ref{tab:values} we find that
$R = n_{el} e^2/\epsilon E_F \sim 0.2$, excluding the incoherent LL.

For a linear spectrum such as in the SWNTs the criterion above does not 
apply. Indeed, for a linear dispersion $E^{\text{kin}} \propto n_{el}$ and 
so the ratio $R = E^{\text{pot}} / E^{\text{kin}}$ is independent of density.
Electron interactions then primarily modify the dielectric constant 
$\epsilon$ and so $K_c$. This leads to a much weaker dependence of $R$ on the interaction
strength, and makes the LL description valid for, in principle, arbitrary 
electron-electron interactions. For SWNTs we have the estimate\cite{egger:1997,egger:1998}
$\epsilon \approx 1.4$ leading to 
$R = e^2 / \pi \hbar v_F \epsilon \approx 0.6$, allowing us to exclude
the incoherent LL as well.


\subsection{Validity of RKKY approximation and bounds on $T^*$}
\label{sec:validity_RKKY}

The RKKY approximation is valid under two conditions.
First, as it is a perturbative expansion in powers of $A_0$, we must verify 
that higher perturbative orders remain smaller than the lower perturbative orders.
Related to this we must examine that the energy scale $k_B T^*$ obtained from 
the RKKY interaction does not violate bounds imposed by the original
Hamiltonian \eqref{eq:H}.
Second, the separation of time scales between the electron and nuclear spin 
systems must be guaranteed, in order to be allowed to interpret the RKKY 
interaction $J'_q$ as instantaneous for the nuclear spins.

\subsubsection{Upper bound on $T^*$}

The perturbative derivation of the RKKY interaction\cite{RKKY} or equivalently 
its derivation through a
Schrieffer-Wolff transformation\cite{simon:2007,simon:2008} consists
in an expansion in $A_0$. The lowest order is proportional to $A_0$, 
the scale of the second order is set by $J_q \propto A_0^2/E_F$,
higher orders scale in further powers of $A_0/E_F$, and so
the condition of validity of perturbation
theory is usually set equal to the condition $|A_0| / E_F \ll 1$. 
This condition is perfectly met for the SWNTs or the GaAs quantum wires.
Yet, there are a few subtleties. First, in the absence of the feedback, we have 
$J_q = (A_0^2/E_F) \times $(number), and the latter number can become
very large. Its maximum defines indeed the scale $k_B T^*_0$.
Through the Schrieffer-Wolff transformation we have in addition 
eliminated the term in the Hamiltonian, which is linear in $A_0$,
and so there is no longer a proper ``first order'' expression to which 
we can compare $J_q$. Hence, the validity of the RKKY scale $k_B T^*_0$ 
must be checked in a different way.

If we entirely neglect the electron Hamiltonian, i.e.
$H= \sum_i A_0 \bS_i \cdot \bI_i$, we can ask which maximal energy scale 
can be obtained from the hyperfine Hamiltonian for the nuclear spins. 
Obviously this scale is obtained by polarizing all electrons such that
$H = \sum_i A_0 (n_{el}/n_I) S\hat{\be} \cdot \bI_i$,
with $S=1/2$, $n_{el}/n_I$ the ratio of electron to nuclear spin densities,
and $\hat{\be}$ an arbitrary unit vector (that may or may not be position-dependent).
For temperatures smaller than the field
$B_0^* = |A_0| S n_{el}/n_I$ all nuclear spins are aligned along $\hat{\be}$ because a mismatch with the fully 
polarized electrons costs the on-site energy $B_0^*$. 
Essential for this argument is that flipping a nuclear spin out of its alignment
costs only energy from the hyperfine interaction. The electron system
is assumed to be energetically unaffected by this process (the hyperfine interaction
conserves the total of nuclear and electron spins, and so the electron spin changes
as well).
Otherwise said, the electron Hamiltonian
$H_{el}$ is independent of the electron polarization. 
Within this framework the RKKY coupling between the nuclear and electron spins
corresponds just to a more sophisticated way of treating the nuclear spin
fluctuations. 
Since the electron state has no influence by assumption,
the maximal energy scale set by $B_0^*$ cannot be overcome by any characteristic temperature  
obtained through the RKKY interaction. 
This argument therefore applies directly to the case when we neglect the feedback 
between the nuclear spins and the LL. The condition then becomes
\begin{equation} \label{eq:validity_RKKY_1}
	k_B T^*_0 \le B_0^* = S A_0 \frac{n_{el}}{n_I}.
\end{equation}
With the values from Table \ref{tab:values} we can verify that this condition is indeed
satisfied. 

The situation is of course very different if both systems are tightly bound together. 
In this case, one pays not only the energy $B_0^*$ but also the energy resulting from the
modification of the electron state. Through the feedback between both systems,
this extra cost in energy is roughly taken into account through the renormalized hyperfine coupling 
constant $A^*$ [see Eq. \eqref{eq:A^*}]. 
If we assume that $A^*$ fully describes the maximal electron response to the hyperfine
coupling, then the scales obtained from the modified RKKY description again cannot overcome this scale.
Hence, we have the modified condition
\begin{equation} \label{eq:B^*}
	k_B T^* \le B^* \equiv S A^* \frac{n_{el}}{n_I}.
\end{equation}
We must interpret this inequality with caution though.
Only one half of the low-energy electron modes
contribute to the renormalization of $A^*$. The Coulomb interaction between the remaining
electrons, which in fact substantially modifies the RKKY interaction and determines $T^*$, 
is not taken into account. Yet, this modified RKKY interaction is a direct consequence of the 
strong coupling to the nuclear system as well, and so the scale $B^*$ may still require further adjustment. 
A hint for this is seen for instance in Fig. \ref{fig:T_Kc_GaAs}, where $k_B T^*$ exceeds 
$B^*$ for $K_c < 0.5$. 

Nonetheless we use Eq. \eqref{eq:B^*} as an upper bound to $k_B T^*$, 
since we do not know if such an extrapolation beyond $B^*$ remains valid
within the RKKY framework. 
However, we interpret Eq. \eqref{eq:B^*} as assuring the validity of the theory, 
and not necessarily as a maximal upper bound on $k_B T^*$.

If we look again at Table \ref{tab:values} we see that $k_B T^* \approx B^*$ 
for the selected values of GaAs quantum wires and SWNTs [within the $O(1)$ uncertainty]. 
Hence, the fluctuations of the gapless modes
stabilize the nuclear order up to the scale $B^*$ set by the gapped modes.
This equality of the scales for the parameters of Table \ref{tab:values}
is actually a coincidence, as can be seen from Fig. \ref{fig:T_Kc_GaAs}.

Note that if $B^*$ is controlled by the cutoff $\xi = L$,
increasing the system length also increases $B^*$ because more electrons 
are involved in the feedback effect. 
In SWNTs, for instance, $\xi=L$ and $B^* \propto (L/a)^{1-g} = (L/a)^{0.4}$.
This means that doubling $L$ corresponds to an increase of $B^*$ by about 1.3.
Since $T^*$ is independent of $L$ such a control of the bound $B^*$
may be quite useful.

\subsubsection{Separation of time scales}

The RKKY interaction requires that the electron response
to a change of the nuclear spin configuration is instantaneous, and so 
a strict separation of time scales between both systems is mandatory. 
The dynamics of 
the nuclear spins is described by the Hamiltonian \eqref{eq:H_n^eff_pol} and 
consists of two parts. The precession of the nuclear spins in the magnetic field
generated (self-consistently) by the polarized electrons and the renormalized
coupling constant $A^*$, and the dynamics from the RKKY interaction $J'_q$.
The former leads to an energy scale $A^* \mean{\bS} \sim (A^*)^2/E_F$, 
which needs to be compared with $E_F$. This results in the condition 
\begin{equation} \label{eq:validity_RKKY_3}
	A^* \ll E_F.
\end{equation}
Notice that this argument is very different from the previous argument
leading to Eqs. \eqref{eq:validity_RKKY_1} and \eqref{eq:B^*},
as it requires the physical, fully self-consistent averages,
not a maximal energy condition.
From Table \ref{tab:values} we see that condition \eqref{eq:validity_RKKY_3} is met
for SWNTs and GaAs quantum wires. Since $A^*/E_F$ measures essentially the proportion 
of electron spin polarization (see Sec. \ref{sec:electron_polarization}), it means 
that only a very small fraction of the electrons is polarized.

On the other hand, the time scale set by $J'_q$ can be identified with the dynamics of 
the fluctuations it describes, and so with the magnon dynamics.
We therefore compare the maximal magnon velocity with the Fermi velocity $v_F$.
For temperatures $T \lesssim T^*$, the maximal magnon
velocity $v_m$ is obtained by the slope of $\omega_q$ at the 
momentum cutoff $q = \pi/L$. We can then use the $T=0$ expression for $J_q$,
as it sets an upper bound to the slope, and obtain from Eqs. \eqref{eq:J_q_T=0}, \eqref{eq:omega_q},
and \eqref{eq:J'_q}
\begin{equation}
	\frac{v_m}{v_F}
	\sim 
	\frac{(2-2g') \sin(\pi g')(2/\pi)^{3-2g'} v_F}{16\pi^2 v_-}
	\frac{I}{N_\perp} \frac{A_0^2}{\Delta_a^2} 
	\left(\frac{L}{a}\right)^{3-2g'}.
\end{equation}
The first 3 factors are small and can overcome the large last factor.
Indeed, for SWNTs we have $v_m/v_F \sim 10^{-9}/ N_\perp$, and for 
GaAs quantum wires (with $L=$ 10 $\mu$m) $v_m/v_F \sim 10^{-3} / N_\perp$.
The already small prefactors are in addition strongly suppressed by the
number $N_\perp$ of nuclear spins in the direction across the 1D system.
The small values for $v_m$ mean that $q=\pi/L$ lies already 
in a region where $\omega_q$ is essentially flat. This is, in fact,
the same criterion we have used for the determination of $T^*$.

The necessary separation of time scales is therefore fulfilled by the
systems under consideration.


\subsection{Stability of the planar magnetic order}
\label{sec:stability_planar_order}

We have observed in Sec. \ref{sec:feedback_nucl} that quite generally 
$|J^{\prime z}_q| < |J^{\prime x,y}_q|$ because
of the smaller exponent $g'_z$ and the larger prefactor of $J^{\prime z}_q$.
Naively, this means that the nuclear spin system can gain RKKY energy by 
forming an Ising-like configuration along the spin $z$ direction. 
An alignment in this direction, however, would destroy the feedback
described above, and so destroy the net energy gain from the planar $(x,y)$ order 
in both the electron and spin systems. 
To keep this feedback and the energy gain active, a deviation from the planar order
of the nuclear spins is not possible.

Indeed, the narrow minimum of $J^{\prime z}_q$ at $q=2k_F$
implies that if there is a magnetization $m_z$
along the spin $z$ direction it has only $q=\pm 2k_F$ Fourier components, 
because they are energetically most favorable. 
The reality of the expectation value of each nuclear spin then imposes that 
we can write the ground state expectation values in the form
$\mean{\tilde{\bI}_i} = I N_\perp m_{2k_F} [ \cos(2k_F r)\hat{\be}_x +  \sin(2k_F r) \hat{\be}_y]  
+ I N_\perp m_z \sin(2k_F r + \eta) \hat{\be}_z$, where $0 \le m_z \le 1$ is the magnetization along
the $z$ direction and $\eta$ an arbitrary phase. For $m_z \neq 0$ we see that
a full polarization $|\mean{\tilde{\bI}_i}| = I N_\perp$ is no longer possible in general.
Instead, the maximally possible polarization is determined by the condition 
$m_{2k_F}^2 + m_z^2 = 1$. Choosing such a $m_{2k_F}$ also minimizes the nuclear spin energy
for a fixed $m_z$, so that we can consider the latter condition as being fulfilled when 
seeking the absolute ground state energy.

Using the bosonization approach\cite{giamarchi:2004} in the same way as 
in Sec. \ref{sec:feedback_nucl} we see that the new $m_z$ component
leads to an Overhauser field for the electron 
system of the form $B_z \sin(\sqrt{2}\phi_c)\sin(\sqrt{2}\phi_s)$ with $B_z = I A_0 m_z$
which is relevant under the RG, plus oscillating terms which can be neglected. 
This new term competes against the term
$B_{xy} \cos(\sqrt{2K}\phi_+) = B_{xy} \cos(\sqrt{2}(\phi_c + \theta_s))$ (with $B_{xy} = I A_0 m_{2k_F}$)
in the RG, because it involves the field $\phi_s$ which is conjugate to 
$\theta_s$. By the uncertainty principle, the pinning of both fields is impossible, 
and generally the term with the larger amplitude, $B_{xy}$ or $B_z$, dominates the 
RG flow to the strong coupling fixed point.\cite{giamarchi:2004}

To our knowledge there is no method allowing a precise evaluation of this RG flow.
The following estimate, however, is sufficient for our needs. 
At the cutoff scale determining $B^*_{xy}$, both $B_{xy}$ and $B_z$ have flown
to strong coupling, 
although through the competition we have $B^*_{xy}(m_z \neq 0) < B^*_{xy}(m_z=0)$.
The scaling dimension of $B_z$ is $2-K_c/2-K_s/2$ and so for $K_s=1$ identical to the 
scaling dimension of $B_{xy}$ [Eq. \eqref{eq:B_xy_RG}]. Hence, the initial
ratio $B_z/B_{xy} = m_z/m_{2k_F}$ remains, up to small corrections, constant throughout the 
RG flow. This allows us to estimate the decrease of $B^*_{xy}$ to be proportional to
$m_z$, which leads to a cost in the electron ground state energy
per lattice site
on the order of $\Delta E^{\text{cost}} = B^*_{xy}-B^*_{xy}(m_z) \sim m_z B^*_{xy}$ [with $B^*_{xy} = B^*_{xy}(m_z=0)$].
This cost must be compared with the gain in nuclear spin energy per lattice site, given by 
$\Delta E^{\text{gain}} \sim I^2 m_z^2 |J^{\prime z}_{2k_F} - J^{\prime x,y}_{2k_F}| 
\sim I^2  m_z^2 |J^{\prime z}_{2k_F}|$.
Since $\Delta E^{\text{gain}} \sim m_z^2$ and $\Delta E^{\text{cost}} \sim m_z$
the relation $\Delta E^{\text{gain}} < \Delta E^{\text{cost}}$ is most likely for small $m_z$
and definitely always fulfilled if 
$I^2 |J^{\prime z}_{2k_F}| < B^*_{xy}$.
Since close to $T^*$ [see Table \ref{tab:values}] we have
$B_{xy}^* \gg k_B T^* \sim I^2 |J^{\prime z}_{2k_F}|$, we conclude that 
any $m_z \neq 0$ is energetically highly unfavorable and so the helical
order in the spin $(x,y)$ plane is stable.

With this argument we also see that the second magnon band derived in 
Eq. \eqref{eq:omega_q_2}, $\omega_q^{(2)} = 2 I (J_{q}^z - J_{2k_F}^x) / N_\perp$,
which has $\omega_q^{(2)} < 0$ at $q \sim 2k_F$, is of no importance. 
Any macroscopic occupation of these negative energy states, which would normally
signify an instability of the assumed ground state, is energetically forbidden. 
The remaining $\omega_q^{(2)}> 0$ describe $2k_F$ fluctuations in the $z$ direction.
Their effect is the same as the fluctuations in the $(x,y)$ plane described
by the first magnon band $\omega^{(1)}_q$ in Eq. \eqref{eq:omega_q_1},
yet they involve large momenta $\sim 2k_F$. Due to this we can
neglect this second magnon band entirely in this theory.

Let us finally note that a pure $m_z$ magnetization (with the $(x,y)$ component $m_{2k_F} = 0$)
would lead to a similar feedback effect as the helical magnetization 
and open up a gap in the $\phi_c + \phi_s$ 
sector. This would lead to a spatially oscillating Ising-like average magnetization
in the $z$ direction. In contrast to the $(x,y)$ helical magnetization, the condition
$|\mean{\bI_i}| = N I$ is then fulfilled only when $\cos(2k_F r_i) = 1$. The resulting
nuclear magnetic energy lies therefore much above the energy from the $(x,y)$ magnetization; 
the minimum is only about half as deep as in the latter case,
and thus such a state is not assumed by the system in the ground state.


\subsection{Anisotropic hyperfine coupling}
\label{sec:anisotropic}

The strong feedback between the nuclear spins and the electrons occurs
only if there is a nonzero Overhauser field on every cross-section 
through the 1D conductor. This was ensured by the ferromagnetic locking 
of the nuclear spins by the coupling to the single transverse electron mode.
Anisotropy in the hyperfine interaction can perturb this situation, and
a reinvestigation of this very important first assumption on the nuclear spins
becomes mandatory.

\begin{figure}
	\includegraphics[width=0.7\columnwidth]{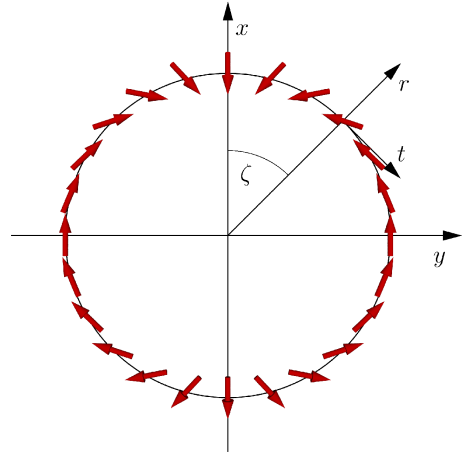}
	\caption{%
	Illustration of the nuclear spin configuration minimizing the energy of 
	Eq. \eqref{eq:H_i||} for the case of a complete electron polarization $S$ 
	pointing upwards in the figure. The figure shows how the sign of the 
	hyperfine constant changes when going around the nanotube cross-section.
	The overall magnetization of this configuration is $m = -0.17$ 
	(along $x$, normalized to $-1<m<1$).
	See also Ref. \onlinecite{fischer:2009}.
	\label{fig:spins_rotating}
	}
\end{figure}

Here we focus on the case of carbon nanotubes, in which indeed anisotropy
is present through the dipolar interaction between the electron and  
nuclear spins on the curved surface.\cite{fischer:2009} 
Rotational symmetry imposes that if
anisotropy is present, it occurs between the radial direction ($r$), the tangential
direction on the circular cross section ($t$), and the direction along the 
tube axis ($c$). Writing the nuclear and electron spin operators in this 
\emph{local} basis,
$\bI_i = (I^r_i, I^t_i,I^c_i)$ and 
$\bS_i = (S^r_i, S^t_i,S^c_i)$, the hyperfine Hamiltonian can be written as
\begin{equation}
	H = \sum_i 
	\left[ A^r I^r_i S^r_i + A^t I^t_i S^t_i + A^c I^c_i S^c_i \right],
\end{equation}
where $i$ runs over the 3D nuclear spin lattice.
We assume henceforth that\cite{fischer:2009} $A^r = -2 A^t = -2 A^c \equiv 2 A_0 > 0$,
and neglect the $A^c$ term as it turns out to be smaller than the
couplings in the plane spanned by the $r$ and $t$ components.
Let us identify this plane with the spin $(x,y)$ plane and rewrite the 
local component in an global spin basis as
$S^r_i = S^x_i \cos(\zeta_i) + S^y_i \sin(\zeta_i)$,
$S^t_i = - S^x_i \sin(\zeta_i) + S^y_i \cos(\zeta_i)$, 
and analogously for the $I_i$ operators, where
$\zeta_i$ is the polar angle of site $i$ on the circular 
cross-section (see Fig. \ref{fig:spins_rotating}).
The Hamiltonian can then be written in the form
\begin{align}
	&H = A_0 
\nonumber\\
	&\times\sum_i
	\bigl( S_i^x , S_i^y \bigr)
	\Bigl[1 + 3\cos(2\zeta_i) \sigma_z + 3\sin(2\zeta_i) \sigma_x \Bigr]
	\begin{pmatrix} I_i^x \\ I_i^y \end{pmatrix},
\end{align}
with $\sigma_{x,z}$ the Pauli matrices.
As in Sec. \ref{sec:confinement_1D} we assume that the electrons are confined in
a single transverse mode, which allows us to sum over the transverse components
of the indices $i$ for the nuclear spins. We see then that the electron spin 
couples only to the ferromagnetic component of the total nuclear spin on the circle,
$I^\alpha_{0,i_{||}} = \sum_{i_\perp} I^\alpha_{i_\perp,i_{||}}$, and the $2\zeta_i$ 
modes 
$I^\alpha_{c,i_{||}} = \sum_{i_\perp} \cos(2\zeta_i) I^\alpha_{i_\perp,i_{||}}$ 
and 
$I^\alpha_{s,i_{||}} = \sum_{i_\perp} \sin(2\zeta_i) I^\alpha_{i_\perp,i_{||}}$.
Since $S^\alpha_{i_\perp,i_{||}} = S^\alpha_{i_\perp}/N_\perp^2$ we have
\begin{equation}
	H = \frac{A_0}{N_\perp^2} \sum_{i_{||}} \sum_{\alpha = x,y}
	S^\alpha_{i_{||}}\left[ I^\alpha_{0,i_{||}} + 3\sigma_z I^\alpha_{c,i_{||}} + 3\sigma_x I^\alpha_{s,i_{||}} \right].
\end{equation}
The coupling to the electron spin, therefore, acts simultaneously on
these three nuclear spin modes only. For the feedback it is essential that 
these modes carry a finite magnetization. This is actually the case as we can 
see by assuming an electron spin at a cross section $i_{||}$ polarized in 
the $S^x = +1/2$ direction. 
The Hamiltonian $H_{i_{||}}$ for the nuclear spins on this cross-section then becomes
\begin{equation} \label{eq:H_i||}
	H_{i_{||}} = \frac{A_0 S}{N_\perp^2}  \sum_{i_\perp} 
	\left(I_{i_{||},i_\perp}^x, I_{i_{||},i_\perp}^y\right)
	\begin{pmatrix}1+3\cos(2\zeta_{i_\perp})\\ 3\sin(2\zeta_{i_\perp}) \end{pmatrix}.
\end{equation}
The energy is minimized when at each position $(i_{||},i_\perp)$ the nuclear spin is opposite 
to the vector on the far right in Eq. \eqref{eq:H_i||}. 
This configuration is shown in Fig. \ref{fig:spins_rotating}
and has a net magnetization along the $x$ direction of
$m = -0.17$. Hence, the coupling to the single transverse electron mode enforces a
nuclear Overhauser field of the same type as in the isotropic case. 

Through the reduced magnetic field, we see from Eq. \eqref{eq:Delta} that 
the gap $B^*_{xy}$ is smaller compared with the isotropic value by $(0.17)^{1/(2-g)} \approx 0.2$. 
This affects the feedback only if the smaller gap
becomes comparable to $k_B T^*$. Since, however, $B^*_{xy} \sim 10$ $\mu$eV $\sim 100$ mK 
for the isotropic case (see Table \ref{tab:values}), 
even with this reduction, it remains on a scale that is larger
than $T^* \sim 10$ mK. Yet it also lowers the limiting scale $B^*$ below $k_B T^*$,
and so the true $T^*$ may be a bit lower as well.
However, since $B^*$ depends directly on $L$ for SWNTs [see Sec. \ref{sec:validity_RKKY}], 
choosing a longer sample length
will push $B^*$ up to higher values, and so $T^*$ may keep its original value.


\section{Renormalization above $T^*$}
\label{sec:above_T^*}

For temperatures above $T^*$ thermal fluctuations destroy the nuclear helimagnet,
but there remains the separation of time scales between the nuclear and electron systems.
This implies that any random fluctuation at time $t$ into a nonzero Fourier component
$m_{2k_F}(t)$ of the nuclear magnetization
triggers a renormalization of the electronic properties in the same way as before 
and leads to an instantaneous Overhauser field $B^*_{xy}(t) = I m_{2k_F}(t) A^*(t)/2$.

Any detection of this Overhauser field would have to rely on the measurement of nonvanishing
quantities such as $(B_{xy}^*)^2$. 
We emphasize though that the renormalization occurs only for the 
$q=\pm 2k_F$ Fourier modes of the magnetization.
Close to $T^*$ thermal fluctuations may still occur preferably in the vicinity 
of the minimum of $J_q$ at $q = \pm 2k_F$.
However, as $T$ is raised away from $T^*$ the relative weight of the $q=\pm 2k_F$ modes
with respect to the other $(N-2)$ Fourier modes rapidly drops and approaches
$1/N$ at high temperatures. Except for a $T$ in the close vicinity
of $T^*$ the renormalized Overhauser field is probably not detectable.

On the other hand, as long as $k_B T$ remains below the maximally possible 
$B_{xy}^* = I |A^*|/2$, the $m_{2k_F}$ fluctuations sporadically open a gap
for one half of the electron modes whenever $B_{xy}^*(t)$ exceeds $k_B T$.
This temporarily freezes out of one half of the conduction channels,\cite{braunecker:2009EPAPS}
and so reduces the time averaged electric conductance of the system. 
The reduction is given by a temperature dependent 
factor $f$ between $1/2$ and $1$,
with the limits $f=1/2$ at $T \to T^*$ and $f=1$ when crossing through 
$k_B T \sim I |A^*|/2$.
Note that if the correlation length $\xi$ is given by the system length $L$,
increasing $L$ also increases $A^*$ [see Eq. \eqref{eq:A^*}], 
and so the onset of $f<1$ occurs already at higher temperatures.


\section{Validity of the one-band description for carbon nanotubes}
\label{sec:nanotubes}

The band structure of armchair carbon nanotubes consists of two Dirac 
cones (or two valleys) centered at the momenta\cite{saito:1998} 
$K = \pm 2\pi/3a$ (folded into the first Brillouin zone).
Hence, the LL theory depends not only on spin and $L$ or $R$ movers but also on 
an index $\alpha = 1,2$ labeling the two Dirac cones. We have neglected 
this additional quantum number $\alpha$ in the previous sections with the argument
that, once the feedback is included, the results obtained from the single band (single cone) 
description employed so far are quantitatively the same. 
In this section this shall be explicitly shown.


\subsection{Luttinger liquid theory for nanotubes}

Without electron-electron interactions SWNTs of the armchair type 
have the effective low energy Hamiltonian\cite{kane:1997a}
\begin{equation}
	H_0 = \int dr \sum_{\alpha\sigma}
	\left[
		\psi_{R\alpha\sigma}^\dagger (-i\nabla) \psi_{R\alpha\sigma}
		-
		\psi_{L\alpha\sigma}^\dagger (-i\nabla) \psi_{L\alpha\sigma}
	\right],
\end{equation}
where $\psi_{\ell\alpha\sigma}$ is the electron operator of $\ell = L,R$
movers in cone $\alpha$ with spin $\sigma$, and 
$\psi_\sigma = \sum_{\ell\alpha} \psi_{\ell\alpha\sigma}$ is the full
electron operator.
This theory can be bosonized within each cone in the usual way\cite{egger:1997,kane:1997b,egger:1998} by 
introducing the boson fields $\phi_{\nu \alpha}$ and $\theta_{\nu \alpha}$
with $\nu = c,s$. In particular $-\nabla \phi_{c\alpha} \sqrt{2}/\pi = \rho^f_{c\alpha}$,
where $\rho^f_{c\alpha} = \sum_{\ell\sigma} \psi_{\ell\alpha\sigma}^\dagger \psi_{\ell\alpha\sigma}$
is the forward-scattering part of the density operator in cone $\alpha$.
The basic bosonization identity \eqref{eq:psi_phi} is now enriched by the index
$\alpha$:
\begin{equation}
	\psi_{\ell\alpha\sigma} 
	= \frac{\eta_{\ell\alpha\sigma}}{\sqrt{2\pi a}}
	\e^{i [ \ell k_F + \alpha k_v ] r 
	    + \frac{i\ell}{\sqrt{2}}[\phi_{c\alpha}-\ell\theta_{c\alpha} + \sigma(\phi_{s\alpha}-\ell\theta_{s\alpha})]},
\end{equation}
where $k_F$ is measured from the crossing point of the two branches of the corresponding
Dirac cone, and so 
the position of these Dirac cones in the Brillouin zone has to be taken into account
through the additional phase factors $\pm i k_v r$ with $k_v = 4\pi/3a$ 
(modulo any reciprocal lattice vector). 

The electron-electron interactions considered here are density-density (Coulomb)
interactions and can be split into a part depending only on the forward-scattering 
densities $\rho^f_{c\alpha}$ and a term depending on backscattering $\ell \to -\ell$,
inter-cone scattering $\alpha \to -\alpha$, and Umklapp scattering.
The latter processes lead to a gap in the system, but with a magnitude that is
exponentially suppressed with an increasing diameter of the tube.\cite{egger:1998} 
At realistic temperatures this gap is only important for very narrow tubes
and can be excluded here. 
Indeed, from Ref. \onlinecite{egger:1998} the dominating gap is 
determined by 
$m_b = \omega_0 \exp(- \pi \hbar v_F / \sqrt{2} b)$,
where $\omega_0 = 7.4$ eV is the bandwidth of the $\pi$ electrons,
and $b \approx 0.1 \ a e^2/R$ is the inter-valley scattering amplitude
with $e$ the electron charge and $R$
the radius of the SWNT. For $(n,n)$ armchair SWNTs 
we have\cite{kane:1997a} $R = |\mathbf{C}|/2\pi = \sqrt{3} a n / 2\pi$,
with $\mathbf{C}$ the chiral vector (see also Sec. \ref{sec:confinement_1D}). 
Putting in numbers (see Table \ref{tab:values})
we find $m_b \sim \omega_0 \exp(-2.2 n)$, and for $n = 13$ 
we have $m_b \sim 3$ peV $\sim$ 30 nK.
We notice, however, that $b$ has an order 1 uncertainty,\cite{egger:1998} and can become
larger for very well screened interactions.
LL physics at the millikelvin temperatures considered here is valid
for gaps that lie below these temperatures, which therefore
requires a sufficiently wide nanotube and
a not too short screening length (yet short enough to allow a local
description for the interaction).
In addition, the experiments of Ref. \onlinecite{bockrath:1999}
show no indication for such a gap down to a temperature $T = 1.6$ K,
confirming the LL picture.

The system is then describable by a LL where only the 
forward scattering interaction part remains of importance. It is of the form
\begin{align}
	V 
	&= \int dr dr' V(r-r') [\rho^f_{c1}(r)+\rho^f_{c2}(r)][\rho^f_{c1}(r')+\rho^f_{c2}(r')]
\nonumber\\
	&\approx \int dr U [\nabla\phi_{c1}(r) + \nabla\phi_{c2}(r)]^2,
\end{align}
where $U$ can be related to a charging energy and $V(r)$ is sufficiently screened to allow a 
local description (yet see the preceding paragraph).

The interacting Hamiltonian can then be diagonalized through boson
fields that describe the symmetric $S$ and antisymmetric $A$ parts of 
density fluctuations between the $\alpha=1,2$ cones,
$\phi_{\nu S}=(\phi_{\nu 1}+\phi_{\nu 2})/\sqrt{2}$ and $\phi_{\nu A}=(\phi_{\nu 1}-\phi_{\nu 2})/\sqrt{2}$.
Similar definitions hold for the $\theta_{\nu\alpha}$ fields. 
The electron-electron interaction acts only on the symmetric $(c,S)$ sector. Therefore the 
LL theory remains that of a noninteracting system in the $(c,A), (s,S)$, and $(s,A)$ 
sectors (more precisely, it remains that of almost a noninteracting system due to slight renormalizations
through the backscattering interaction). The bosonic Hamiltonian for the nanotube then takes
the form
\begin{equation}
	H = \sum_{\nu P}\int \frac{dr}{2\pi} 
	\left[\frac{v_{\nu P}}{K_{\nu P}} (\nabla \phi_{\nu P})^2 + v_{\nu P} K_{\nu P} (\nabla\theta_{\nu P})^2 \right],
\end{equation}
where $P = S,A$. Here $K_{c A} \approx K_{s S} \approx K_{s A} \approx 1$, while $K_{c S}$ is
strongly renormalized by the interaction and typically is about\cite{egger:1997,kane:1997b,egger:1998}
 $K_{c S} \approx 0.2$.
The velocities are given by $v_{\nu P} = v_F / K_{\nu P}$.

Correlation functions computed with this theory are consequently described by 
power laws with exponents composed of all four LL parameters $K_{\nu P}$. 
Of particular interest for us are the backscattering amplitudes appearing in the
spin operators $S^x$ and $S^y$, for instance for spin-flip backscattering $R \to L$,
\begin{align} 
	\psi^\dagger_{L\alpha\up}\psi_{R\alpha\dw}
	&\sim \e^{2i k_F r} \e^{\frac{i}{\sqrt{2}} (\phi_{L\alpha\up} + \phi_{R\alpha\dw})}
	= \e^{2i k_F r} \e^{i \sqrt{2} (\phi_{c\alpha} +\theta_{s\alpha})}
\nonumber \\
	&= \e^{2i k_F r} \e^{i ( \phi_{cS} + \alpha \phi_{cA} + \theta_{sS} + \alpha \theta_{sA})},
\label{eq:app_bs_1}
\end{align}
for inter-cone scattering $\alpha=1 \to 2$,
\begin{align}
	\psi^\dagger_{L2\up}\psi_{L1\dw}
	&\sim 
	\e^{2i k_v r} 
	\e^{\frac{i}{\sqrt{2}}(\phi_{L2\up}-\phi_{L1\dw})}
\nonumber \\
	&= 	
	\e^{2i k_v r} 
	\e^{i (-\phi_{cA} - \theta_{cA} + \phi_{sS} + \theta_{sS})},
\label{eq:app_bs_2}
\\
	\psi^\dagger_{R2\up}\psi_{R1\dw}
	&\sim 
	\e^{2i k_v r} 
	\e^{\frac{-i}{\sqrt{2}}(\phi_{R2\up}-\phi_{R1\dw})}
\nonumber\\
	&= 	
	\e^{2i k_v r} 
	\e^{-i (-\phi_{cA} + \theta_{cA} + \phi_{sS} - \theta_{sS})},
\label{eq:app_bs_3}
\end{align}
and for the combinations $L \to R$ together with $\alpha = 1 \to 2$ or $\alpha = 2 \to 1$,
\begin{align}
	\psi^\dagger_{L2\up}\psi_{R1\dw}
	&\sim 
	\e^{2i (k_F + k_v) r} 
	\e^{\frac{i}{\sqrt{2}}(\phi_{L2\up}+\phi_{R1\dw})}
\nonumber \\
	&= 	
	\e^{2i (k_F + k_v) r} 
	\e^{i (\phi_{cS} - \theta_{cA} - \phi_{sA} + \theta_{sS})},
\label{eq:app_bs_4}
\\
	\psi^\dagger_{L2\dw}\psi_{R1\up}
	&\sim 
	\e^{2i (k_F + k_v) r} 
	\e^{\frac{i}{\sqrt{2}}(\phi_{L2\dw}+\phi_{R1\up})}
\nonumber \\
	&= 	
	\e^{2i (k_F + k_v) r} 
	\e^{i (\phi_{cS} - \theta_{cA} + \phi_{sA} - \theta_{sS})},
\label{eq:app_bs_5}
\\
	\psi^\dagger_{L1\up}\psi_{R2\dw}
	&\sim 
	\e^{2i (k_F - k_v) r} 
	\e^{\frac{i}{\sqrt{2}}(\phi_{L1\up}+\phi_{R2\dw})}
\nonumber \\
	&= 	
	\e^{2i (k_F - k_v) r} 
	\e^{i (\phi_{cS} + \theta_{cA} + \phi_{sA} + \theta_{sS})},
\label{eq:app_bs_6}
\\
	\psi^\dagger_{L1\dw}\psi_{R2\up}
	&\sim 
	\e^{2i (k_F - k_v) r} 
	\e^{\frac{i}{\sqrt{2}}(\phi_{L1\dw}+\phi_{R2\up})}
\nonumber \\
	&= 	
	\e^{2i (k_F - k_v) r} 
	\e^{i (\phi_{cS} + \theta_{cA} - \phi_{sA} - \theta_{sS})}.
\label{eq:app_bs_7}
\end{align}
All these expressions enter the susceptibility and lead to power-law divergences
at the momenta $2k_F$, $2k_v$ and $2(k_F \pm k_v)$. The operators involving backscattering 
$L \leftrightarrow R$ depend all on $\phi_{cS}$ 
and 3 of the other fields, leading to the same exponents
$g = (K_{cS} + 3)/4 \approx 0.8$. 
Eqs. \eqref{eq:app_bs_2} and \eqref{eq:app_bs_3} do not depend on 
$\phi_{cS}$ and so lead to correlators with a larger exponent, and so a 
shallower minimum for the RKKY interaction. 

This means that the RKKY interaction $J_q$ between the nuclear spins has 3 equal minima
and so the helical order is in principle not a simple helimagnet, but involves the superposition
of the 3 spatial frequencies. Yet this, once again, neglects the feedback on the electron
system, which selects only the $2k_F$ frequency as relevant for the feedback and for lowering
the ground state energy, as we shall see next.


\subsection{Feedback from the nuclear Overhauser field}

Let us assume that the Overhauser field consists of the spatial frequencies
$2k_F$ and $2(k_F \pm k_v)$ associated with the minima of the susceptibility. For completeness
we keep also here the local minimum at $2k_v$, and so write the coupling of the 
nuclear magnetic field to the electron spin as
\begin{equation}
	\sum_{i=1}^4 B_i \left[ S^x(r) \cos(2k_i r+\eta_i) + S^y(r) \sin(2k_i r+\eta_i) \right],
\end{equation}
with $k_1 = k_F, k_2 = k_v, k_3 = k_F + k_v$, and $k_4 = k_F - k_v$, 
the corresponding amplitudes $B_i \propto m_{k_i}$ proportional to the $k_i$ 
components of the magnetization,
and 
$\eta_1,\dots,\eta_4$ arbitrary angles, which we can set to zero because they 
can be absorbed by suitable shifts of the boson fields. 
Using Eqs. \eqref{eq:app_bs_1} -- \eqref{eq:app_bs_5} to express $S^{x,y}$ 
in terms of the boson fields, we are led to the interaction
\begin{align}
	&B_{1} \Bigl[ 
		\cos(\sqrt{2}(\phi_{c1}+\theta_{s1})) + \cos(\sqrt{2}(\phi_{c2}+\theta_{s2}))
	\Bigr]
\nonumber \\
	&+ 
	2 B_2 \cos(\phi_{cA} - \phi_{sS}) \sin(\theta_{cA} - \theta_{sS})
\nonumber \\
	&+
	2 B_3 \cos(\phi_{cS} - \theta_{cA}) \cos(\phi_{sA} - \theta_{sS} )
\nonumber \\
	&+
	2 B_4 \cos(\phi_{cS} + \theta_{cA}) \cos(\phi_{sA} + \theta_{sS} )
\nonumber \\
	&+ \text{oscillating terms}.
\end{align}
The terms proportional to $B_1$ are precisely the combinations $\phi^+ \propto \phi_c + \theta_s$
used previously, and lead to a gap in each Dirac cone separately. 
The terms proportional to $B_2$ involve fields that are conjugate to each other, 
and so this interaction remains always 
critical and can be neglected. 
The arguments proportional to $B_3$ and $B_4$ are relevant and open gaps
in the corresponding boson fields. Their effect is, however, quite different from
that of the $B_1$ part. 

Indeed, for the $B_1$ part, the opening of the gap in the 
$\phi^+ \propto \phi_c + \theta_s$ channels leaves still the combination
$\phi_c - \theta_s$ gapless [see Eq. \eqref{eq:hboso}]. 
Both combinations appear in the susceptibilities $\chi_{x,y}$ 
and correspond to the processes $L\up \leftrightarrow R\dw$ and $L\dw \leftrightarrow R\up$.
Hence, even with the gap for $\phi_c + \theta_s$ there remains (at $T=0$) a 
power-law singularity at $2k_F$ in $\chi_{x,y}$ 
due to $\phi_c - \theta_s$ with the consequences
discussed in the previous sections. This is much different for the $B_3$ and $B_4$ processes. 
Because of the mixing between the Dirac cones, the processes $L\up \leftrightarrow R\dw$ 
and $L\dw \leftrightarrow R\up$ have the different momentum transfers
of $2(k_F + k_v)$ and $2(k_F - k_v)$, respectively. 
The opening of a gap for one of those processes
entirely eliminates the steep minimum of $\chi_{x,y}(q)$ at the corresponding momentum
$q = 2(k_F \pm k_v)$, and so eliminates this energy minimum
of the nuclear spins. 
Consequently, the nuclear spins will reorient themselves into one of the 
remaining energy minima, i.e. the amplitudes $B_3$ and $B_4$ vanish.
The only remaining minima are then those of $B_1$. 
With vanishing $B_3$ and $B_4$, however, the corresponding gaps vanish as well. 
Hence, even though the $B_{3,4}$ energy minima 
exist, they cannot be populated by the nuclear spin modes $\tilde{\bI}_q$
because this immediately opens gaps and costs energy.
This is very similar to the stabilization effect discussed in Sec. \ref{sec:stability_planar_order}.

Therefore, any nuclear spin order with $2(k_F \pm k_v)$
is unstable, and the only stable minimum for the feedback is
the nuclear helical order at $2k_F$.


\subsection{Validity of the single-band model}

Since the $\cos(\sqrt{2}(\phi_{c\alpha} + \theta_{s\alpha}))$ interaction terms in each Dirac cone
$\alpha=1,2$ are highly RG-relevant, we can evaluate their effect in each cone
separately. Transforming into the $\alpha=1,2$ basis,
the Hamiltonian takes the form,
\begin{align}
	&H = \int\frac{dr}{2\pi} \sum_{\nu\alpha} 
	\biggl[
		\frac{v_{\nu\alpha}}{K_{\nu\alpha}} (\nabla\phi_{\nu\alpha})^2 
		+ 
		v_{\nu\alpha}K_{\nu\alpha} (\nabla\theta_{\nu\alpha})^2
\nonumber\\
	&+ \frac{B}{a} \cos\left(\sqrt{2}(\phi_{c\alpha} + \theta_{s\alpha})\right)
	+ v_F \bigl(K_{cS}^{-2}-1\bigr) \nabla\phi_{c1} \nabla\phi_{c2}
	\biggr],
\label{eq:H_nanotube}
\end{align}
for $\nu = c,s, \alpha = 1,2$,  and 
with $K_{c\alpha} = \sqrt{K_{cS}^2/(1+K_{cS}^2)} \approx K_{cS} = 0.2$, $K_{s\alpha} = 1$, 
and $v_{\nu\alpha} = v_F/K_{\nu\alpha}$.
Without the last Dirac cone mixing term this Hamiltonian is identical to two 
copies of the single-band Hamiltonian \eqref{eq:H_pm}. 
The coupling between the Dirac cones
is, however, marginal in the same sense as the gradient products in the last
term of Eq. \eqref{eq:H_pm} and has only a tiny influence on the opening of
the gap by the cosine term in each cone separately. Hence, for the evaluation 
of the gap we can safely neglect this coupling. 
The size of the gap is then determined by replacing $K_c$ and $K_s$
in Sec. \ref{sec:feedback_el} by $K_c = K_{c1} = K_{c2} \approx 0.2$ and
$K_s = 1$.

The resulting low-energy Hamiltonian then depends only on the gapless fields
$\phi_{-,\alpha}$ and $\theta_{-,\alpha}$, defined as in 
Eqs. \eqref{eq:phi_c} -- \eqref{eq:theta_s}, and is given by
\begin{align}
	H_-  = \int &\frac{dr}{2\pi} 
	\biggl\{
		\sum_{\alpha = 1,2} 
		v_-
		\left[ (\nabla\phi_{-,\alpha})^2 + (\nabla\theta_{-,\alpha})^2 \right]
\nonumber\\
	&
	+ v_F \bigl(K_{cS}^{-2}-1\bigr) \frac{K_c}{K_s K} 
		\nabla\phi_{-,1} \nabla\phi_{-,2}
	\biggr\}
\nonumber\\
	= \int &\frac{dr}{2\pi} 
	v_-
	\biggl\{
		\sum_{\alpha = 1,2} 
		\left[ (\nabla\phi_{-,\alpha})^2 + (\nabla\theta_{-,\alpha})^2 \right]
\nonumber\\
	&
	+ 2\gamma \nabla\phi_{-,1} \nabla\phi_{-,2}
	\biggr\},
\label{eq:H_-}
\end{align}
with 
$\gamma = v_F [K_{cS}^{-2}-1] K_c/2 v_- K_s K = (1-K_{cS}^2)/2(1+2K_{cS}^2) \approx 0.44$.
This Hamiltonian is diagonalized by the symmetric and antisymmetric combinations
$\phi_{S} = (\phi_{-,1}+\phi_{-,2})/\sqrt{2}$ and 
$\phi_{A} = (\phi_{-,1}-\phi_{-,2})/\sqrt{2}$, and similarly for the $\theta$ fields.
We obtain
\begin{equation}
\label{eq:H_-_2}
	H_- = \int \frac{dr}{2\pi} \sum_{P=S,A} \left[ 
		\frac{v_{P}}{K_{P}} (\nabla \phi_{P})^2
		+
		v_{P} K_{P} (\nabla \theta_{P})^2
	\right],
\end{equation}
with $K_{S} = \sqrt{1+\gamma} \approx 1.20$, $K_{A} = \sqrt{1-\gamma} \approx 0.75$ and 
$v_{P} = v_-/K_{P}$.

The single-band model used in the previous sections can be applied if it produces the same results
as the model \eqref{eq:H_-_2}.
The essential quantities are the susceptibilities $\chi_{x,y,z}$ evaluated with the
remaining gapless fields for the $2k_F$ backscattering of Eq. \eqref{eq:app_bs_1}.
In the single-band model the LL constant is $K_- = 1$.
From Eq. \eqref{eq:app_bs_1} we see that 
only the combinations $\phi_S \pm \phi_A$ and $\theta_S \pm \theta_A$ appear
(the prefactors are the same as in the single-band model up to negligible corrections).
The $\phi_S \pm \phi_A$ lead to exponents depending on $(K_S + K_A)/2 \approx 0.97$ 
and the $\theta_S \pm \theta_A$ lead to exponents depending on $(K_S^{-1}+K_A^{-1})/2 \approx 1.08$.
Since both values are very close to 1, the 2-band theory leads indeed to the same conclusions
as the single-band theory.

In fact, the single-band theory can be interpreted as neglecting the coupling between
the Dirac cones, which is the approximation used in Ref. \onlinecite{braunecker:2009}.
As we have seen above, with the feedback this becomes very accurate
because the relative coupling strength $\gamma$ between the two cones
[Eq. \eqref{eq:H_-}] is much reduced. Indeed, corrections 
to the decoupled systems appear only at $\gamma^2$ (from expanding $K_S+K_A$ or $K_S^{-1}+K_A^{-1}$). 
The value of $\gamma \approx 0.44$ is small
enough compared with the original $(K_{cS}^{-2}-1) \approx 24$ in Eq. \eqref{eq:H_nanotube}
so that the coupling between the Dirac cones has a negligible effect.


\section{Conclusions}
\label{sec:conclusions}

We have shown in this paper that the hyperfine interaction between a lattice of 
nuclear spins and Luttinger liquid leads to order in both systems in the form of
a nuclear helimagnet and a helical spin density wave for half of the
electron modes. 
A strong feedback between the electrons and nuclear spins stabilizes this
combined order up to temperatures that are within experimental reach, even though the
hyperfine interaction generally is very weak. 
The feedback is a direct consequence of Luttinger liquid behavior
and is absent for noninteracting electrons or Fermi liquids.

This leads to several remarkable effects that should be detectable experimentally
and that may be used for a direct proof of Luttinger liquid physics:
(i) 
The helical magnetization $m_{2k_F}$ resulting from the nuclear spin ordering follows 
the modified Bloch law of Eq. \eqref{eq:m_T_intro}.
(ii) 
The helical electron spin density wave resulting from the renormalization triggered by 
the nuclear Overhauser field has an excitation gap proportional to 
$m_{2k_F}$, or to some power of $m_{2k_F}$ depending on the Luttinger liquid
parameters [see Eq. \eqref{eq:B_T}]. Measuring this electronic 
excitation gap is a direct way of measuring $m_{2k_F}$.
(iii) The pinning of one half of the electron conduction modes causes
a reduction of the electric conductance by the factor 2
when cooling down through the cross-over temperature $T^*$.
(iv) Finally, the strong binding of the nuclear helimagnet
to the electron modes leads to anisotropy in the electron spin susceptibility 
between the $(x,y)$ plane of the nuclear helimagnet and the orthogonal
$z$ direction.

We refer the reader to Sec. \ref{sec:results} for 
a complete summary of our results and the main conclusions.
The physical mechanism for magnetic ordering described here was worked out for 
two experimentally available systems, $^{13}$C single wall nanotubes, and 
GaAs-based quantum wires. 
We note, however, 
that we expect similar physics for a large class of Kondo-lattice-type systems 
coupled to a Luttinger liquid defined by the conditions listed in Sec. \ref{sec:results}.

\begin{acknowledgments}
We thank B. I. Halperin and M. J. Schmidt for helpful discussions and critical 
remarks, and A. Yacoby and G. Barak for providing the values characterizing the 
GaAs quantum wires.
This work was supported by the Swiss NSF, NCCR Nanoscience (Basel),
and DARPA QUEST.
\end{acknowledgments}


\appendix


\section{Electron spin susceptibility}
\label{sec:susceptibility}

In this appendix we evaluate the electron spin susceptibilities for the cases
without and with the feedback from the nuclear Overhauser field. 


\subsection{Without feedback from Overhauser field}
\label{sec:susc}

Without the Overhauser field the electron system is described by a 
standard LL. 

We define the static electron spin susceptibility in the 
1D tight-binding basis as [see Eq. \eqref{eq:susc_def}]
\begin{equation} \label{eq:def_chi_ij}
	\chi_{ij}^{\alpha\beta}
	= \frac{-i}{a^2} \int_0^\infty dt \e^{-\eta t} \mean{[ S_i^\alpha(t) , S_j^\beta(0) ]},
\end{equation}
with an infinitesimal $\eta > 0$. 
For the further calculation it is more convenient to pass to the continuum description
where the tight binding operators $S_i^\alpha(t)/a$ are replaced by the fields $S^\alpha(r,t)$,
\begin{equation} \label{eq:def_chi_rr'}
	\chi^{\alpha\beta}(r-r')
	= -i\int_0^\infty dt \e^{-\eta t} \mean{[ S^\alpha(r,t) , S^\beta(r',0) ]}.
\end{equation}
From the conservation of total spin we have
$\chi^{\alpha\beta} = \chi_\alpha \delta_{\alpha\beta}$.

The spin operators split into forward scattering,
$S^\alpha_f(r)$, 
and backscattering parts, $S^\alpha_b(r)$. While the forward scattering contribution to the 
susceptibility remains regular, the backscattering contribution has (at zero
temperature) a singularity at momenta $q = \pm 2k_F$. Since this singular
behavior dominates the physics described in this work, we thus keep only the backscattering part.
We then define 
\begin{equation}
	\chi^>_\alpha(r,t) = -i \mean{S_b^\alpha(r,t) S_b^\alpha(0,0)},
\end{equation}
such that $\chi_\alpha(r)$ is the $\omega \to 0 + i\eta$ Fourier transform
of 
$\chi_{\alpha}(r,t) = \vartheta(t) [ \chi_\alpha^>(r,t) - \chi_\alpha^>(-r,-t)] 
= 2 \vartheta(t) \mathrm{Im}[i \chi_\alpha^>(r,t)]$, with $\vartheta$ the step function.
The operators $S_b^\alpha$ are given by 
\begin{equation}
	S_b^\alpha 
	= \frac{1}{2} \sum_{\sigma,\sigma' = \up,\dw}
	\sigma^{\alpha}_{\sigma\sigma'}
	\left[
		\psi_{L\sigma}^\dagger \psi_{R\sigma'} + 
		\psi_{R\sigma}^\dagger \psi_{L\sigma'}
	\right],
\end{equation}
with $\sigma^\alpha$ the Pauli matrices.
Using the bosonization identities \eqref{eq:psi_phi} and \eqref{eq:phi_phi_theta}
we see that these spin operators can be written in the form\cite{giamarchi:2004}
$S_b^\alpha = [ O_{SDW}^\alpha + (O_{SDW}^\alpha)^\dagger]/2$, where
\begin{align}
\label{eq:O_SDW_x}
	O_{SDW}^x
	&= 
	\frac{\e^{-2ik_F r}}{2\pi a} \e^{i \sqrt{2} \phi_c} 
	\left[ \e^{i \sqrt{2} \theta_s}+ \e^{-i \sqrt{2}\theta_s}\right],
\\
\label{eq:O_SDW_y}
	O_{SDW}^y
	&= 
	i\frac{\e^{-2ik_F r}}{2\pi a} \e^{i \sqrt{2} \phi_c} 
	\left[ \e^{i \sqrt{2} \theta_s}- \e^{-i \sqrt{2}\theta_s}\right],
\\
\label{eq:O_SDW_z}
	O_{SDW}^z
	&= 
	\frac{\e^{-2ik_F r}}{2\pi a} \e^{i \sqrt{2} \phi_c} 
	\left[ \e^{i \sqrt{2} \phi_s}- \e^{-i \sqrt{2}\phi_s}\right].
\end{align}
We have omitted in the latter expressions the Klein factors because they 
drop out in the averages.
We then find
\begin{align}
	&\chi^>_{x}(r,t) 
\nonumber\\
	&= \frac{-i\cos(2k_F r)}{(2\pi a)^2}
	\mean{\e^{i \sqrt{2} (\phi_c(r,t)+\theta_s(r,t))} \e^{-i \sqrt{2} (\phi_c(0) + \theta_s(0))}},
\\
	&\chi^>_{y}(r,t) 
\nonumber\\
	&= \frac{-i\cos(2k_F r)}{(2\pi a)^2}
	\mean{\e^{i \sqrt{2} (\phi_c(r,t)-\theta_s(r,t))} \e^{-i \sqrt{2} (\phi_c(0) - \theta_s(0))}},
\\
	&\chi^>_{z}(r,t) 
\nonumber\\
	&= \frac{-i\cos(2k_F r)}{(2\pi a)^2}
	\mean{\e^{i \sqrt{2} (\phi_c(r,t)+\phi_s(r,t))} \e^{-i \sqrt{2} (\phi_c(0) + \phi_s(0))}}.
\end{align}
The determination of these averages is a standard calculation in the bosonization 
technique. At zero temperature, we obtain\cite{giamarchi:2004}
\begin{align}
	&\chi^>_\alpha(r,t) 
\nonumber\\
	&= \frac{-i \cos(2k_F r)}{(2\pi a)^2}
	\left(\frac{a}{r-v_F t + i \delta}\right)^{g_\alpha} 
	\left(\frac{a}{r+v_F t - i \delta}\right)^{g_\alpha},
\label{eq:chi^>_T=0}
\end{align}
with
$g_x = g_y = (K_c + K_s^{-1})/2$, $g_z = (K_c + K_s)/2$,
and $\delta > 0$ a short distance/time cutoff. For $g_\alpha$ well below $1$
we can choose $\delta$ infinitesimal and proceed analytically as below.
For $g_\alpha \lesssim 1$, the singularities in Eq. \eqref{eq:chi^>_T=0}
become too pronounced and the cutoff plays an increasingly important role.
We then must include a finite $\delta$ and proceed numerically with the further
calculation below and the subsequent evaluation of, for instance, 
the cross-over temperature $T^*$ (see caption of Fig. \ref{fig:T_Kc_GaAs}).
For $g_\alpha \lesssim 0.8$, however, such a cutoff is not required. 
For the explicit analytic expressions below we keep therefore an infinitesimal $\delta$.

The finite temperature expressions are
\begin{align}
	&\chi^>_\alpha(r,t) 
	= \frac{-i \cos(2k_F r)}{(2\pi a)^2}
	\left(\frac{\pi a/ \beta v_F}{\sinh\left(\frac{\pi}{\beta v_F} (r-v_F t + i \delta)\right)}\right)^{g_\alpha} 
\nonumber\\
	&\times
	\left(\frac{\pi a/ \beta v_F}{\sinh\left(\frac{\pi}{\beta v_F} (r+v_F t - i \delta)\right)}\right)^{g_\alpha},
\label{eq:chi^>_T>0}
\end{align}
where $\beta = 1/k_B T$.
Twice the imaginary part of these susceptibilities determines 
$\chi_\alpha(r,t)$. We have at zero temperature
\begin{equation} 
	\chi_\alpha(r,t)
	= 
	- \vartheta(t) \vartheta(vt -|r|) \frac{\sin(\pi g_\alpha)\cos(2k_F r)}{2 \pi^2 a^2}
	\left|\frac{a^2}{r^2-v^2t^2}\right|^{g_{\alpha}},
\label{eq:chi(rt)_T=0}
\end{equation}
and at finite temperature
\begin{align}
	&\chi_\alpha(r,t)
	= 
	- \vartheta(t) \vartheta(vt -|r|) \frac{\sin(\pi g_\alpha)\cos(2k_F r)}{2 \pi^2 a^2}
\nonumber\\
	&\times
	\left|\frac{\pi a/ \beta v_F}{\sinh\left(\frac{\pi}{\beta v_F} (r-v_F t)\right)}\right|^{g_\alpha} 
	\left|\frac{\pi a/ \beta v_F}{\sinh\left(\frac{\pi}{\beta v_F} (r+v_F t)\right)}\right|^{g_\alpha}.
\label{eq:chi(rt)_T>0}
\end{align}
Further Fourier transforming to $(q,\omega)$ space
and taking the $\omega \to 0+i\eta$ limit leads to\cite{giamarchi:2004}
\begin{equation}
	\chi_\alpha(q) 
	= - \frac{\sin(\pi g_\alpha)}{4 \pi^2 v_F}
	\sum_{\kappa = \pm} \left|\frac{2}{a(q + \kappa 2k_F)}\right|^{2-2g_\alpha}
\label{eq:susc_app_T0}
\end{equation}
at zero temperature and
\begin{align}
	&\chi_\alpha(q) 
	= - \frac{\sin(\pi g_\alpha)}{4 \pi^2 v_F} \Gamma^2(1-g_\alpha)
	\left(\frac{\beta v_F}{2\pi a}\right)^{2-2g_\alpha}
\nonumber\\
	&\times
	\sum_{\kappa = \pm} 
	\left|
		\frac{\Gamma\left(\frac{g_\alpha}{2} - i \frac{\beta v_F}{4\pi}(q + \kappa 2k_F)\right)}
		     {\Gamma\left(\frac{2-g_\alpha}{2} - i \frac{\beta v_F}{4\pi}(q + \kappa 2k_F)\right)}
	\right|^{2}
\label{eq:susc_app}
\end{align}
at finite temperature, where $\Gamma$ is Euler's Gamma function.


\subsection{With feedback from the Overhauser field}
\label{sec:susc_fb}

We have seen in Sec. \ref{sec:feedback_el} that the feedback from the nuclear magnetic 
field strongly renormalizes the electron system and opens a gap $B_{xy}^*$ for the 
field $\phi_+ \propto \phi_c + \theta_s$. Because the fluctuations of this mode
are frozen out at $k_B T < B_{xy}^*$, 
the response of the electron system to external perturbations
is modified. In particular, this affects the electron spin susceptibility $\chi_\alpha$,
and here most importantly the exponents $g$, which become smaller and anisotropic.

The calculation of $\chi_\alpha$ in this situation has been carried out in 
Ref. \onlinecite{braunecker:2009EPAPS} (see also Ref. \onlinecite{egger:1996}). 
For completeness we outline this 
calculation here again. 

Let us set $\bar{r} = (r,t)$, 
$\bar{\chi}_\alpha(\bar{r}) = - i 2 (\pi a)^2\chi_\alpha(r,t)$. 
We focus first on $\chi_x$. Using the relations \eqref{eq:phi_c}--\eqref{eq:theta_s} and 
Eq. \eqref{eq:O_SDW_x} we can then write
\begin{align}
	&\bar{\chi}_x(\bar{r})
	= 
	\cos(2k_F r) \Bigl[
		\bmean{\bigl[ \e^{i \sqrt{2K} \phi_+(\bar{r})} \, , \,  \e^{-i\sqrt{2K} \phi_+(0)} \bigr] }
\nonumber\\
	&+
		\bmean{\bigl[ \e^{i \sqrt{2K'}\phi_-(\bar{r}) - i \sqrt{2K''}\phi_+(\bar{r})} 
		              \, , \,
		              \e^{i \sqrt{2K'}\phi_-(0)       + i \sqrt{2K''}\phi_+(0)} 
		       \bigr]}
	\Bigr],
\label{eq:chi_x_in_phi_pm}
\end{align}
with $K' = 4K_c/K_s K$ and $K'' = (K_c-K_s^{-1})/K$, and where we have used 
the invariance of the Hamiltonian \eqref{eq:H_pm} under a simultaneous sign change 
of all the boson fields.

Since the Hamiltonian \eqref{eq:H_pm} is quadratic in the boson fields all these
averages are fully determined by the 2-point correlators\cite{giamarchi:2004}
\begin{equation} \label{eq:phi_-_correlator}
	\mean{\phi_-^*(q,\omega) \phi_-(q,\omega)}
	= \frac{\pi v_-}{(\omega\pm i\eta)^2 - v_-^2 q^2}
\end{equation}
for the massless field and 
\begin{equation} \label{eq:phi_+_correlator}
	\mean{\phi_+^*(q,\omega) \phi_+(q,\omega)}
	= \frac{\pi v_+}{(\omega\pm i\eta)^2 - v_+^2 q^2 - (B_{xy}^*)^2}
\end{equation}
for the massive field.
The sign of the infinitesimal shift $\pm i\eta$ is determined by the time order of 
the operators.
The corresponding $\theta_\pm$ correlators 
(important for $\chi_z$) can be obtained, for instance, through the equations of motion
$\partial_t \phi_\pm(r,t) = v_\pm \nabla \theta_\pm(r,t)$.

The correct treatment of the singularity in Eq. \eqref{eq:phi_-_correlator} 
at $(\omega \pm v_- q) \to 0$ results in the singular power laws (at $T=0$) 
of $\chi_\alpha$ 
of the LL theory and so leads to a contribution to the susceptibility 
entirely equivalent to those of Appendix \ref{sec:susc}.

This singularity is absent in Eq. \eqref{eq:phi_+_correlator} and hence the
power law singularities are broadened to resonances with a width and height determined
by $B_{xy}^*$. The precise evaluation of the expressions is cumbersome.
Approximations can be found in Refs. \onlinecite{voit:1998,wiegmann:1999,dora:2007,dora:2008}.
For our purpose, however, this evaluation is not required because the physics
of the combined electron--nuclear spin system is entirely dominated by the 
singular behavior of the $\phi_-$ correlators at $q=2k_F$, provided that $k_B T < B_{xy}^*$,
which is a necessary condition anyway that is indeed satisfied for the 
considered systems.

To see this in detail, let us expand Eq. \eqref{eq:chi_x_in_phi_pm} in powers
of $\phi_+$. The lowest nonzero term in $\phi_+$ is
\begin{align}
	&\cos(2k_F r) \Bigl[ 
	2K \bmean{\bigl[ \phi_+(\bar{r}) \, , \, \phi_+(0) \bigr]}
\nonumber \\
	&+ 2 K'' \bmean{\bigl[\e^{i\sqrt{2K'}\phi_-(\bar{r})} \phi_+(\bar{r}) \, , \, 
	               \e^{-i\sqrt{2K'}\phi_-(0))} \phi_+(0)\bigr]}
	\Bigr].
\label{eq:phi_+_expansion}
\end{align}
The Fourier transform of the first term can be directly evaluated 
with Eq. \eqref{eq:phi_+_correlator}. In the limit $\omega \to 0$ and $q \to \pm 2k_F$
it tends to a constant $\sim 1/(B_{xy}^*)^2$ and so contributes only insignificantly
to $\chi_x(q)$. The second term leads to a sum of expressions of the type
\begin{equation}
	\cos(2k_F r) \e^{K' \mean{\phi_-(\bar{r})\phi_-(0)}} \mean{\phi_+(\bar{r}) \phi_+(0)}.
\end{equation}
The fields $\phi_-$ are gapless and so lead to power laws of the form
\eqref{eq:susc_app_T0} (replacing $q^2$ by $\omega^2/v_-^2 - q^2$ and the exponent 
by $1-K'$). 
The Fourier transform of the latter expression is then
a convolution of the form
\begin{align}
	&\int dq' d\omega' \left|\frac{1}{\omega'^2-v_-^2q'^2}\right|^{1-K'}
\nonumber\\
	&\times
	\frac{1}{(\omega-\omega')^2 - v_+^2(q_\pm-q')^2 - (B_{xy}^*)^2},
\end{align}
with $q_\pm = q \pm 2k_F$.
In the static limit $\omega \to 0$ and at $q \to \pm 2k_F$, the $\omega'$
integral is dominated by the poles at $\omega'= \pm \sqrt{v_+^2 q'^2 + (B_{xy}^*)^2}$ 
and the integrable weak singularities at $\omega' = \pm v_- q'$.
A potential singular behavior must thus come from the poles. 
If we focus on the pole at $\omega' = \sqrt{v_+^2 q'^2 + (B_{xy}^*)^2}$ we obtain
\begin{equation}
	\sim \frac{1}{(B_{xy}^*)^{2(1-K')}}
	\int dq' \frac{1}{\sqrt{v_+^2 q'^2 + (B_{xy}^*)^2}}.
\end{equation}
The remaining integral leads to an arcsinh, which has an ultraviolet divergence
that has to be cut off at $1/a$. More importantly, however, the result has no
infrared divergence, meaning that this expression remains regular at $q_\pm \to 0$,
with a value determined by $B_{xy}^*$.

From this calculation we conclude that the Fourier transform of
\eqref{eq:phi_+_expansion} remains regular at $\omega \to 0, q_\pm \to 0$.
Since the Hamiltonian is quadratic in the boson operators higher order correlators
are products of the latter results and so remain fully regular.
We have therefore shown that the singular behavior of the susceptibility $\chi_x(q)$
is fully controlled by the massless fields $\phi_-$ only. 
This allows us to entirely neglect the gapped fields and approximate the susceptibility by 
\begin{equation}
	\bar{\chi}_x(\bar{r}) 
	= \cos(2k_F r) \bmean{\bigl[ \e^{i\sqrt{2K'}\phi_-(\bar{r})} \, , \, \e^{-i\sqrt{2K'}\phi_-(0)} \bigr]},
\end{equation}
which is of precisely the same form as the susceptibility of a standard LL
as discussed in Appendix \ref{sec:susc}. The difference is the modified exponent $K'$, the velocity $v_-$,
and an amplitude reduced by $1/2$
because the second term in Eq. \eqref{eq:chi_x_in_phi_pm} depends on $\phi_+$ only and drops out. 
From the results of Appendix \ref{sec:susc} and the Hamiltonian \eqref{eq:H_eff_-}
we obtain, for instance, at zero temperature,
\begin{equation}  \label{eq:chi_x_feedback}
	\chi_x(q)
	= - \frac{1}{2} \frac{\sin(\pi g'_x)}{4 v_-\pi^2} \Gamma^2(1-g'_x)
	\sum_{\kappa = \pm} \left|\frac{2}{a (q + \kappa 2k_F)}\right|^{2-2g'_x},
\end{equation}
with $g'_x = K'/2$.

The evaluation of $\chi_y(q)$ leads to precisely the same result,
$\chi_y(q) = \chi_x(q)$ and in particular $g'_y = g'_x$.

The susceptibility $\chi_z(q)$ is different in that it involves
different boson fields.
From Eq. \eqref{eq:O_SDW_z} we see that $\chi_z$ depends on the 
combination 
$\phi_c + \phi_s = \frac{1}{\sqrt{K}}( K_c \phi_+ - \sqrt{\frac{K_c}{K_s}}\phi_- + \theta_+ + \sqrt{K_c K_s} \theta_-)$,
where we have used Eqs. \eqref{eq:phi_c}--\eqref{eq:theta_s}.
Neglecting the gapped fields $\phi_+$ and $\theta_+$ then leads to
\begin{equation} \label{eq:chi_z_feedback}
	\chi_z(q)
	= - \frac{\sin(\pi g'_z)}{4 v_-\pi^2} \Gamma^2(1-g'_z)
	\sum_{\kappa = \pm} \left|\frac{2}{a (q + \kappa 2k_F)}\right|^{2-2g'_z},
\end{equation}
The prefactor $1/2$ of $\chi_{x,y}$ is here missing and the exponent is
$g'_z = (K_c/K_s + K_c K_s)/2 K$.

The extensions to finite temperatures are obtained in precisely the same
way as in Appendix \ref{sec:susc}.


\section{Real space form of RKKY interaction}
\label{sec:RKKY_real_space}

The real-space form $J^\alpha(r)$ of the RKKY interaction can be found by time integrating
the susceptibility $\chi_\alpha(r,t)$, given in Eq. \eqref{eq:chi(rt)_T>0}.
This integration was evaluated in the limit $|r| \ll \lambda_T$ in 
Ref. \onlinecite{egger:1996}. Here we determine $\chi_\alpha(r)$ and so $J^\alpha(r)$ 
in the general case.
For convenience we shall drop the index $\alpha=x,y,z$ in this section.

The stationary real space form $\chi(r)$ corresponds to the $\omega=0$
Fourier transform of $\chi(r,t)$, which from Eq. \eqref{eq:chi(rt)_T>0}
is determined by the integral
\begin{align}
	&\chi(r)
	= - \frac{\sin(\pi g) \cos(2k_F x)}{2\pi^2 a^2}
	\int_{|r|/v_F}^\infty dt \
\nonumber \\
	&\times
	\left|\frac{\pi a/\beta v_F}{\sinh\left(\frac{\pi}{\beta v_F}(r-v_F t)\right)}\right|^g
	\left|\frac{\pi a/\beta v_F}{\sinh\left(\frac{\pi}{\beta v_F}(r+v_F t)\right)}\right|^g.
\end{align}
Setting $y = t v_F / |r|$, $\xi = \pi |r|/ \beta v_F$, and using the relation
$\sinh(a+b)\sinh(a-b) = \sinh^2(a)-\sinh^2(b)$ we can rewrite this as
\begin{align}
	&\chi(r)
	= - \frac{\sin(\pi g) \cos(2k_F r)}{2\pi^2 a^2} \left(\frac{\pi a}{\beta v_F}\right)^{2g}
	\frac{|r|}{v_F} 
\nonumber \\
	&\times
\int_1^\infty dy \left|\sinh^2(\xi) - \sinh^2(\xi y)\right|^{-g}.
\end{align}
A further change of variable $y \to z = \sinh^2(\xi)/\sinh^2(\xi y)$ leads to
\begin{align}
	&\chi(r)
	= - \frac{\sin(\pi g) \cos(2k_F r)}{2\pi^2 a^2} \left(\frac{\pi a}{\beta v_F}\right)^{2g}
	\frac{|r|/v_F}{2\xi \sinh^{2g}(\xi)}
\nonumber \\
	&\times
	\int_0^1 dz z^{g-1} (1-z)^{-g} [1+ z/\sinh^2(\xi)]^{-1/2},
\end{align}
which is the standard form for the Gaussian hypergeometric function
[Ref. \onlinecite{abramowitz:1964}, \S 15.3.1],
\begin{equation} \label{eq:hypergeom}
	F(a,b;c;z) = \frac{\Gamma(c)}{\Gamma(b)\Gamma(c-b)} \int_0^1 dz z^{b-1} (1-z)^{c-b-1}(1-wz)^{-a}.
\end{equation}
Hence, noting that $v_F \beta = \lambda_T$ is the thermal length,
\begin{align}
	&\chi(r)
	= - \frac{\sin(\pi g) \cos(2k_F x) \lambda_T}{4\pi^2 a^2 v_F} \left(\frac{\pi a/\lambda_T}{\sinh(\pi |r|/\lambda_T)}\right)^{2g}
\nonumber \\
	&\times F\left(1/2, g; 1; -\sinh^{-2}(\pi |r|/\lambda_T)\right).
\label{eq:chi(r)_app}
\end{align}
The hypergeometric function has the asymptotic expansion for $|w| \to \infty$
[Ref. \onlinecite{abramowitz:1964}, \S 15.3.7]
\begin{align}
	&F(a,b;c;w) = 
	\frac{\Gamma(b-a)\Gamma(c)}{\Gamma(b)\Gamma(c-a)} (-w)^{-a} [1 + O(w^{-1})]
\nonumber\\
	&+
	\frac{\Gamma(a-b)\Gamma(c)}{\Gamma(a)\Gamma(c-b)} (-w)^{-b} [1 + O(w^{-1})].
\label{eq:hypergeom_expansion}
\end{align}
For $g > 1/2$ we obtain from this expansion the
asymptotic behavior for $|r| \ll \lambda_T$ [cf. Ref. \onlinecite{egger:1996}],
\begin{equation}\label{eq:chi(r)_asymp_app}
	\chi(r) \sim - \frac{1}{4\pi a v_F} \frac{\Gamma(g-1/2)}{\Gamma(g)\sqrt{\pi}} \cos(2k_F r) 
	\left(\frac{a}{|r|}\right)^{2g-1}.
\end{equation}
For $g<1/2$ the latter equation must be complemented by further corrections
from Eq. \eqref{eq:hypergeom_expansion}.
The RKKY interaction is eventually determined by $J(r) = A^2 a \chi(r)/2$,
with $r$ running over the sites of the nuclear spin lattice.


\section{Magnon spectrum}
\label{sec:magnon_spectrum}

The magnon spectrum about the helical ground state of the nuclear spins can 
be evaluated by mapping the helical state back onto a ferromagnet. This allows
us to use the standard results\cite{spinwave} for spin waves in anisotropic ferromagnets.
We focus on the $+$ helicity in Eq. \eqref{eq:I_gs}. The case for the 
opposite helicity is equivalent.

The mapping is achieved by noting that the ground state \eqref{eq:I_gs}
can be mapped onto a ferromagnetic alignment
along the spin $\hat{\be}_x$ direction by defining the local transformation
$\tilde{\bI}_i = R_i \hat{\bI}_i$ with 
\begin{equation}
	R_i = 
	\begin{pmatrix}
		 \cos(2 k_F r_i) & \sin(2 k_F r_i) & 0 \\
		-\sin(2 k_F r_i) & \cos(2 k_F r_i) & 0 \\
		 0               & 0               & 1
	\end{pmatrix}.
\end{equation}
In the classical ground state we then have $\hat{\bI}_i \equiv (I N_\perp, 0, 0)$.
The RKKY Hamiltonian \eqref{eq:H_eff_n} becomes in terms of these new spin vectors
\begin{equation}
	H_n^\text{eff} = \sum_{ij,\alpha\beta} \frac{\hat{J}_{ij}^{\alpha\beta}}{N_\perp^2} \hat{I}_i^\alpha \hat{I}_i^\beta,
\end{equation}
with $\hat{J}_{ij}^{\alpha\beta} = \sum_{\gamma} R_{i}^{\gamma\alpha} R_{j}^{\gamma\beta} J_{ij}^\gamma$.
Here the RKKY couplings are generally assumed to be anisotropic as 
$J_{ij}^{x} = J_{ij}^{y} \neq J_{ij}^z$, which covers both the isotropic case
without the feedback, and the anisotropic case with the feedback on the electron system.
In Fourier space the latter Hamiltonian becomes
\begin{align}
	&H_n^\text{eff} =
\nonumber\\
	&\frac{1}{N} \sum_{q\neq 0,\alpha\beta}
	\left[
		\frac{J_{q-2k_F}^x}{N_\perp^2} R_-^{\alpha\beta} 
		+
		\frac{J_{q+2k_F}^x}{N_\perp^2} R_+^{\alpha\beta} 
		+
		\frac{J_q^z}{N_\perp^2}        R_0^{\alpha\beta}
	\right] 
	\hat{I}_{-q}^\alpha \hat{I}_q^\beta,
\label{eq:H_n_eff_tilde}
\end{align}
with
\begin{equation}
	R_+ 
	= R_-^* 
	= \frac{1}{2} 
	  \begin{pmatrix} 1 & i & 0 \\ -i & 1 & 0 \\ 0 & 0 & 0 \end{pmatrix},
	R_0 
	= 
	\begin{pmatrix} 0 & 0 & 0 \\ 0 & 0 & 0 \\ 0 & 0 & 1 \end{pmatrix}.
\end{equation}
The magnon description is obtained by replacing the spin operators $\hat{I}^\alpha_q$
with the Holstein-Primakoff\cite{spinwave} boson operators $a_q$. This leads to
$\hat{I}^x_q = (I N_\perp) N \delta_{q,0} - \frac{1}{N} \sum_p a_{p+q}^\dagger a_p$,
$\hat{I}^y_q = (I N_\perp/2)^{1/2} ( a_{-q}^\dagger + a_q )$, and
$\hat{I}^z_q = -i (I N_\perp/2)^{1/2} ( a_{-q}^\dagger - a_q )$.
Neglecting terms involving more than two $a_q$ operators results in a Hamiltonian
that is equivalent to \eqref{eq:H_n_eff_tilde} up to corrections of order $1/I N_\perp$.
Since $N_\perp \gg 1$ this approximation is very accurate.
We then obtain
\begin{equation}
	H = E_0
	+ \frac{2}{N} \sum_{q>0}
	(a_q^\dagger, a_{-q})
	\begin{pmatrix}
		h^{(1)}_q & h^{(2)}_q
		\\
		h^{(2)}_q & h^{(1)}_q
	\end{pmatrix}
	\begin{pmatrix} a_q \\ a_{-q}^\dagger \end{pmatrix}
\end{equation}
with $E_0 = (N_\perp I) (N_\perp I + 1) N (J_{2k_F}^x/N_\perp^2)$,
$h^{(1)}_q = (-2J_{2k_F}^x + J_{2k_F-q}^x + J_q^z)/N_\perp^2$,
and 
$h^{(2)}_q = (J_{2k_F-q}^x-J_q^z)/N_\perp^2$.
Finally diagonalizing this Hamiltonian leads to two magnon bands
with dispersions
\begin{align}
\label{eq:omega_q_1}
	\omega_q^{(1)} 
	&= 2 I (J_{2k_F-q}^x - J_{2k_F}^x) / N_\perp,
\\
\label{eq:omega_q_2}
	\omega_q^{(2)} 
	&= 2 I (J_{q}^z - J_{2k_F}^x) / N_\perp.
\end{align}
The first magnon band is gapless and dominates the fluctuations
about the nuclear ground state. 
The second magnon band is gapped at $q = 0$ with a gap of the size
$\sim |J^{x}_{2k_F}|$. Since quite generally through the feedback
$|J^{z}_{2k_F}| > |J^{x}_{2k_F}|$ we have, however, $\omega_{q}^{(2)} < 0$
at $q \sim 2k_F$. This normally leads to an instability of the ground state
and so normally means that the assumption of the ordered ground state is 
not valid. However, as shown in Sec. \ref{sec:stability_planar_order}, in the
present case such a destabilization would destroy the feedback that stabilizes
the order and so causes the anisotropy $J^z_q \neq J^x_q$. This destruction 
would have a very high cost in ground state energy. Therefore the occupation 
of the $\omega_q^{(2)} < 0$ modes by a macroscopic magnon number is not possible.
The remaining $\omega_q^{(2)}> 0$ modes can then be captured by the same treatment
as the $\omega_q^{(1)}$ modes. Yet as they involve large momenta $\sim 2k_F$, 
they are neglected in the present treatment.


\section{Magnetization for systems with small level spacing}
\label{sec:small_level_spacing}

For systems with a large enough length $L$ such that the consistency relation
$\Delta_L \gg k_B T^*_0$ (or $T^*$) put forward in Sec. \ref{sec:finite_systems}  
no longer holds, we need to reexamine carefully the derivation of Eqs. \eqref{eq:m_T*_0} and \eqref{eq:m_T*}
for the magnetization $m_{2k_F}$. 
This situation is indeed met for typical GaAs quantum wires (see Table \ref{tab:values}) 
where $\Delta_L \sim k_B T^*_0$. 
In this case the singularity in the magnon 
occupation number in Eq. \eqref{eq:m} may be controlled by the cutoff at 
$q = \pi /L$, i.e. by $\Delta_L$ instead of the $L$-independent scale $k_B T^*_0$ (or $k_B T^*$).
Here we show, however, that this singularity dominates the magnetization
and hence modifies the cross-over temperature only for $L$ that lie many 
orders of magnitude above any realistic length. The result \eqref{eq:m_T*},
therefore, remains valid and robust also for $\Delta_L > k_B T^*_0$,
and in particular also for $\Delta_L \sim k_B T^*$.

To see this we focus on the $q \to 0$ part only in the magnon occupation 
number in Eq. \eqref{eq:m} and neglect the contributions 
$\omega_q \sim 2 I |J_{2k_F}|/N_\perp$ that have led to Eqs. \eqref{eq:m_T*_0} and \eqref{eq:m_T*}
before. For simplicity we discuss only the case without feedback, and note that the conclusion
is the same for the case with the feedback.
For $q < \pi/\lambda_T$ we have 
$\omega_q < 2 I |J_{2k_F}|/N_\perp$ and the dispersion $\omega_q$ can be well
approximated by a parabola, $\omega_q/k_B T = D q^2$, with
$D 
= (k_B T)^{-1} d^2\omega_q/dq^2|_{q=0}
= 2 I \bar{C} v_F^2 (k_B T)^{2g-5}/N_\perp$, and
$\bar{C} = C(g,v_F) \Gamma^4(g/2)[\pi^2 - 2 \sin^2(\pi g/2)\psi_1(g/2)]/16\pi^4$.
Here $\psi_1$ is the polygamma function.

We then ask under which conditions the following approximation holds:
\begin{equation}
	\frac{2}{I N_\perp N} \sum_{q>0} \frac{1}{\e^{\omega_q/k_B T} - 1}
	\overset{?}{\approx}
	\frac{2a }{I N_\perp N} \int_{\pi/L}^{\pi/\lambda_T} \frac{dq}{2\pi} \frac{1}{\e^{D q^2} - 1}.
\end{equation}
At $T < T^*_0 (T^*)$ the integrand at $q = \pi/\lambda_T$ is small and we can push
the upper integral boundary to infinity.
Setting then $x = \sqrt{D} q$ and $X = \sqrt{D} \pi / L = \omega_{\pi/L} / k_B T$ 
we have to evaluate
\begin{align}
	Q(X) 
	&= \int_{X}^\infty \frac{dx}{\e^{x^2}-1}
	= \int_{0}^\infty dx \left[\frac{1}{\e^{x^2}-1}- \frac{1}{x^2}\right] 
\nonumber\\
	&+\int_X^\infty \frac{dx}{x^2}
	- \int_{0}^X dx \left[\frac{1}{\e^{x^2}-1}- \frac{1}{x^2}\right] 
\nonumber\\
	&= - 1.294 + \frac{1}{X} + \frac{X}{2} + \mathcal{O}(X^2).
\end{align}
At the considered temperatures $T<T^*_0$, $X$ is 
a small number, and we can keep the $1/X$ contribution only.
This leads to
\begin{align}
	m_{2k_F} 
	&\overset{?}{\approx} 1 - \frac{2 a}{I L N_\perp \sqrt{D} (\pi/\lambda_T)}
	\frac{1}{\sqrt{\omega_{\pi/L} / k_B T}}
\\
	&=1 - \frac{2 a k_B T}{IL N_\perp \omega_{\pi/L}}
	=1 - \left(\frac{T}{T^*_L}\right)^{5-2g},	
\end{align}
where
\begin{equation}
	k_B T_L^* = 
	\left[\bar{C} I^2 \pi^2 / La \right]^{\frac{1}{5-2g_x}}
\end{equation}
is an $L$-dependent characteristic temperature. This temperature defines the 
cross-over temperature if $T_L^* < T^*_0$. Putting in numbers corresponding
to GaAs quantum wires (see Table \ref{tab:values})
we see, however, that $T_L^* \gg T^*_0$. For $T<T^*_0$,
therefore, the $T_L^*$ dependent contribution to $m_{2k_F}$ is negligible.
Only at entirely unrealistic lengths $L \sim$ 1 m we obtain $T_L^* \sim T^*_0$.
For any realistic sample, therefore, $T^*_L$ and so $L$ do not affect magnetization
and cross-over temperature even for $\Delta_L \lesssim k_B T^*_0$ and
Eq. \eqref{eq:m_T*_0} remains valid. As stated above the same holds for the
case with feedback where $T^*_0$ is replaced by $T^*$ and the magnetization is given by
Eq. \eqref{eq:m_T*}.


\vfill


\end{document}